\documentstyle[twoside,epsf,epsfig]{article}

\catcode`\@=11
\long\def\@makefntext#1{
\protect\noindent \hbox to 3.2pt {\hskip-.9pt  
$^{{\eightrm\@thefnmark}}$\hfil}#1\hfill}		

\def\thefootnote{\fnsymbol{footnote}}
\def\@makefnmark{\hbox to 0pt{$^{\@thefnmark}$\hss}}	
	
\def\ps@myheadings{\let\@mkboth\@gobbletwo
\def\@oddhead{\hbox{}
\rightmark\hfil\eightrm\thepage}   
\def\@oddfoot{}\def\@evenhead{\eightrm\thepage\hfil
\leftmark\hbox{}}\def\@evenfoot{}
\def\sectionmark##1{}\def\subsectionmark##1{}}

\oddsidemargin=\evensidemargin
\renewcommand{\thefootnote}{\fnsymbol{footnote}}

\renewcommand\section{\@startsection {section}{1}{\z@}%
                                   {-3.5ex \@plus -1ex \@minus -.2ex}%
                                   {2.3ex \@plus.2ex}%
                                   {\tenbf}}
\renewcommand\subsection{\@startsection{subsection}{2}{\z@}%
                                     {-3.25ex\@plus -1ex \@minus -.2ex}%
                                     {1.5ex \@plus .2ex}%
                                     {\bf}}
\renewcommand\subsubsection{\@startsection{subsubsection}{3}{\z@}%
                                     {-3.25ex\@plus -1ex \@minus -.2ex}%
                                     {1.5ex \@plus .2ex}%
                                     {\tenrm}}
\newcommand{\nonumsection}[1] {\vspace{12pt}\noindent{\tenbf #1}
	\par\vspace{5pt}}

\newcounter{appendixc}
\newcounter{subappendixc}[appendixc]
\newcounter{subsubappendixc}[subappendixc]
\renewcommand{\thesubappendixc}{\Alph{appendixc}.\arabic{subappendixc}}
\renewcommand{\thesubsubappendixc}
	{\Alph{appendixc}.\arabic{subappendixc}.\arabic{subsubappendixc}}

\renewcommand{\appendix}[1] {\vspace{12pt}
        \refstepcounter{appendixc}
        \setcounter{figure}{0}
        \setcounter{table}{0}
        \setcounter{lemma}{0}
        \setcounter{theorem}{0}
        \setcounter{corollary}{0}
        \setcounter{definition}{0}
        \setcounter{equation}{0}
        \renewcommand{\thefigure}{\Alph{appendixc}.\arabic{figure}}
        \renewcommand{\thetable}{\Alph{appendixc}.\arabic{table}}
        \renewcommand{\theappendixc}{\Alph{appendixc}}
        \renewcommand{\thelemma}{\Alph{appendixc}.\arabic{lemma}}
        \renewcommand{\thetheorem}{\Alph{appendixc}.\arabic{theorem}}
        \renewcommand{\thedefinition}{\Alph{appendixc}.\arabic{definition}}
        \renewcommand{\thecorollary}{\Alph{appendixc}.\arabic{corollary}}
        \renewcommand{\theequation}{\Alph{appendixc}.\arabic{equation}}
        \noindent{\tenbf Appendix \theappendixc #1}\par\vspace{5pt}}
\newcommand{\subappendix}[1] {\vspace{12pt}
        \refstepcounter{subappendixc}
        \noindent{\bf Appendix \thesubappendixc. {\kern1pt \bfit #1}}
	\par\vspace{5pt}}
\newcommand{\subsubappendix}[1] {\vspace{12pt}
        \refstepcounter{subsubappendixc}
        \noindent{\rm Appendix \thesubsubappendixc. {\kern1pt \tenit #1}}
	\par\vspace{5pt}}

\newcommand{\textlineskip}{\baselineskip=13pt}
\newcommand{\smalllineskip}{\baselineskip=10pt}

\newcommand{\copyrightheading}[1]
	{\vspace*{-2.5cm}\smalllineskip{\flushleft
	{\footnotesize International Journal of Modern Physics D, #1}\\
	{\footnotesize \copyright\kern2pt World Scientific Publishing
	 Company}\\
	 }}

\newcommand{\publisher}[2]{{\begin{center}\footnotesize\smalllineskip 
	Received #1\\
	Revised #2
	\end{center}
	}}

\def\abstracts#1#2#3{{
	\centering{\begin{minipage}{4.5in}\footnotesize\baselineskip=10pt
	\parindent=0pt #1\par 
	\parindent=15pt #2\par
	\parindent=15pt #3
	\end{minipage}}\par}} 

\renewenvironment{thebibliography}[1]
        {\frenchspacing
	 \ninerm\baselineskip=11pt
         \begin{list}{\arabic{enumi}.}
        {\usecounter{enumi}\setlength{\parsep}{0pt}     
	 \setlength{\leftmargin 12.7pt}{\rightmargin 0pt}
         \setlength{\itemsep}{0pt} \settowidth
	{\labelwidth}{#1.}\sloppy}}{\end{list}}

\newcounter{itemlistc}
\newcounter{romanlistc}
\newcounter{alphlistc}
\newcounter{arabiclistc}

\newenvironment{romanlist}
	{\setcounter{romanlistc}{0}
	 \begin{list}{$($\roman{romanlistc}$)$}
	{\usecounter{romanlistc}
	 \setlength{\parsep}{0pt}
	 \setlength{\itemsep}{0pt}}}{\end{list}}

\newcommand{\fcaption}[1]{
        \refstepcounter{figure}
        \setbox\@tempboxa = \hbox{\footnotesize Fig.~\thefigure. #1}
        \ifdim \wd\@tempboxa > 5in
           {\begin{center}
        \parbox{5in}{\footnotesize\smalllineskip Fig.~\thefigure. #1}
            \end{center}}
        \else
             {\begin{center}
             {\footnotesize Fig.~\thefigure. #1}
              \end{center}}
        \fi}

\newcommand{\tcaption}[1]{
        \refstepcounter{table}
        \setbox\@tempboxa = \hbox{\footnotesize Table~\thetable. #1}
        \ifdim \wd\@tempboxa > 5in
           {\begin{center}
        \parbox{5in}{\footnotesize\smalllineskip Table~\thetable. #1}
            \end{center}}
        \else
             {\begin{center}
             {\footnotesize Table~\thetable. #1}
              \end{center}}
        \fi}

\def\@citex[#1]#2{\if@filesw\immediate\write\@auxout
	{\string\citation{#2}}\fi
\def\@citea{}\@cite{\@for\@citeb:=#2\do
	{\@citea\def\@citea{,}\@ifundefined
	{b@\@citeb}{{\bf ?}\@warning
	{Citation `\@citeb' on page \thepage \space undefined}}
	{\csname b@\@citeb\endcsname}}}{#1}}

\newif\if@cghi
\def\cite{\@cghitrue\@ifnextchar [{\@tempswatrue
	\@citex}{\@tempswafalse\@citex[]}}
\def\citelow{\@cghifalse\@ifnextchar [{\@tempswatrue
	\@citex}{\@tempswafalse\@citex[]}}
\def\@cite#1#2{{$\null^{#1}$\if@tempswa\typeout
	{IJCGA warning: optional citation argument 
	ignored: `#2'} \fi}}

\def\pmb#1{\setbox0=\hbox{#1}
	\kern-.025em\copy0\kern-\wd0
	\kern.05em\copy0\kern-\wd0
	\kern-.025em\raise.0433em\box0}

\def\fnm#1{$^{\mbox{\scriptsize #1}}$}
\def\fnt#1#2{\footnotetext{\kern-.3em
	{$^{\mbox{\scriptsize #1}}$}{#2}}}

\def\fpage#1{\begingroup
\voffset=.3in
\thispagestyle{empty}\begin{table}[b]\centerline{\footnotesize #1}
	\end{table}\endgroup}

\def\runninghead#1#2{\pagestyle{myheadings}
\markboth{{\protect\footnotesize\it{\quad #1}}\hfill}
{\hfill{\protect\footnotesize\it{#2\quad}}}}
\font\tenrm=cmr10
\font\tenit=cmti10 
\font\tenbf=cmbx10
\font\bfit=cmbxti10 at 10pt
\font\ninerm=cmr9

\font\eightrm=cmr8

\def\qed{\hbox{${\vcenter{\vbox{	          
   \hrule height 0.4pt\hbox{\vrule width 0.4pt height 6pt
   \kern5pt\vrule width 0.4pt}\hrule height 0.4pt}}}$}}

\renewcommand{\thefootnote}{\fnsymbol{footnote}}  

\newcommand\lsim{\mathrel{\rlap{\lower4pt\hbox{\hskip1pt$\sim$}}
    \raise1pt\hbox{$<$}}}
\newcommand\gsim{\mathrel{\rlap{\lower4pt\hbox{\hskip1pt$\sim$}}
    \raise1pt\hbox{$>$}}}

\begin{document}

\setlength{\textheight}{7.8truein}    

\runninghead{Cosmic Strings, Loops, and Linear Growth of Matter Perturbations} {J.\ H.\ P.\ Wu et al.}

\normalsize\textlineskip
\thispagestyle{empty}
\setcounter{page}{1}

\copyrightheading{}		

\vspace*{0.88truein}

\fpage{1}
\centerline{\bf COSMIC STRINGS, LOOPS, AND}
\vspace*{0.035truein}
\centerline{\bf LINEAR GROWTH OF MATTER PERTURBATIONS}
\vspace*{0.37truein}
\centerline{\footnotesize Jiun-Huei Proty Wu
}
\vspace*{0.015truein}
\centerline{\footnotesize\it Astronomy Department,
  University of California, Berkeley,
  601 Campbell Hall,}
\baselineskip=10pt
\centerline{\footnotesize\it Berkeley, CA 94720-3411, USA}
\vspace*{10pt}
\centerline{\footnotesize Pedro P.\ Avelino}
\vspace*{0.015truein}
\centerline{\footnotesize\it Centro de Astrof{\' \i}sica, Universidade do Porto,
  Rua das Estrelas s/n,}
\baselineskip=10pt
\centerline{\footnotesize\it  4150-762, Porto, Portugal}
\baselineskip=10pt
\centerline{\footnotesize\it 
  Dep. de F{\' \i}sica da Faculdade de Ci\^encias da Univ. do Porto,
  Rua do Campo Alegre 687,}
\baselineskip=10pt
\centerline{\footnotesize\it  4169-007, Porto, Portugal}
\vspace*{10pt}
\centerline{\footnotesize E.\ P.\ S.\ Shellard}
\vspace*{0.015truein}
\centerline{\footnotesize\it 
Department of Applied Mathematics and Theoretical Physics,
University of Cambridge, Silver Street,}
\baselineskip=10pt
\centerline{\footnotesize\it  Cambridge CB3 9EW, UK}
\vspace*{10pt}
\centerline{\footnotesize  Bruce Allen}
\vspace*{0.015truein}
\centerline{\footnotesize\it 
Department of Physics, University of Wisconsin--Milwaukee
P.O. Box 413, Milwaukee,}
\baselineskip=10pt
\centerline{\footnotesize\it Wisconsin 53201, U.S.A.}
\vspace*{0.225truein}
\publisher{(received date)}{(revised date)}

\baselineskip 22pt 

\vspace*{0.21truein}
\abstracts{
We describe a detailed study of string-seeded structure formation 
using high resolution numerical simulations in open universes
and those with a non-zero cosmological constant. 
We provide a semi-analytical model which can reproduce these 
simulation results including the effect from small loops chopped 
of by the string network. A detailed study of cosmic string network 
properties regarding structure formation is also given, including 
the correlation time,
the topological analysis of the source spectrum,
the correlation between long strings and loops,
and
the evolution of long-string and loop energy densities.
For models with $\Gamma\,=\,\Omega h\,=\,0.1$--$0.2$ and a cold dark 
matter background, we show that the linear density fluctuation power
spectrum induced by cosmic strings has both an amplitude at $8 h^{-1}$Mpc, 
$\sigma_8$, and an overall shape which are consistent within 
uncertainties with those currently inferred from galaxy surveys.
The cosmic string scenario with hot dark matter 
requires a strongly scale-dependent bias in order
to agree with observations.
}{}{}



\setcounter{footnote}{0}
\renewcommand{\thefootnote}{\alph{footnote}}

\section{Introduction}
\label{introduction}
\noindent
One of the outstanding problems in cosmology today is developing
a more precise understanding of structure formation in the universe,
that is, the origin of galaxies and other large scale structures.
Existing theories for the structure formation of the Universe
fall into two categories, based
either upon the amplification of quantum fluctuations in a scalar field
during inflation,
or upon a symmetry breaking phase transition in the early Universe
which leads to the formation of topological defects.
While techniques for computing density perturbations for the former
are well established,\cite{BonEfs,Efs,LidLyt}
only little quantitative work exists for the latter
due to calculational difficulties in modeling nonlinear effects,
especially for cosmic string models.\cite{VilShe}

The cosmic string scenario predated inflation as
a realistic structure formation model,\cite{VilShe}
 but it has proved computationally
much more challenging to make robust predictions with which
to confront observations.
Only until recently,
significant progress in understanding
cosmic strings as seeds for large-scale 
structure and Cosmic Microwave Background Radiation (CMBR) anisotropies
has been achieved.\cite{PST,AveShe2,AveShe4,AveShe3,AveCal1,against,BattyeLam,ACDKSS,ConHin}
In this paper,
we concentrate on local cosmic strings and
describe a self-consistent method
based on a fluid approximation
to study string-seeded structure formation with either a cold or hot
dark matter (CDM or HDM) background.
The primary quantities of interest for comparing theories to observations are
the power spectra of fluctuations, both for the mass density and the CMBR.
In the work presented here, we concentrate only on the former
but we will briefly discuss the latter.
The present work relies on high-resolution numerical simulations of a
cosmic string
network,\cite{AS1} with a dynamic range extending from well before the
matter-radiation transition through to deep into the matter era.
Several important properties of local cosmic strings are revealed.
The resulting power spectrum of linear density perturbations, ${\cal P}(k)$,
and the mass fluctuation amplitude at $8 h^{-1}$Mpc, $\sigma_8$,
are calculated.
A semi-analytic model is also introduced. It can reproduce the ${\cal P}(k)$
of our high-resolution simulations.
We further investigate the dependence of ${\cal P}(k)$ on the curvature $K$
and the cosmological constant $\Lambda$.
In particular,
for models with $\Gamma\,=\,\Omega h\,=\,0.1$--$0.2$ and a CDM
background, we show that
both $\sigma_8$ and the overall shape of ${\cal P}(k)$
are consistent within uncertainties with 
those currently inferred from galaxy surveys.
The HDM scenario with cosmic
strings seems to require a large scale-dependent biasing
in order to be consistent with observations.

The framework developed here
is also suitable for investigating any other matter sources behaving stiffly,
whose evolution is largely independent of the background matter and any
inhomogeneities in the universe.
This
direct numerical approach marks a considerable quantitative advance by 
incorporating important aspects of the relevant physics not included in 
previous treatments. As such, the cosmic string power spectra presented here
should be the most reliable to date.

The structure of the paper is as follows:
In section \ref{perequ}, we investigate the linear perturbation equations
for cosmic defect models
in a flat $\Lambda$-universe.
We then argue that this can be extended
to closed and open universes for cosmic defect models.
Section \ref{persou} starts by investigating the topology of generic sources.
Then several important string properties are described, 
including the slow relaxation of the long-string and loop energy 
densities from the radiation to the matter era,
the topology of string network,
the string network correlation time, and
the correlation between long strings and loops, etc.
In section \ref{appsch},
we describe the approximation schemes invoked in this work,
including the compensation of the source into the background,
the inclusion of HDM,
a semi-analytical model which can accurately reproduce the simulation results,
and a simple extrapolation scheme which can generalize the simulation
results from a flat $\Lambda=0$ model to open and flat $\Lambda$-models
with any desired Hubble parameter.
This simple scheme is then numerically verified with high accuracy.
The unimportance of the late-time non-scaling behavior of strings
in open or $\Lambda$-models is addressed.
In section \ref{resdis}, we present the main results.
Empirical formulae describing our main results are presented here.
A brief discussion regarding the CMBR anisotropies is also given.
Finally, a conclusion is given in section \ref{conclu}.
We have defined the conventions of the background cosmology and
given the analytical solutions for open, flat, close and $\Lambda$-models in
Appendix A. 
Appendix \ref{conpow}  defines the conventions of our power spectrum
and variance calculations.
Through out the paper, we will use $h=0.7$,
where the Hubble parameter is defined in the usual way
$H=100h {\rm km s^{-1} Mpc^{-1}}$.
A generalization of results from this choice to any other $h$ is provided
in section \ref{depome}.


\section{{Perturbation Equations}}
\label{perequ}

\noindent
We consider density perturbations about a flat FRW model which are causally
sourced by an evolving external-source network with energy-momentum tensor
$\Theta_{\alpha \beta}({\bf x}, \eta)$.
In the synchronous gauge with a cosmological constant, the
linear evolution of the radiation and CDM perturbations,
$\delta_{\rm r}$ and $\delta_{\rm c}$ respectively, are given
by (modified from Ref.\cite{VeeSte})
\begin{eqnarray}
  \ddot \delta_{\rm c} + {\dot a \over a} \dot \delta_{\rm c} -
  {3 \over 2}\Big({\dot a \over a}\Big)^2 \,
  \left(
    \Omega_{\rm c}\delta_{\rm c} + 2 \Omega_{\rm r}\delta_{\rm r}
  \right)
  = 4 \pi G \Theta_+,
  \label{two} \\
  \ddot \delta_{\rm r} - {1 \over 3} \nabla^2 \delta_{\rm r}
  - {4 \over 3}
  \ddot \delta_{\rm c} = 0,~~~~~~~~~~
  \label{three}
\end{eqnarray}
where $a(\eta)$ is the scale factor,
$\Theta_+ =  \Theta_{00} +\Theta_{ii}$,
a dot represents a derivative with respect to the conformal time $\eta$,
and
\begin{equation}
  \label{omega_cr}
  \Omega_{\rm c} =
  {a \over 1 + a + Ba^2 + Ca^4},
  \quad
  \Omega_{\rm r} =
  {1 \over 1 + a + Ba^2 + Ca^4},
\end{equation}
with $B$ and $C$ defined in (\ref{aABC}).
Eqns.~(\ref{two}) and (\ref{three}) are exact when $K=0$,
i.e.\ $\Omega_{K0}=0$. However, they are still a very good approximation 
even if $K \neq 0$ because most perturbations on scales of interest 
are generated  when the curvature-associated perturbation terms are 
small.

It proves useful to split these linear perturbations into
initial (I) and subsequent (S) parts,\cite{VeeSte}
\begin{equation}
\delta_N ({\bf x}, \eta) = \delta_N^{\rm I}({\bf x}, \eta)
+ \delta_N^{\rm S}({\bf x}, \eta)\,,\qquad N={\rm c, r}\,,
\label{IS}
\end{equation}
with initial conditions
\begin{eqnarray}
  \delta_N^{\rm I}(\eta_{\rm i})=\delta_N(\eta_{\rm i}),
  \quad
  \dot{\delta}_N^{\rm I}(\eta_{\rm i})=\dot{\delta}_N(\eta_{\rm i}),
  \label{inic_deltaI}
  \\
  \delta_N^{\rm S}(\hat{\eta})=\dot{\delta}_N^{\rm S}(\hat{\eta})=0,
  \quad {\rm with} \, \eta>\hat{\eta}>\eta_{\rm i} \,.
  \label{inic_deltaS}
\end{eqnarray}
In the case of topological defects,
the initial perturbations $\delta^{\rm I}({\bf x}, \eta)$ depend on the defect
configuration at the defect formation time $\eta_{\rm i}$,
because ultimately the
formation of defects creates under-densities in the initially homogeneous
background out of which they are carved.
The subsequent perturbations $\delta^{\rm S}({\bf x}, \eta)$ are those
actively generated by the defects themselves
for $\hat{\eta}>\eta_{\rm i}$.
Because defects induce iso-curvature perturbations,
$\delta^{\rm I}({\bf x}, \eta)$
must compensate  $\delta^{\rm S}({\bf x}, \eta)$ on comoving scales
$|{\bf x}-{\bf x}'|>\eta$ to prevent acausal fluctuation growth on
super-horizon scales, as we shall discuss later.

The system of eqns.~(\ref{two}) and (\ref{three}) can be solved for the
subsequent perturbations  $\delta^{\rm S}({\bf x},\eta)$
by using a
discretized version of the integral equation with Green functions:
\begin{equation}
  \delta^{\rm S}_N({\bf x},\eta)
  = 4 \pi G \int_{\eta_{\rm i}}^{\eta} \!\! d\hat{\eta}
  \int d^3x'\: {\cal G}_N (X;\eta,\hat{\eta}) \Theta_{+}({\bf x'},\hat{\eta}),
  \qquad N={\rm c,r}\,,
  \label{delta_S_N}
\end{equation}
where $X=|{\bf x} -{\bf x'}|$.
The easiest method of obtaining the Green function solutions is to go to
Fourier space and solve the resulting homogeneous system of
ordinary differential equations
numerically with appropriate initial conditions.
Since the Green functions depend only on
the modulus of $X=|{\bf x}-{\bf x'}|$, it follows that their Fourier
amplitudes must only depend on the modulus of ${\bf k}$. We will use
a tilde~$\;\widetilde{ }\;$~to denote the Fourier transform of a function.
With the change of variable $y=1+(\sqrt{2}-1)\eta/\eta_{\rm eq}$ and using
(\ref{delta_S_N}), in Fourier space
eqns.~(\ref{two}) and (\ref{three}) with $K=\Lambda=0$ become
\begin{eqnarray}
  (1-y^2){\widetilde{\cal G}}_{\rm c}''-2y{\widetilde{\cal G}}_{\rm c}'+
  \left[ 6 -
    \frac{12{\widetilde{\cal G}}_{\rm r}/{\widetilde{\cal G}}_{\rm c}}{1-y^2}
  \right]{\widetilde{\cal G}}_{\rm c} = 0\ ,
  \label{G_one}\\
  \widetilde{\cal G}_{\rm r}''
  - {4\over 3}\widetilde{\cal G}_{\rm c}''
  + {4 k^2 \over 3A^2} \widetilde{\cal G}_{\rm r} = 0\ ,
  \label{G_two}
\end{eqnarray}
with $A$ defined by (\ref{aABC}), and
initial conditions at $\eta=\hat{\eta}$ (or $\hat{y}=y(\hat{\eta})$) satisfying
\begin{equation}
  \label{inic_G}
  \widetilde{\cal G}_{\rm r}=\widetilde{\cal G}_{\rm c}=0,\
  \widetilde{\cal G}'_{\rm r}=\frac{4}{3}\widetilde{\cal G}'_{\rm c}
  =\frac{8}{3A} \,.
\end{equation}
Here, a prime represents a derivative with respect to $y$
(though only in these equations).

\begin{figure}[htbp]
\vspace*{13pt}
  \centering\epsfig{figure=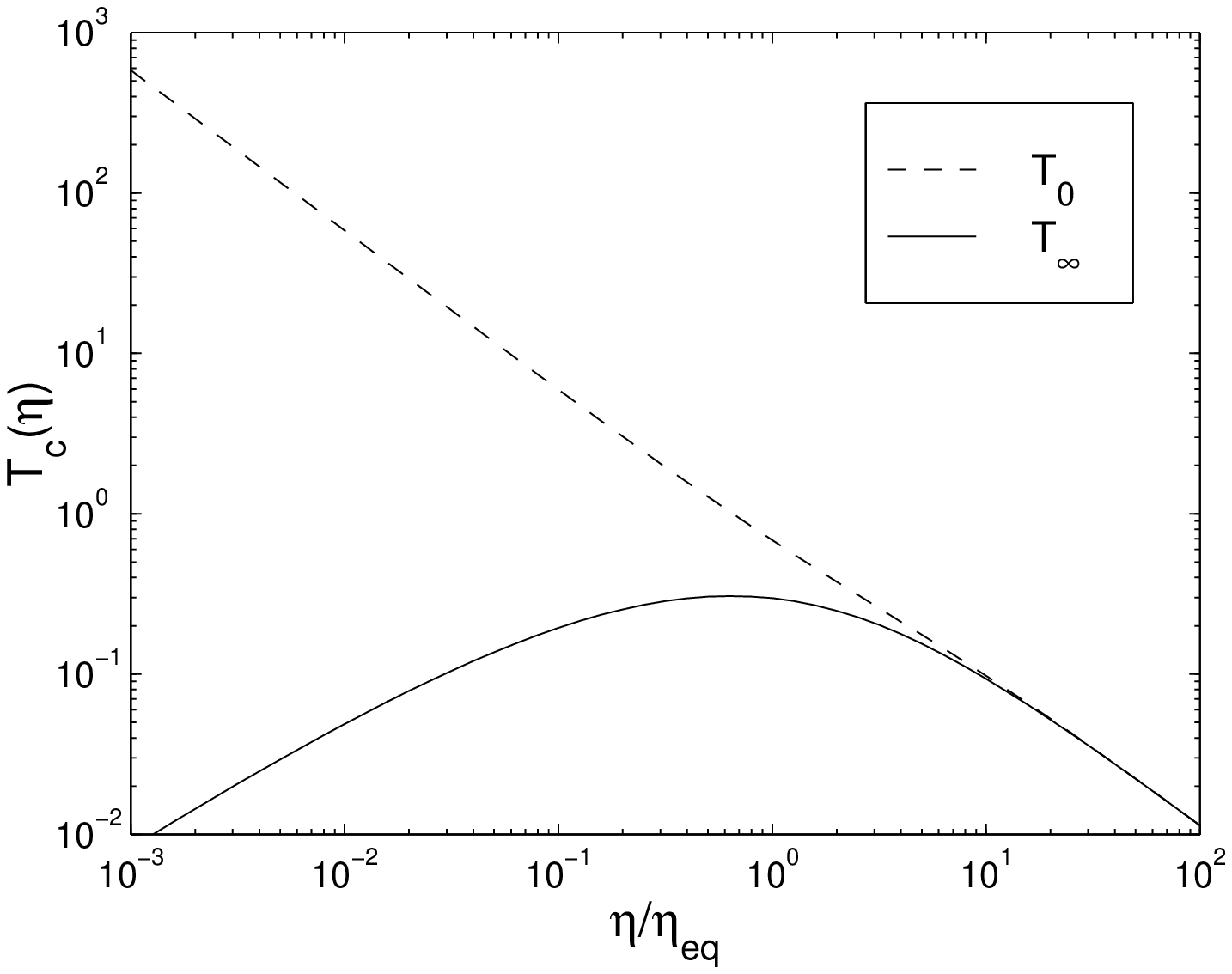, width=4in}\\
\vspace*{13pt}
  \fcaption
  {The evolution of the functions
    $T_{\rm c}(\eta) \equiv \widetilde{T}_{\rm c}(k; \hat{\eta})$
    under the two limits:
    $k\eta\ll 1$ (dashed line), and $k\eta\gg 1$ (solid line).
    }
  \label{T_0infty}
\end{figure}

\begin{figure}[htbp]
\vspace*{13pt}
  \centering\epsfig{figure=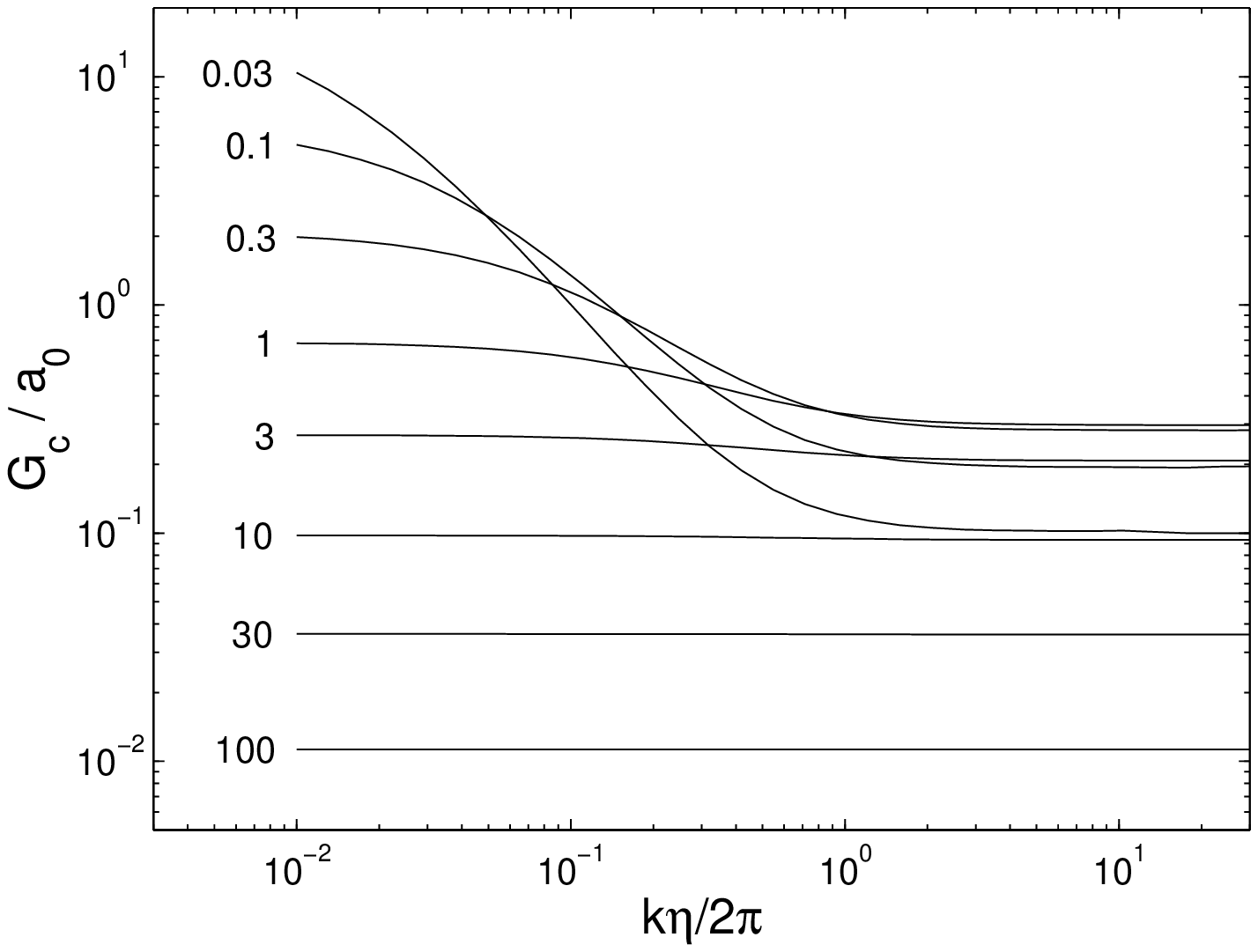, width=4in}\\
\vspace*{13pt}
  \fcaption
  {The numerical solutions of
    $\widetilde{\cal G}_{\rm c}(k ; \eta_0, \hat{\eta})$.
    In the plot, $\eta \equiv \hat{\eta}$, and
    $G_{\rm c} \equiv \widetilde{\cal G}_{\rm c}(k ; \eta_0, \hat{\eta})$.
    The numbers attached to each lines indicate $\hat{\eta}/\eta_{\rm eq}$.
    }
  \label{Gkfig}
\end{figure}

Under certain limits, (\ref{G_one}) and (\ref{G_two}) can be solved analytically.
When $k\eta \gg 1$, the radiation component will oscillate many times per
expansion time and will have little net effect on the matter.
So we can set $\widetilde{\cal G}_{\rm r}=0$ in this case
and thus (\ref{G_one}) decays to a homogeneous
associated Legendre equation, whose solution is the linear combination
of the two associated Legendre polynomials $P_2$ and $Q_2$.
When $k\eta \ll 1$, by (\ref{G_two}) and (\ref{inic_G}) we know
$\widetilde{\cal G}_{\rm r}/\widetilde{\cal G}_{\rm c} = 4/3$.
So (\ref{G_one}) decays to
a homogeneous associated Legendre equation, whose solution is the linear
combination of the two associated Legendre polynomials $P_2^{-4}$ and $Q_2^4$.
One can therefore solve (\ref{G_one}) and (\ref{G_two}) with (\ref{inic_G})
for $\widetilde{\cal G}_{\rm c}$ under these limits to get:
\begin{equation}
  \label{G_c}
  \widetilde{\cal G}_{\rm c}(k ; \eta, \hat{\eta})=
  \left\{
    \begin{array}{ll}
      \frac{2}{A}(\hat{y}^2-1)
      [{\cal A}(\hat{y}){\cal B}(y)-{\cal B}(\hat y){\cal A}(y)] \,, &
      k\eta \gg 1 \,,\\
      \frac{2}{A}(y^2-1)^{-1}
      [{\cal C}(\hat{y}){\cal D}(y)-{\cal D}(\hat y){\cal C}(y)] \,, &
      k\eta \ll 1 \,,
    \end{array}
  \right.
\end{equation}
where
\begin{eqnarray}
  {\cal A}(y) & = & \left(\frac{3y^2-1}{4}\right)
  \log{\left(\frac{y+1}{y-1}\right)}-\frac{3y}{2}\,,
  \label{G_cA} \\
  {\cal B}(y) & = & \frac{3y^2-1}{2}\,,
  \label{G_cB} \\
  {\cal C}(y) & = & \frac{y}{y^2-1}\,,
  \label{G_cC} \\
  {\cal D}(y) & = & \frac{(y-1)^3(y^2+4y+5)}{5(y+1)}\,.
  \label{G_cD}
\end{eqnarray}
We notice that the scale factor $a(\eta)=y^2-1$.
Because we are only interested in the matter perturbations at
late times (i.e.\ today $\eta_0$) and we know
from (\ref{G_one}) that $\widetilde{\cal G}_{\rm c} \propto a/a_{\rm eq}=a$
when $\eta/\eta_{\rm eq} \rightarrow \infty$,
what we actually need is the function:
\begin{equation}
  \label{T_limit}
  \widetilde{T}_{\rm c}(k; \hat{\eta})
  \equiv \lim_{\eta/\eta_{\rm eq} \rightarrow \infty}
  \frac{1}{a} \widetilde{\cal G}_{\rm c}(k; \eta, \hat{\eta})\, .
\end{equation}
Hence for $k \eta \rightarrow 0 \;{\rm and}\; \infty$,
we obtain from (\ref{G_c}) that
\begin{eqnarray}
  \widetilde{T}_\infty(y(\hat{\eta})) & = &
  \frac{3}{A}(\hat{y}^2-1) {\cal A}(\hat{y})\,,\quad
   k \eta \gg 1\,,
  \label{T_infty} \\
  \widetilde{T}_0(y(\hat{\eta})) & = &
  \frac{2}{5A} {\cal C}(\hat{y})\,,\quad
   k \eta \ll 1\,.
  \label{T_0}
\end{eqnarray}
Fig.~\ref{T_0infty} shows how (\ref{T_infty}) and (\ref{T_0}) evolve with time.
Fig.~\ref{Gkfig} shows the numerical solutions of
$\widetilde{\cal G}_{\rm c}(k ; \eta_0, \hat{\eta})$.
It confirms the asymptotic behaviors in these two regimes:
on super-horizon scales ($k \ll 2\pi/\eta$),
$\widetilde{\cal G}_{\rm c}(k ; \eta_0, \hat{\eta})$
scales as $\hat{\eta}^{-1}$,
which is indicated by (\ref{T_0});
on sub-horizon scales ($k \gg 2\pi/\eta$),
the growth of matter perturbations has a maximum at
$\hat{\eta} \sim \eta_{\rm eq}$,
as seen in (\ref{T_infty}) (see also Fig.~\ref{T_0infty}).
We note that on sub-horizon scales, when going from
$\hat{\eta} = 0.05 \eta_{\rm eq}$ to $\hat{\eta} = 5 \eta_{\rm eq}$,
$\widetilde{\cal G}_{\rm c}(k ; \eta_0, \hat{\eta})$
changes only within a factor of $2$.
Adding the fact that 
cosmic defects seed matter perturbations only on sub-horizon modes
because on super-horizon scales they are compensated by the
background and therefore can not create density perturbations,
we see from Fig.~\ref{T_0infty} and \ref{Gkfig}
that defects induce power into matter perturbations mainly
during the radiation-matter transition regime.
This is generically different from inflationary models,
in which matter perturbations are seeded during inflation
in the deep radiation regime
when all the modes are well outside the horizon.

\begin{figure}[htbp]
\vspace*{13pt}
  \centering\epsfig{figure=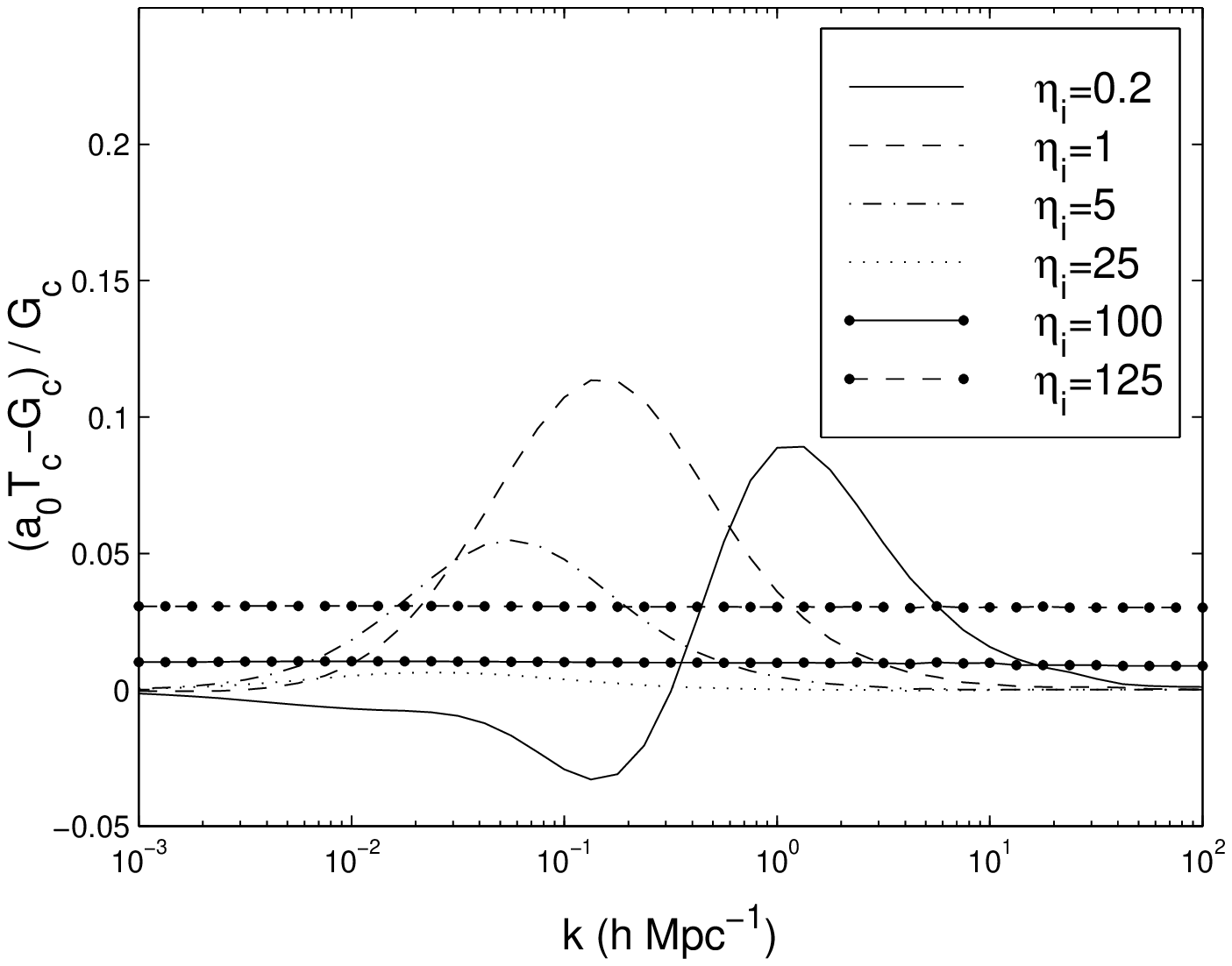, width=4in}\\
  \centering\epsfig{figure=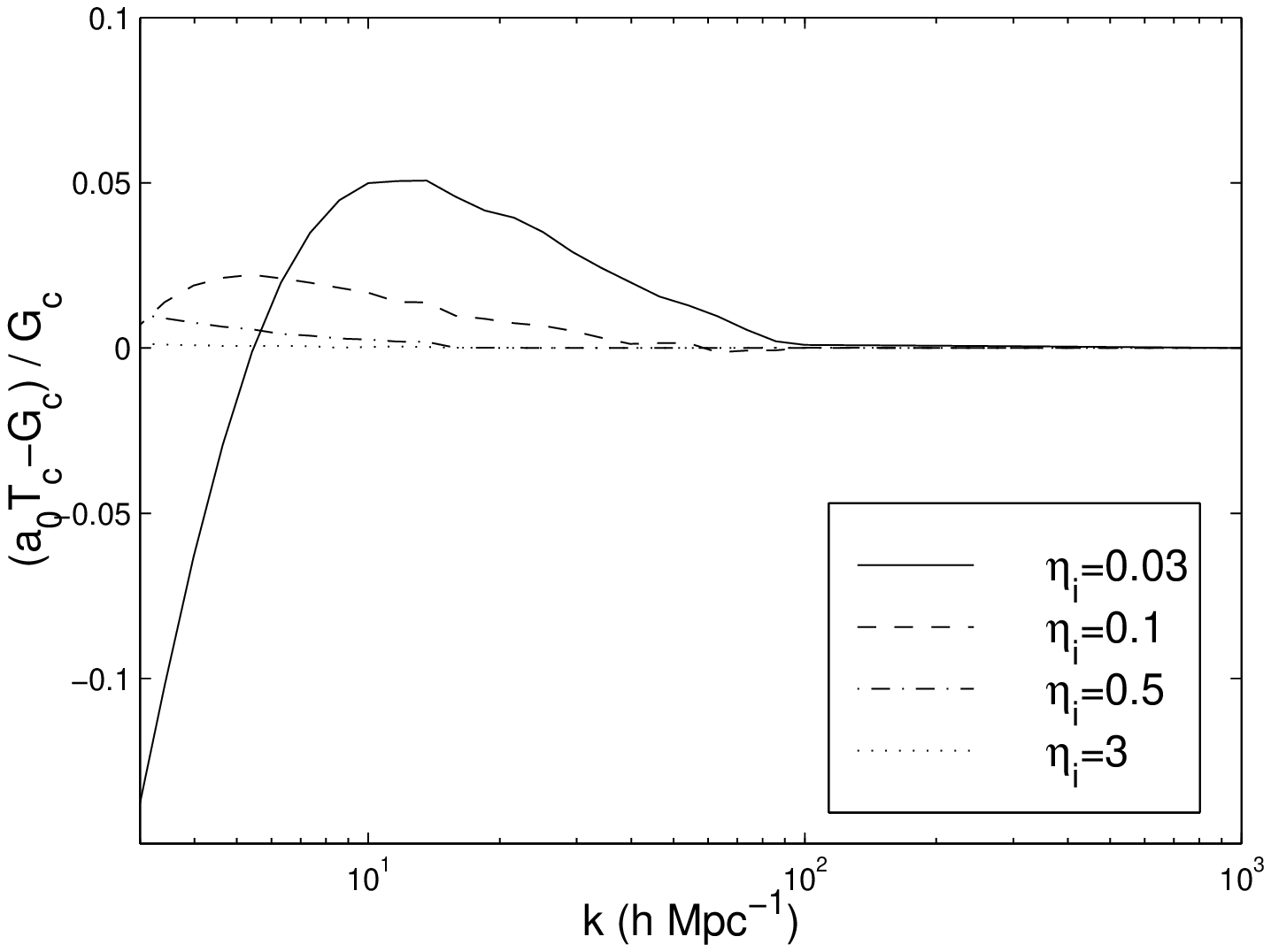, width=4in}\\
\vspace*{13pt}
  \fcaption
  {The accuracy of eq.~(\ref{T_k}).
    $T_{\rm c}$ is equivalent to $\widetilde{T}_{\rm c}(k; \hat{\eta})$
    in (\ref{T_k}),
    while $G_{\rm c} \equiv \widetilde{\cal G}_{\rm c}(k; \eta_0, \hat{\eta})$
    is numerically calculated from (\ref{two}) and (\ref{three})
    with initial conditions (\ref{inic_G}).
    In the legends, $\eta_{\rm i}\equiv \hat{\eta}/\eta_{\rm eq}$.
    }
  \label{T_acc}
\end{figure}

Now we can obtain an approximated solution for the Green function
by combining these two modes:
\begin{equation}
  \label{T_k}
  \begin{array}{lll}
    \widetilde{\cal G}_{\rm c}(k; \eta_0, \hat{\eta}) &
    \approx 
    a_{\scriptscriptstyle 0}
    \widetilde{T}_{\rm c}(k; \hat{\eta})  \vspace*{2mm}\\
    &
    \approx 
    a_{\scriptscriptstyle 0}
    \left\{
      \widetilde{T}_{\infty}+
      (\widetilde{T}_0-\widetilde{T}_{\infty})
      \left[
        1 + (\alpha k\eta_{\rm eq})^{4/3}
      \right]^{-1}
    \right\}   \vspace*{2mm}\\
    &
    {\rm with}\ 
    \left\{
      \begin{array}{ll}
        \alpha=0.75, & {\rm for\ 0.03<\hat{\eta}/\eta_{\rm eq}<125
          \ and\ 3 {\it h}Mpc^{-1} < k},\\
        \alpha=0.5,  & {\rm for\ 0.2<\hat{\eta}/\eta_{\rm eq}}<125.
      \end{array}
    \right.
  \end{array}
\end{equation}
The combination factor in (\ref{T_k}) is a numerically verified guess
and its accuracy is plotted in Fig.~\ref{T_acc}.
We note that the late-time positive departure of the accuracy
(e.g.\ $\hat{\eta}=100,\,125\eta_{\rm eq}$, see Fig.~\ref{T_acc})
is caused by neglecting the negative terms associated with ${\cal B}(\hat{y})$
and ${\cal D}(\hat{y})$ in (\ref{G_c})
when taking the limit (\ref{T_limit}) to get (\ref{T_infty}) and (\ref{T_0}).

Finally, we obtain the subsequent perturbations in Fourier space today
by integrating
the Green function with the external source throughout the dynamical range
within which the external source is present:
\begin{equation}
  \label{delta_T}
  \widetilde{\delta}_{\rm c}^{\rm S}({\bf k}, \eta_0)
  =
  4\pi G
  \int_{\eta_{\rm i}}^{\eta_0}
    \widetilde{\cal G}_{\rm c}(k; \eta_0, \hat{\eta})\widetilde{\Theta}_+({\bf k},\hat{\eta})
    d\hat{\eta}.
\end{equation}
Here the Green function $\widetilde{\cal G}_{\rm c}(k; \eta_0, \hat{\eta})$
can be either (\ref{T_k}), or numerically obtained from 
eqns.~(\ref{G_one}) and (\ref{G_two}) with initial conditions (\ref{inic_G}).

\begin{figure}[htbp]
\vspace*{13pt}
  \centering\epsfig{figure=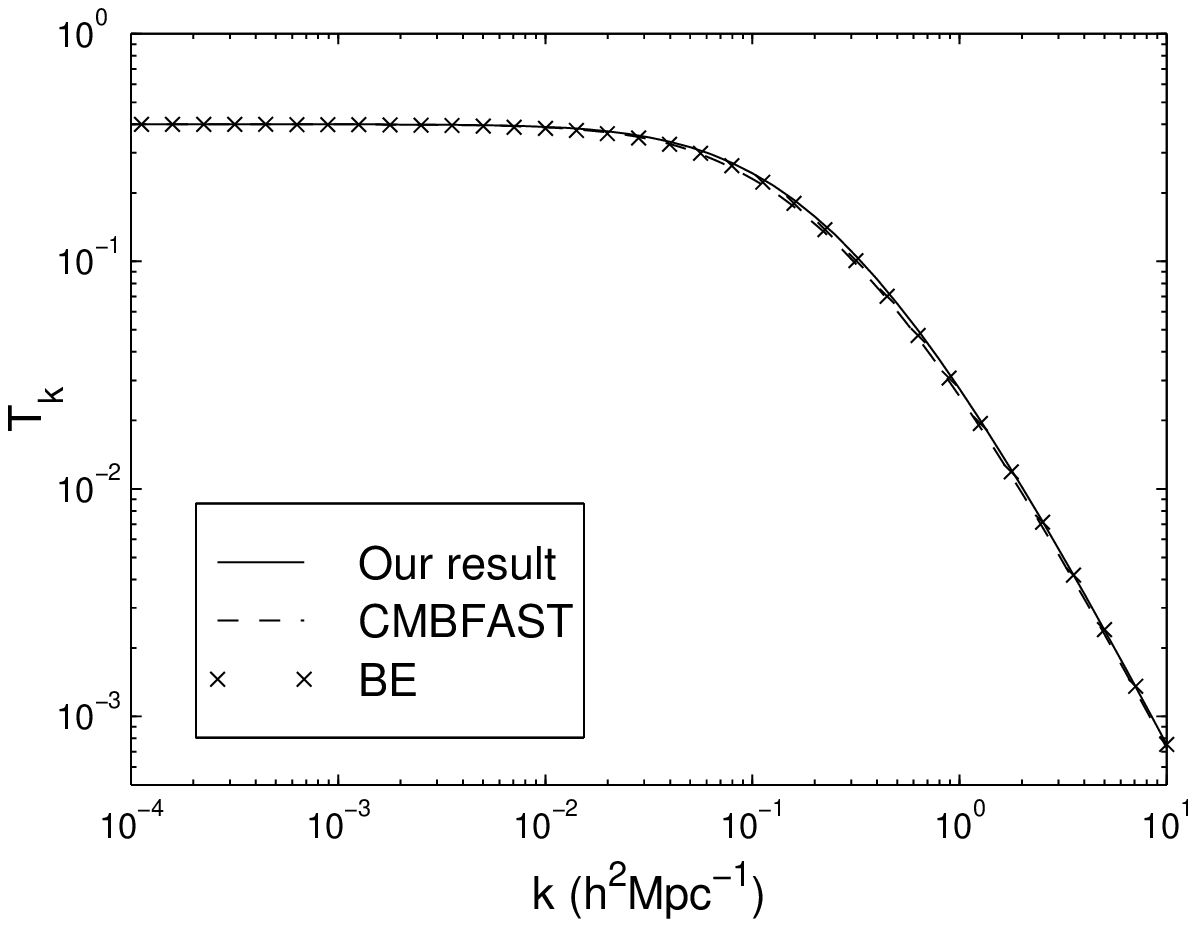, width=3.8in}\\
\vspace*{13pt}
 \fcaption
  {The transfer function of standard CDM model.
    The solid line is our result using the fluid approximation
    (\ref{two}) and (\ref{three});
    the dashed line is from CMBFAST,\cite{cmbfast}
    with $\Omega_{\rm c0}=0.97$, $\Omega_{\rm b}=0.03$, and $h=0.75$;
    the crosses show the fit by Bond and Efstathiou (BE),\cite{BE}
    with the same parameter choice.
    }
  \label{transTfig}
\end{figure}

We also tested the performance of
the fluid approximation (\ref{two}) and (\ref{three}),
by computing the ``transfer function'' 
for primordial density perturbations in the absence of a source ($T_k$), 
defined as the ratio of the growing mode coefficient
deep in the matter era to that deep in the radiation era.
In Fig.~\ref{transTfig}, we compare our result with other previous work for the 
case of a flat universe with no cosmological constant.
We see that although we have used the fluid approximation to model only the
matter and radiation components, our result is only slightly larger
with a maximum deviation being $5\%$ from the fit given by
Bond and Efstathiou (BE),\cite{BE}
and $8\%$ from the result from CMBFAST,\cite{cmbfast}
on the scale $k\sim 0.5 h^2{\rm Mpc}^{-1}$. We have also verified that the 
performance of the fluid approximation is very weakly dependent on the 
choice of cosmological parameters.

\section{{The Perturbation Source---Cosmic Strings}}
\label{persou}
\subsection{{\bfit Topological analysis}}
\label{funa}
\begin{figure}[htbp]
\vspace*{13pt}
  \centering\epsfig{figure=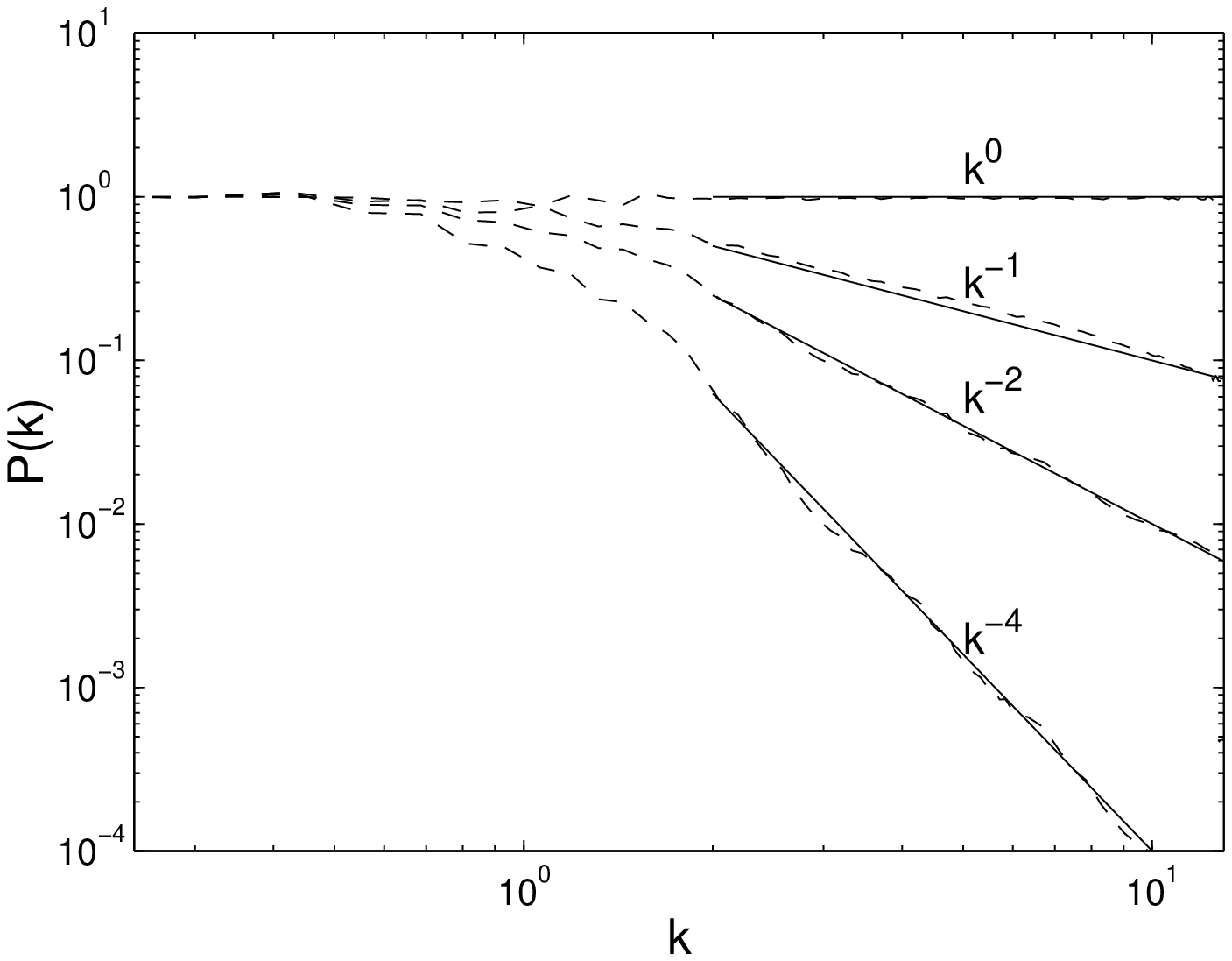, width=4in}\\
\vspace*{13pt}
  \fcaption
  {The power spectra of randomly distributed
    zero, one, two and three dimensional objects
    (dashed lines, downwards).
    Their characteristic sizes have been normalized to $2\pi$,
    and we have also used
    (\ref{deconvolve}) to deconvolve the lattice effect.
    These spectra agree very well on small scales
    with those solid lines, which have exact
    slopes of $0$, $-1$, $-2$ and $-4$ respectively.
    On large scales, they all give white noise ${\cal P}(k)\propto k^0$.
    }
  \label{k0123}
\end{figure}

\noindent
The types of external sources we will come across can be roughly classified
into four categories, according to their topology:
zero-, one-, two-, or three-dimensional objects.
For the first three types, Ref.\cite{newpicture} gave the forms
of their power spectra by Fourier transforming
a set of randomly distributed points, straight filaments, and spherical shells:
\begin{eqnarray}
  {\cal P}_0(k)&=& \frac{V^2}{8\pi^3}\ ,
  \label{P_0}\\
  {\cal P}_1(k)&=& \frac{V^2}{8\pi^3}\frac{\tan^{-1}(kc_*)}{kc_*}\ ,
  \label{P_1}\\
  {\cal P}_2(k)&=& \frac{V^2}{8\pi^3}\frac{1}{(1+4k^2R_*^2)}\ ,
  \label{P_2}
\end{eqnarray}
where $V$ is the volume of the box to be Fourier transformed, and
we have taken the distributions $p(c)=c^2\exp(-c/c_*)/2c_*^3$
for filament lengths
and $p(R)=R^2\exp(-R/R_*)/2R_*^3$ for shell radii. 
This leads to
$\langle c\rangle = 2\sqrt{3}\,c_*$ and
$\langle R\rangle = 2\sqrt{3}\,R_*$.
Here, we further consider three-dimensional objects.
The power spectrum of a set of randomly distributed uniform 
spheres, with a radius
distribution of $p(r)=r^6\exp(-r/r_*)/720r_*^7$ (which gives an rms 
radius $\langle r \rangle = 2\sqrt{14}\,r_*$) is:
\begin{equation}
  \label{P_3}
  {\cal P}_3(k) = \frac{V^2}{8\pi^3}\frac{(1+0.8 k^2 r_*^2)}{(1+4k^2r_*^2)^3}\ .
\end{equation}
We can see that for sufficiently small $k$,
they all give a white-noise power spectrum ${\cal P}(k)=V^2/8\pi^3\propto k^0$.
This is because on scales sufficiently above their characteristic scales
$2\sqrt{3}c_*$, $2\sqrt{3}R_*$ and $2\sqrt{14}r_*$,
all objects are essentially point-like.
On the other hand,
at larger $k$, they give the characteristic behaviors of zero-, one-, two-
and three-dimensional objects:
${\cal P}(k)\propto k^0, k^{-1}, k^{-2}$ and $k^{-4}$ respectively.
In Fig.~\ref{k0123},
we show the power spectra of one realization of a set of 
randomly distributed
zero-, one-, two-, and three-dimensional objects.
They agree very well with the analytical behaviors of
(\ref{P_0}), (\ref{P_1}), (\ref{P_2}) and (\ref{P_3}).
The reason for having a steeper small-scale slope
in the power spectrum of higher-dimensional objects
with the same overall mean density is that
the mass is more diluted in this case.

\subsection{{\bfit Cosmic strings}}
\label{coss}
\noindent
The external source under consideration is a network of cosmic strings.
For local
gauge strings, the microphysical width of the string will be many orders of
magnitude smaller than its typical curvature radius, so that we can take the
zero thickness limit.
Therefore, the cosmic strings which dynamically source
the subsequent perturbations $\delta^{\rm S}({\bf x},\eta)$ in (\ref{IS})
have
spacetime trajectories which we can represent as $x^\mu_{\rm s} = (\eta, {\bf
x}_{\rm s}(\sigma,\eta))$, where $\sigma$ is a space-like parameter labeling
points along the string.
The stress energy tensor of the string source
is then given by\cite{VilShe}
\begin{equation}
  \Theta_{\alpha \beta}({\bf x}, \eta) = \mu
  \int d \sigma (\epsilon {\dot x_{\rm s}}^{\alpha} {\dot x}_{\rm s}^{\beta} -
  \epsilon^{-1} {x_{\rm s}'}^{\alpha} {x_{\rm s}'}^{\beta})
  \delta^3 ({\bf x} - {\bf x}_{\rm s} (\sigma, \eta)),
  \label{stret}
\end{equation}
where $\mu$ is the string linear energy density,
a prime  represents a derivative with respect to $\sigma$,
$\epsilon = [{\bf x_{\rm s}'}^2 / ({1 - {\bf {\dot x_{\rm s}}}^2})]^{1/2}$,
and we have also taken that ${\dot {\bf x}_{\rm s}} \cdot  {{\bf x_{\rm
s}}'}=0$.
It is then straightforward to compute $\Theta_+$ in (\ref{two}) as
\begin{equation}
\Theta_+({\bf x},\eta)=\Theta_{00}+\Theta_{ii}=
2 \mu \int d \sigma \epsilon
{\bf {\dot x_{\rm s}}}^2 \delta^3 ({\bf x} - {\bf x_{\rm s}} (\sigma, \eta)).
\label{theta_plus}
\end{equation}
This source term can be calculated directly from the string
network which was evolved using the Allen-Shellard (AS) string
simulation,\cite{AS1} 
and it is in good agreement with the other high resolution simulation of
Bennett-Bouchet.\cite{BB1,BB2}
Large-scale parallelized simulations were performed
on the COSMOS supercomputer, a Silicon Graphics Origin2000 with 20 Gbytes main
memory.  Dynamic ranges exceeding one thousand expansion times were feasible
because of the simulation size and by using a modified `point-joining'
algorithm which maintained fixed comoving resolution.
Because sampling points are used to trace cosmic strings,
we invoke the ``cloud in cell'' (CIC) method
to assign these sampling points of strings onto the comoving grids
and then Fast Fourier Transform them.
We then multiply this transformed quantity on grids by
the following sharpening function
to compensate for the smoothing effect 
induced by the CIC method:\cite{Villumsen}
\begin{equation}
  W({\bf k})=\frac{(k/2)^2}{\sin^2(k_1/2)+\sin^2(k_2/2)+\sin^2(k_3/2)}\,,
  \label{deconvolve}
\end{equation}
where ${\bf k}=(k_1, k_2, k_3)$ and $k=|{\bf k}|$.
This scheme has been verified against
sets of zero, one, two and three dimensional
objects as we have seen in Fig.~\ref{k0123}.

We will divide the string network into two parts:
the long strings and small loops.
We define the latter by taking those close loops whose lengths are smaller
than one tenth of the horizon size, and
we find that the resulting energy evolution
in either the long strings or the loops is not very sensitive to this threshold
because the loop sizes are typically at least two orders below the horizon
size and there are not many large loops around the threshold.
We will first investigate the properties of long-string network
in this subsection, and
leave the investigation of loops to the next subsection.

\begin{figure}[htbp]
\vspace*{13pt}
  \centering\epsfig{figure=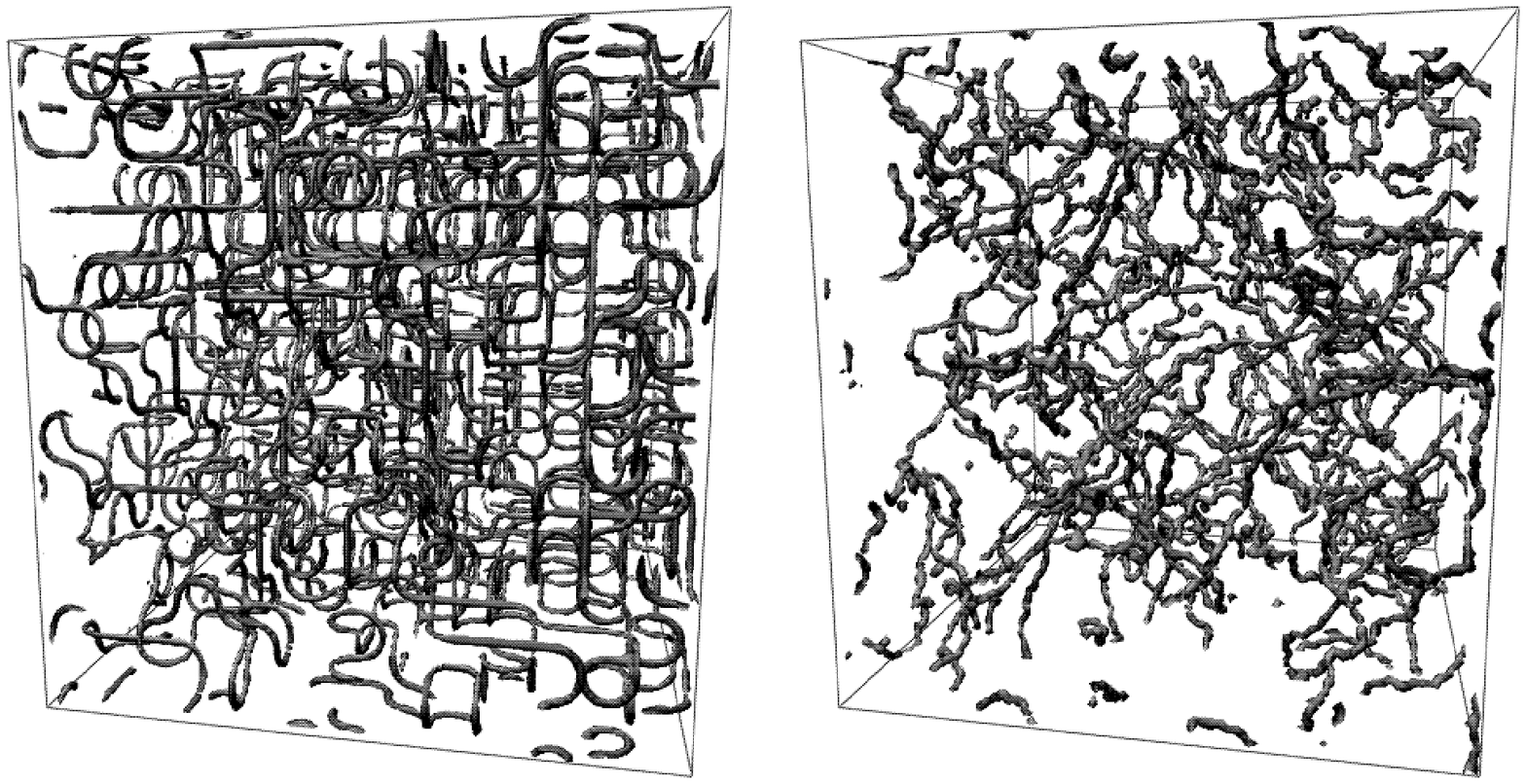, width=5in}\\
  \centering\epsfig{figure=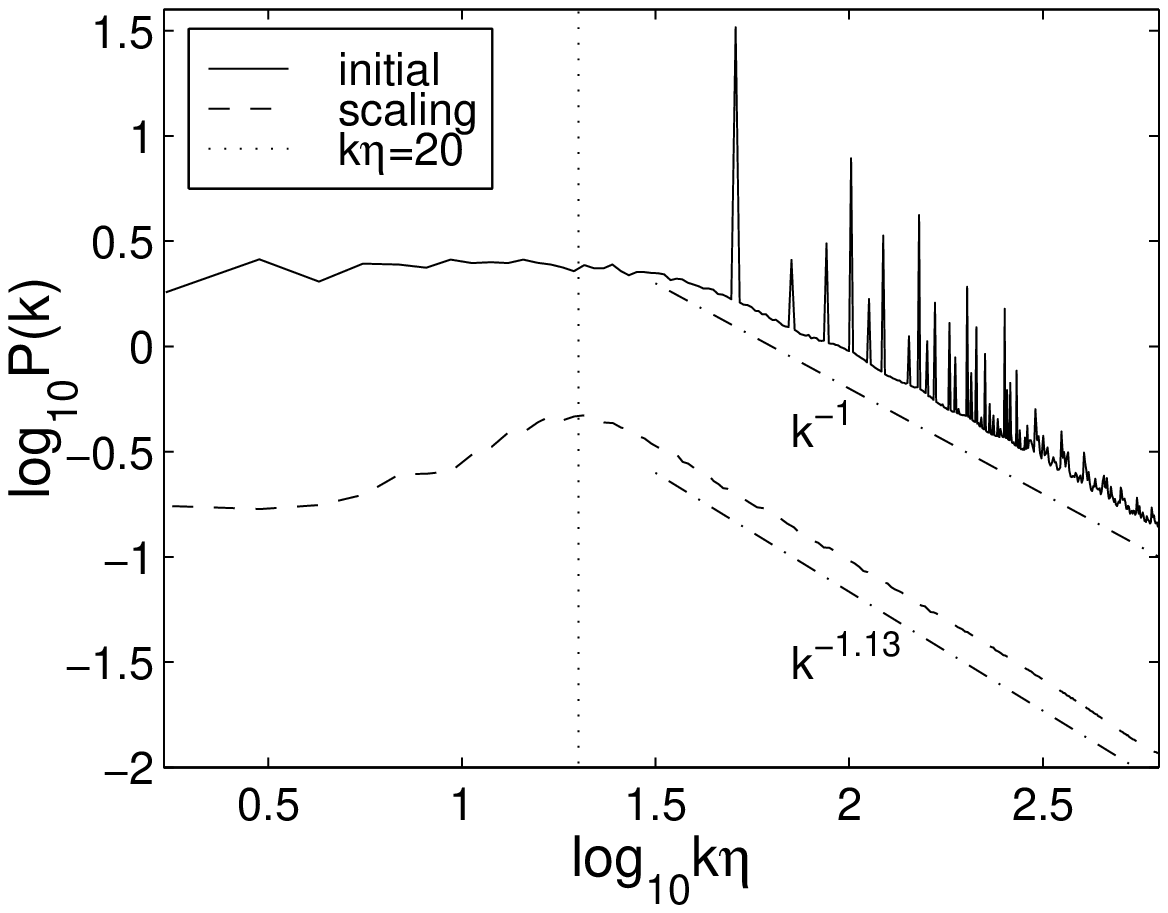, width=5in}\\
\vspace*{13pt}
\fcaption
{The top-left is 
  an initial network of cosmic strings,
  the top-right a scaling long-string network,
  and the bottom their power spectra.
  The first peak in the initial spectrum appears at $k_\xi=16\pi/\eta$,
  which accurately corresponds to the initial comoving correlation length
  $\xi_{\rm com}=\eta / 8$ set by hand.
  The turnover scale in the scaling regime is clearly $k_\xi=20/\eta$,
  and the slope of the spectrum turns from $-1$ in the initial network
  to $-1.13$ in the scaling regime.
  The normalization here is arbitrary.
  }
\label{string_net}
\end{figure}

Fig.~\ref{string_net} shows the initial and scaling
long-string configurations, and their power spectra.
Because the initial network is a random walk with a
correlation length $\xi$ along the $x$, $y$ or $z$ directions,
its properties are very similar to that of a sampling function
(the periodic Dirac Delta)
\begin{equation}
  \Pi(x)=\sum_{n=-\infty}^{\infty} \delta(x-n\xi) \;,
\end{equation}
where $n$ is an integer.
We know its Fourier transform is also a sampling function.
Hence, the power spectrum of this string network has periodic peaks,
whose interval length relates to the correlation length as
$\Delta k=2\pi/\xi$.
Due to the 3-dimensional structure,
there are also other minor peaks corresponding to
$2\sqrt{2}\pi/\xi$, $2\sqrt{3}\pi/\xi$, etc.
With $\xi=\eta/8$, we also note that
at small scales ($k \gsim 16\pi/\eta$),
the power spectra have an overall slope of $-1$
because strings are line-like objects;
at large scales ($k \lsim 16\pi/\eta$),
the power spectra turn flat
because of the point-like property
(see subsection \ref{funa} for the topological analysis).

On the other hand,
the scaling string network in Fig.~\ref{string_net} shows different properties.
Its power spectrum is much smoother than the previous case because
the periodicity of strings has no preferred directions (i.e.\ the $x$, $y$ and
$z$ directions in the previous case) and therefore
the periodic peaks degenerate to only one broad peak here,
corresponding to the mean spacing of strings.
This smoothness feature of the power spectrum can be employed to ensure
that when the simulation of structure formation starts,
the string network is already scaling so that 
the initial configuration of the network will have no effect
on the subsequent result.
We also note that instead of turning from $k^0$ to $k^{-1}$ very smoothly
at the correlation scale as we have seen from randomly distributed filaments
(see Fig.~\ref{k0123}),
the power spectrum has a broad peak on that scale.
This illustrates the fact that in the scaling regime,
a real string network is not like random walk, but
has a configuration in which all strings remains a roughly constant distance
(which is about the correlation length) from each other,
so that the power on this scale is amplified to form a broad peak.
We verify that this broad peak appears at $k_\xi \approx 2 \pi/ \xi 
\approx 20/\eta $ persistently from deep in the radiation era through to deep 
into the matter era, with an amplitude of about three times larger than the 
large-scale white noise. Here $\xi \approx \eta/3$ is the comoving mean 
distance among long strings.
Therefore, any previous work based on modeling cosmic strings with random 
filaments would have underestimated the resulting power spectrum by a maximum 
factor of three (e.g.~Ref.\cite{against,ABR2}),
because such a broad peak in a real string network appears persistently with
an amplitude of about three times larger than the large-scale white noise.

We further notice in Fig.~\ref{string_net} that
the slope of the spectrum in the scaling regime is not exactly $-1$
on small scales, but $-1.13$.
This is because strings are very wiggly on small scales and therefore
have an overall dimension of slightly greater than one
so as to steepen the small-scale slope to $-1.13$.

Another important property of the string network is the correlation time.
Because the strings' configurations are uncorrelated when separated
by a sufficiently long time,
we can define a correlation time $\eta_{\rm c}(k,\eta)$ as the time period above
which the strings' configurations are uncorrelated, i.e.\ 
the unequal time correlator (UETC)
\begin{equation}
  \label{eta_c}
  \langle
  \widetilde{\Theta}_+({\bf k},\eta)
  \widetilde{\Theta}_+({\bf k},\eta')
  \rangle \approx 0
  \quad {\rm for}\; |\eta-\eta'|>\eta_{\rm c} \,.
\end{equation}
We see that on smaller scales inside the horizon
the correlation time is shorter, and
that on the horizon scale
the correlation time $\eta_{\rm c}$ is about the order
of the conformal time $\eta$,
although the UETC is dominated within a dynamic range $\eta'=(1\pm 0.2)\eta$,
taking the half-maximum threshold.
In a quantitative analysis,
an accurate calculation of UETC's requires a dynamic range of at least $4$ in conformal time.
This is equivalent to a dynamic range of $16$ in radiation-era physical time,
and $64$ in matter-era physical time.
So far in the literature,
only our simulations have achieved this criterion for local cosmic strings,
while keeping a highest resolution equivalent to
$1000^3$ points in a comoving box.
Each of our simulations typically has a dynamic range of
more than 10 in conformal time.
We also find
in the UETC $\langle
  \widetilde{\Theta}_+({\bf k},\eta)
  \widetilde{\Theta}_+({\bf k},\eta')
  \rangle$
that
a hot spot locates at around $\eta'=\eta$ and $k \approx 20/\eta$,\cite{WASB}
which is exactly the correlation wavenumber $k_\xi$ we have previously seen
in the string source power spectrum.
If we take a slice through $\eta'=\eta$ in
the UETC $\langle
  \widetilde{\Theta}_+({\bf k},\eta)
  \widetilde{\Theta}_+({\bf k},\eta')
  \rangle$,
then the profile of this slice
gives the scaling spectrum in Fig.~\ref{string_net},
with a broad peak corresponding to the hot spot in the UETC.

\begin{figure}[htbp]
\vspace*{13pt}
  \centering\epsfig{figure=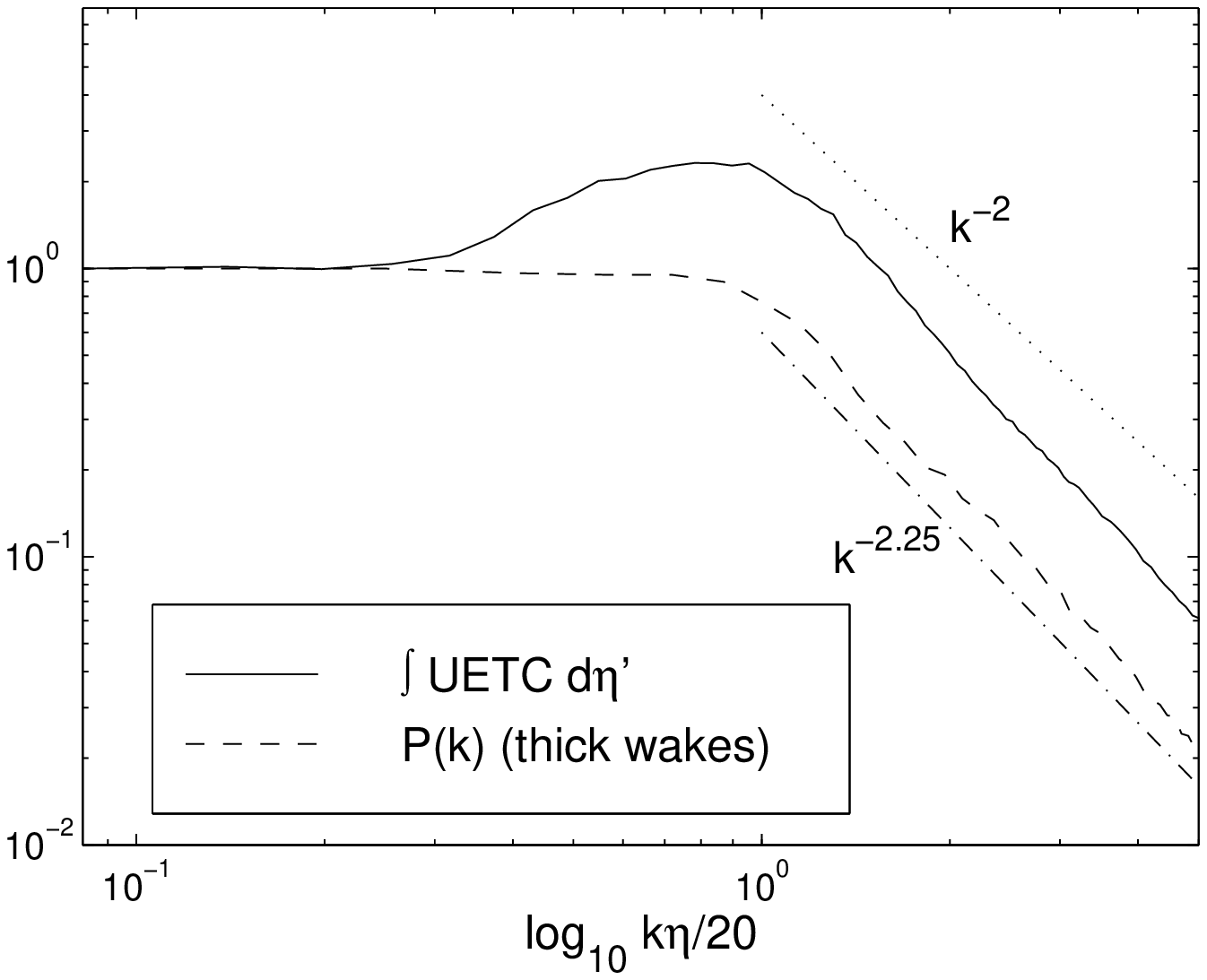, width=4in}\\
\vspace*{13pt}
  \fcaption
  {The integrated UETC,
    $\int \langle\widetilde{\Theta}_+({\bf k},\eta)
    \widetilde{\Theta}_+({\bf k},\eta')\rangle d\eta'$ (solid line),
    and the power spectrum of wakes with a thickness of $1/10$ of their
    correlation length (dashed line).
    The normalization is arbitrary.
    The dotted and dot-dashed lines have exact slopes of
    $-2$ and $-2.25$ respectively.
    }
  \label{integ_UETC}
\end{figure}

Fig.~\ref{integ_UETC} shows the integration of the UETC over time.
This quantity reflects the topology of the path swept out by strings.
Since a two-dimensional object will have a small-scale slope of $-2$,
we expect such a quantity to behave in the same way.
However, as we can see from the figure, the small-scale slope is not
exactly $-2$, but $-2.25$. This is due to the string wiggliness which is 
reflected in the wiggliness of the wakes they generate.
We investigated the level of such a wiggliness in wakes, and verified that 
the wiggles of cosmic string wakes are typically about one tenth of the 
correlation length.

\begin{figure}[htbp]
\vspace*{13pt}
  \centering\epsfig{figure=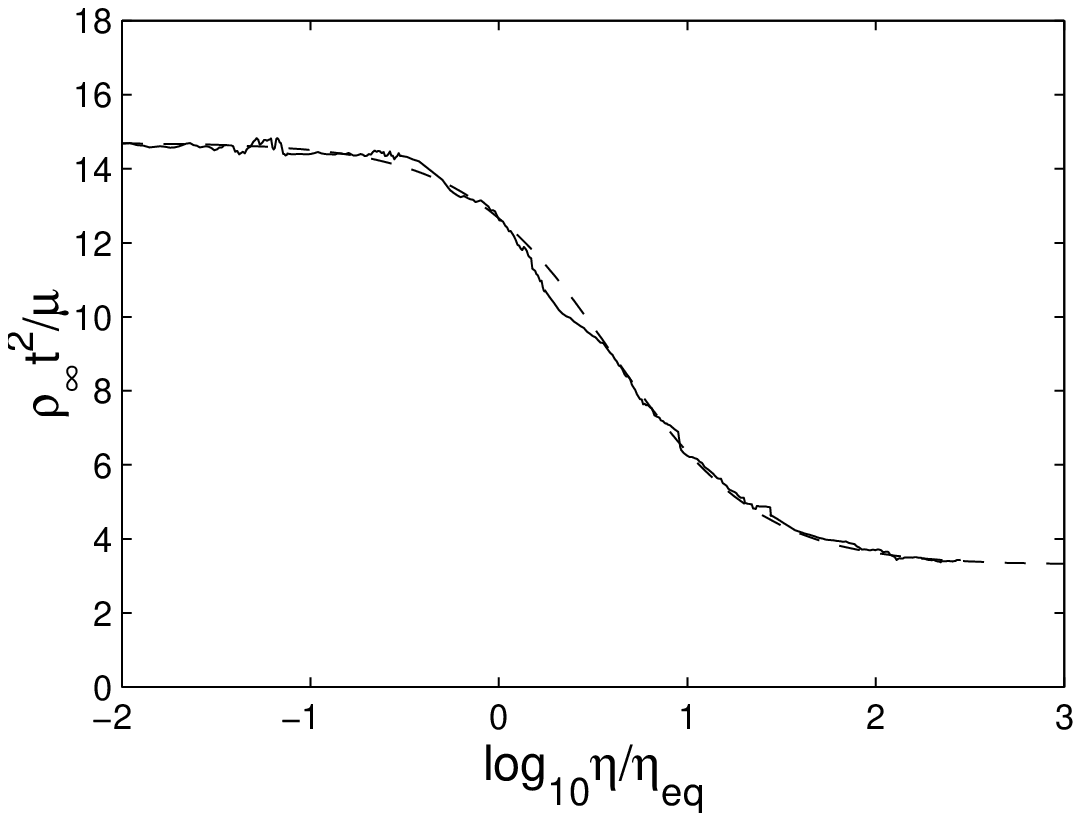, width=3.8in}\\
\vspace*{13pt}
  \fcaption
  {The evolution of long-string energy density
    observed from high-resolution simulations (solid),
    and the analytic fit (\ref{rho_str_ana}) (dashed).
    It is clear that
    $\rho_\infty t^2/\mu$ drops very slowly after $\eta_{\rm eq}$,
    and at $\eta_{\rm rec}$
    it is still about $1.5$ times larger than its matter-era asymptotic value.
    }
  \label{str_density}
\end{figure}

Another important aspect regarding the cosmic string evolution is the
radiation-matter transition process.
The key observation was the
very slow relaxation to the matter era scaling density.
Fig.\ \ref{str_density} shows the evolution of long-string energy density
$\rho_\infty$.
A good analytic fit to the evolution of
$\zeta(\eta)=\rho_\infty t^2/\mu$ is found to be
\begin{equation}
  \zeta(\eta)=14.7-
  \frac{11.4}{1+\left(4\eta_{\rm eq}/\eta\right)^{1.1}}\;.
  \label{rho_str_ana}
\end{equation}
When compared with simulations of higher resolutions,
we find that our values here are about $10\%$ higher,
while the overall shape remains unchanged.
This is due to a lower kink density in our case,
shortening the mean length of strings.
We can see from Fig.\ \ref{str_density} that the relaxation
process extended well beyond recombination
(for $h=0.7$, $\log_{10}(\eta_{\rm rec}/\eta_{\rm eq}) \approx 0.76$
and $\log_{10}(\eta_0/\eta_{\rm eq}) \approx 2.4$).
This has important implications for
large-scale structure and CMBR anisotropies as we shall discuss.
Fig.~\ref{str_density} also shows that the long string energy density at 
the radiation-matter transition is closer to the radiation era asymptotic 
value than previously thought.

\subsection{{\bfit Cosmic string loops}}
\label{cosl}
\noindent
The role of small loops produced by the string network has evolved from 
a potential one-to-one correspondence between loops and
cosmological objects,\cite{Vil}
through to a completely subsidiary role relative to the wakes
swept out by long strings.\cite{SilZel} This dethronement of loops
was a result of numerical studies which showed that the average loop 
size $\bar\ell = \alpha t$ was much smaller than the horizon, 
$\bar\ell <\!< d_{\rm H}$;\cite{AS1,BenBou2}
they may even be as small as the lengthscale set by gravitational 
back-reaction $\alpha \sim 10^{-4}$, a value appropriate 
for GUT-scale strings.\cite{VilShe}  
Add the high ballistic 
loop velocities observed $\bar v\approx c/\sqrt{2}$ 
and it was not 
surprising that these tiny loops have been assumed to be more or less
uniformly  distributed and hence a negligible source relative to the
long-string network.\cite{newpicture}
Nevertheless, small loops always make up a significant 
fraction of the total string energy density at any one time and, as we
will demonstrate later, loop-induced inhomogeneities are considerable 
if their lifetime
is not much smaller than the Hubble time.  By properly 
incorporating these loop perturbations, we will show that their contribution 
relative to the long-string wakes is almost comparable and also highly 
correlated with these wakes.

To investigate loops, we keep all the loops generated by the long-string
network, and
model those which are smaller than a fixed fraction of the horizon 
size as relativistic point masses.
The effects of the evaporation of these loops into 
gravitational waves and the damping of loop motion due to expansion 
are also included.
Note that here we assume the standard scenario with an 
evolving `infinite' string network, but there
are alternative models in which the initial string configuration consists 
entirely of loops (e.g., Ref.\cite{Loops}). 

\begin{figure}[htbp]
\vspace*{13pt}
\centering 
\leavevmode\epsfxsize=4in \epsfbox{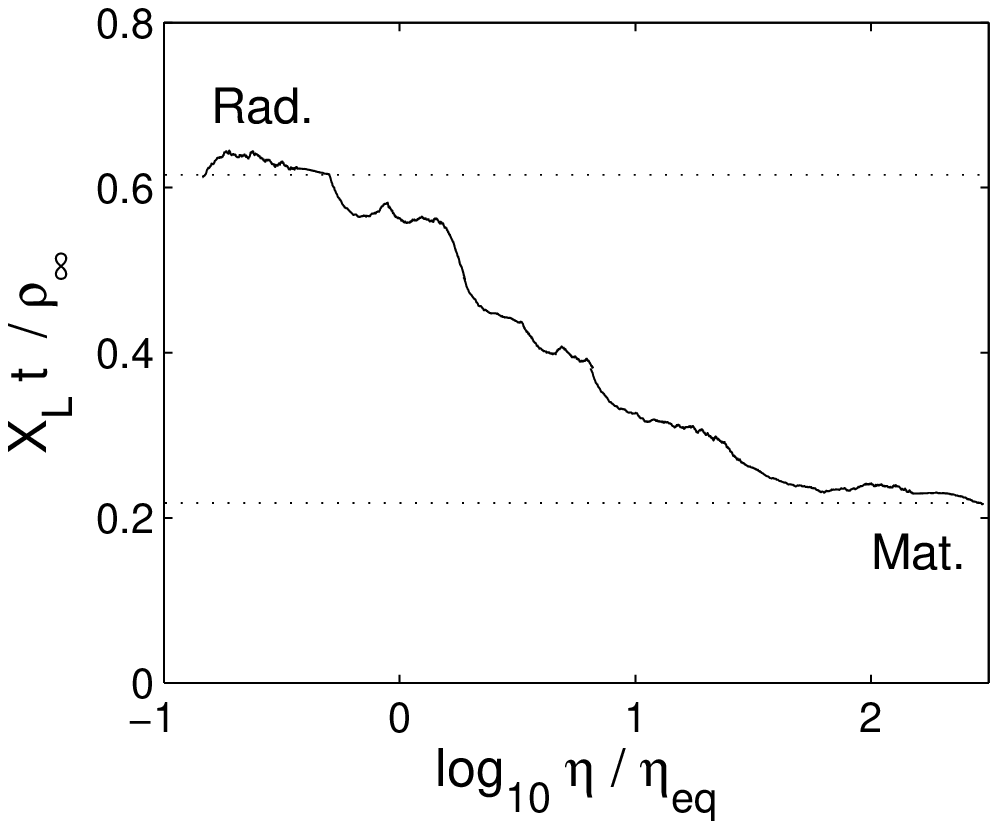}\\ 
\vspace*{13pt}
 \fcaption
  {Evolution of $X_L(t)$.
    The dotted lines are the
    asymptotic values in the radiation and the matter eras in (\ref{one}).}
  \label{figure1}
\end{figure}

To begin with, let us consider some analytical properties of loops.
First,
the Nambu equations of motion  for cosmic strings in an expanding
universe can be averaged to yield:
\begin{equation}
  {{d \rho_\infty} \over dt}+2H(1+{\langle v^2 \rangle}) \rho_\infty= - X_L, 
  \label{rhotwo}
\end{equation}
where
$\langle v^2 \rangle$ is the mean square velocity of strings
and 
$X_L$ is the transfer rate of energy density from long strings into loops.
In the scaling regime the 
long-string energy density should scale with the background
energy density  evolving as
\begin{equation}
  {{d \rho_\infty} \over dt} = -2{{\rho_\infty} \over t}.
  \label{rhoone}
\end{equation}
Substituting this into (\ref{rhotwo}) to eliminate $d \rho_\infty/dt$ gives
\begin{equation}
  {{t X_L} \over \rho_\infty}=
  \left\{
    \begin{array}{ll}
      (1-{\langle v^2_{\rm r} \rangle}) \sim 0.6
      & {\rm in\ radiation\ era},\\
      {2  \over 3} (1-2 {\langle v^2_{\rm m} \rangle}) \sim 0.2
      & {\rm in\ matter\ era},
    \end{array}
  \right.
  \label{one}
\end{equation}
where ${\langle v^2_{\rm r} \rangle} \gsim {\langle v^2_{\rm m} \rangle} \sim 0.6$.\cite{AS1,BenBou2}
Both (\ref{rhoone}) and (\ref{one}) provide a check for the scaling behavior of
long strings and loops in the cosmic string network simulations.
Figure \ref{figure1} shows the evolution of $X_L$.
We can see that the expected amount of energy
was converted into loops in our simulations so that
$X_L$ has the correct asymptotic behavior given by (\ref{one}).
However,
the typical loop-size
(and consequently their lifetime) 
does not approach scaling so rapidly and is therefore larger than 
physically expected for most of the duration 
in the simulations.\cite{AS1,BenBou2} 
To overcome this problem we need to rescale the 
loop lifetime in an appropriate way.

The loops produced by a cosmic string network
will decay into gravitational 
radiation, with a roughly constant decay rate $\Gamma G \mu^2$,
where $\mu$ is the string linear energy density.
Typically $\Gamma=50-100$ with
an average $\langle \Gamma \rangle \sim 65$.\cite{SchQua,AllShe}
In very high-resolution simulations,
we observe a loop-production scale on which the loops are most effectively
generated by the network, although this is only approached towards the 
end of simulations used here.  For 
both analytical and numerical simplicity, we make the reasonable 
assumption that the loop production as `monochromatic', that is, all loops 
formed at the same time will have the same mass in the limiting 
scale-invariant distribution.  
We can write the initial rest mass of a loop formed at time $t_*$ as 
\begin{equation}
  M_L^*=\alpha\mu t_* \equiv f \Gamma G \mu^2 t_*\ .
  \label{inimass}
\end{equation}
Here, the parameter $f = \alpha/\Gamma G \mu$ is 
expected to be of order unity if the size of the loops formed at the time $t$ 
is determined by gravitational radiation back-reaction,
which smoothes strings on scales smaller than $\Gamma G \mu t$. 
Thus the uncertainty in the average mass and therefore the lifetime of loops
formed at a given time
is quantified by the choice of the parameter $f$.
We also find that
the initial rms velocity of loops observed from the simulations
is $\langle v^2_*\rangle^{1/2} \gsim 0.7c$ throughout all the regimes.

In the simulations we only impose the monochromatic assumption through $f$ on 
the loop decay lifetimes. 
Thus we have the rest mass of a loop formed at time $t_*$ evolving as
\begin{equation}
  M_L(t_*, t)= M_L^* W(t_*, t)\ ,
  \label{massdecay}
\end{equation}
where  
\begin{equation}
  W(t_*, t)=
  \left\{
    \begin{array}{ll}
      1- {t-t_* \over \tau(t_*)} &
      {\rm for \ } t_* \le t \le t_*+\tau(t_*) \,, \\
      0 & {\rm otherwise} \,.
    \end{array}
  \right.
\label{four}
\end{equation}
Here $\tau(t_*)\approx ft_*$ is the lifetime of loops produced at time $t_*$
($f=2$, $3$ implies the decay occurs in one horizon time in the radiation 
and matter eras
respectively).
The evolution of the loop energy density is then given by:
\begin{eqnarray}
    \rho_L (t)
    & = &
    \int_0^{t} X_L(t') {\left[{{a(t')} \over {a(t)}}\right]}^3 W(t', t) dt'
    \nonumber
    \\
    & \propto &
    \left\{
      \begin{array}{ll}
        f/t^2        & {\rm for\ } f \ll 1,\\
        \sqrt{f}/t^2 & {\rm for\ } f \gg 1 {\rm \ (radiation\ era),}\\
        (\ln{f})/t^2 & {\rm for\ } f \gg 1 {\rm \ (matter\ era),}
      \end{array}  
    \right.
    \label{rhol}
\end{eqnarray}
where we have used the scaling behavior
(\ref{rhoone}) and (\ref{one}).
Consequently, the scaling of the power spectrum induced by loops in $f$
should interpolate between
$f^2$ and $f$ (radiation era) or $(\ln{f})^2$ (matter era).
We notice in (\ref{rhol}) that we have ignored the effect of loop velocity
redshifting
due to the expansion of the Universe, which causes a change in the
effective mass.
Because loops are formed with relativistic velocities,
we expect this damping mechanism to have the strongest effect for
$f \gg 1$, but to be negligible for  $f \ll 1$.

If a loop formed at time $t_*$ has an initial physical 
velocity ${\bf v}_*$, its trajectory in physical space
accounting for the expansion of the Universe is then given by:
\begin{equation}
  {\bf x}(t)
  = {\bf x}(t_*) +
  a(t) \int_{t_*}^t {{\bf A} \over {\sqrt{a(t')^2 + A^2}}} {dt' \over a(t')}\,,
  \label{five}
\end{equation}
for $t \ge t_*$,
where
${\bf A}=\gamma_* {\bf v}_* a_*$, $A=|{\bf A}|$ and 
$\gamma_*=(1-|{\bf v}_*|^2)^{1/2}$.
Here we have neglected the acceleration of loops
due to the momentum carried away by the gravitational radiation, the 
so-called `rocket effect'.
A numerical calculation for several asymmetric loops shows that
the rate of momentum radiation from an oscillating loop is
\begin{equation}
|{\dot {\bf P}}|=\Gamma_P G \mu^2\ ,
\end{equation}
where $\Gamma_P \sim 10$.\cite{VacVil}
Combining with (\ref{massdecay}),
one can show that this rocket effect will become important 
only when:
\begin{equation}
  {t \over t_*} \gsim
  1+ {f \over 1+ \Gamma_P/(\Gamma \gamma v)} \ ,
\end{equation}
which affects only the final stages of the loop lifetime
as long as $\Gamma_P/(\Gamma \gamma v) < 1$, or equivalently $v \gsim 0.15c$.
For a typical $v_*\sim c/\sqrt2$,
one requires a loop lifetime $\gsim 43t_*$ in the radiation era
and $\gsim 17t_*$ in the matter era to redshift down to
this critical velocity according to (\ref{five}).
Since the values of $f$ we explore here are of order unity,
it is a reasonable approximation to neglect
the transfer of momentum due to gravitational radiation.

\begin{figure}[htbp]
\vspace*{13pt}
  \centering\epsfig{figure=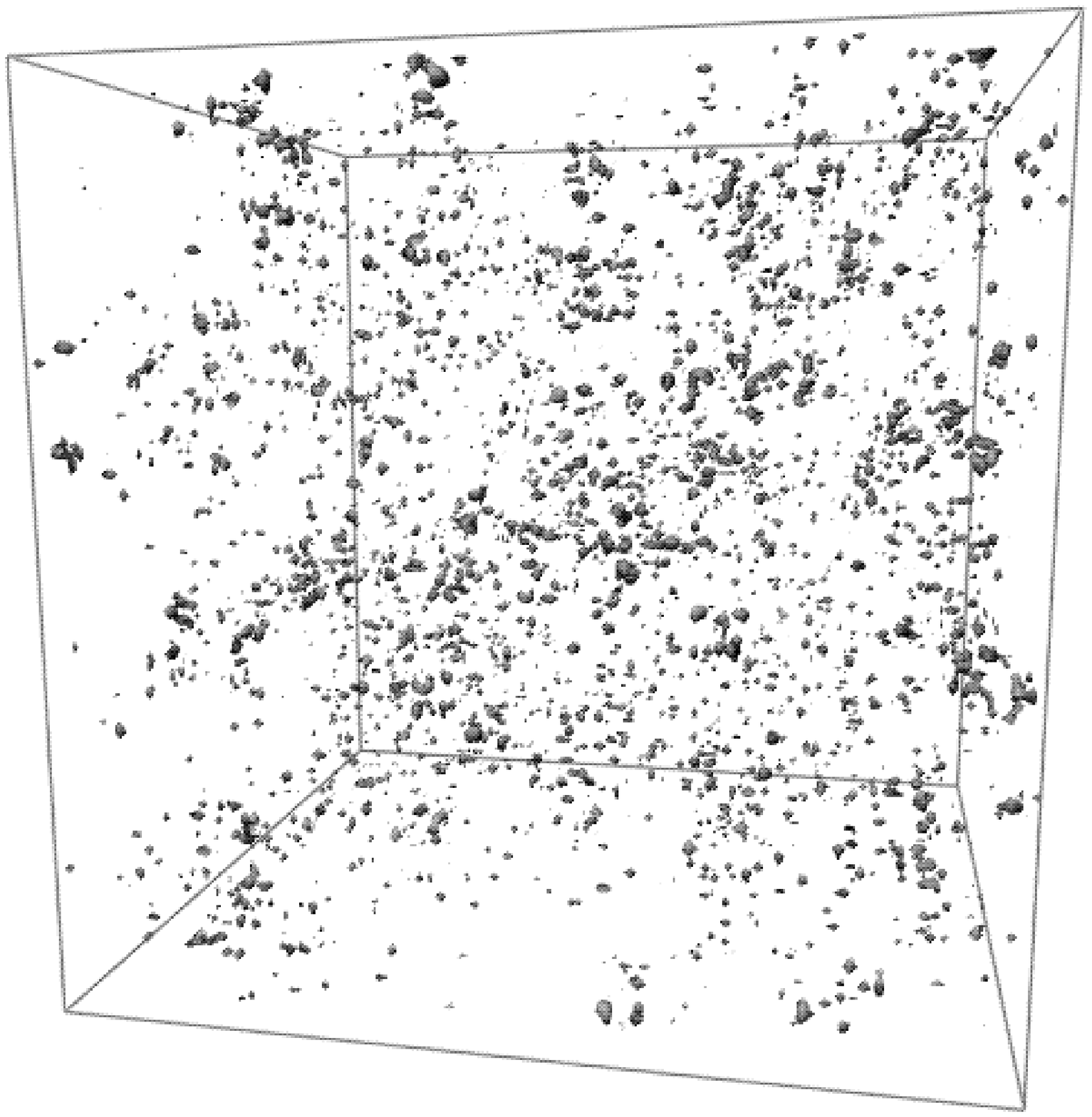, width=4in}\\
  \centering\epsfig{figure=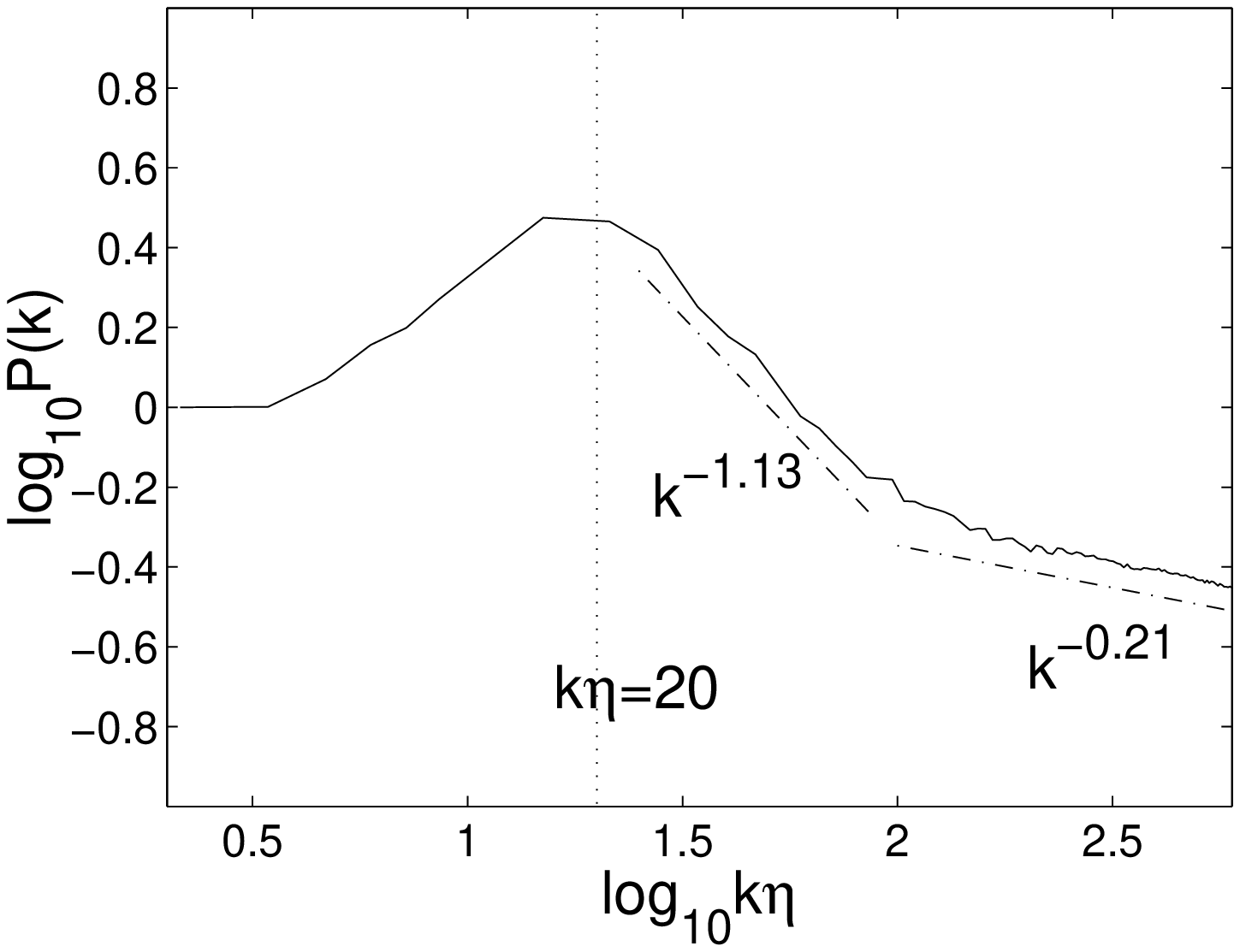, width=3.6in}\\
\vspace*{13pt}
  \fcaption
  {On the top is a snap shot of loop distribution in the scaling regime.
    The size of each point reflects its mass.
    The box size is the same as the horizon.
    On the bottom is their power spectrum with arbitrary normalization.
    The dot-dashed lines have exact slopes of $-1.13$ and $-0.21$.
    The dotted line indicates the correlation wavenumber $k_\xi=20/\eta$
    seen in the long-string network.
    On large scales ($k\eta\lsim 100$), we see loops mimic long strings
    so that their spectrum shape is identical to that of long strings
    (see Fig.~\ref{string_net}).
    The normalization is arbitrary.
    }
  \label{loop_spec}
\end{figure}

With the treatment of (\ref{inimass}) and the effects of (\ref{massdecay})
and (\ref{five}) Fig.~\ref{loop_spec} shows a loop spectrum with $f=1$
in the scaling regime.
As we can see, on small scales ($k\eta\lsim 100$)
the shape of loop spectrum is identical to that
of long strings (see Fig.~\ref{string_net}).
This is because loops trace the paths of long strings
even after they are formed.
We verify this by calculating the correlation coefficient between long strings
and loops in the scaling regime.
This coefficient is always greater than $0.5$ on larger scales.
This phenomenon has also been verified by observing the movies of
string evolution made from the simulations.
On the other hand, on small scales the loop spectrum is not exactly
a white noise. It has a slope of $-0.21$.
This is because even on small scales loops are still not exactly point-like,
but with some clumpy structures,
so that their spectrum slope is steepened as discussed
in subsection~\ref{funa}.
In section \ref{final_loops}, we will see further that this correlation between
the long-string and loop distributions can have a significant effect on
the resulting power spectrum of matter perturbations.

\section{{Approximation Schemes}}
\label{appsch}

\subsection{{\bfit Compensation scale}}
\label{coms}

\noindent
It is a very substantial numerical challenge to evolve the
initial and subsequent perturbations induced by cosmic strings in (\ref{IS})
such that they accurately cancel on super-horizon scales by the present
day $\eta_0$.
Long-string network scaling entails the copious
production of enormous numbers of tiny loops,\cite{AS1,BB1,BB2} 
whose evolving distribution and decay must be
carefully followed as described in subsection~\ref{cosl}.
For the
large dynamic range required for the present study, however, we have
by necessity adopted the compensation factor approximation suggested in
a semi-analytic context in Ref.\cite{AScdm}.  To implement this,
we accurately evolved the string network numerically,
and then multiplied the Fourier transform
of the resulting stress energy $\widetilde\Theta_{+}({\bf k},\eta)$ by
a cut-off function ${\widetilde F}(k, \eta)$ given by
\begin{equation}
  \label{F_comp}
  \widetilde{F}(k, \eta)
  = \left[1+\left(\frac{k_{\rm c}(\eta)}{k}\right)^2\right]^{-1} \,,
\end{equation}
where $k_{\rm c}(\eta)$ is the compensation wave number.
This results in the correct $k^4$ fall-off in the power spectrum at large
wavelengths above the compensation scale $k_{\rm c}^{-1}\sim \eta$.
Thus we obtain
\begin{equation}
  \label{delta_tilde}
  \widetilde{\delta}_{\rm c} ({\bf k}, \eta_0) =
  \widetilde{\delta}_{\rm c}^{\rm I}({\bf k}, \eta_0) +
  \widetilde{\delta}_{\rm c}^{\rm S}({\bf k}, \eta_0) \approx
  4\pi G
  \int_{\eta_{\rm i}}^{\eta}
    \widetilde{\cal G}_{\rm c}(k; \eta_0, \eta')
    \widetilde{\Theta}_+({\bf k},\eta')
    \widetilde{F}(k, \eta')
  d\eta'\,.
\end{equation}

\begin{figure}[htbp]
\vspace*{13pt}
  \centering\epsfig{figure=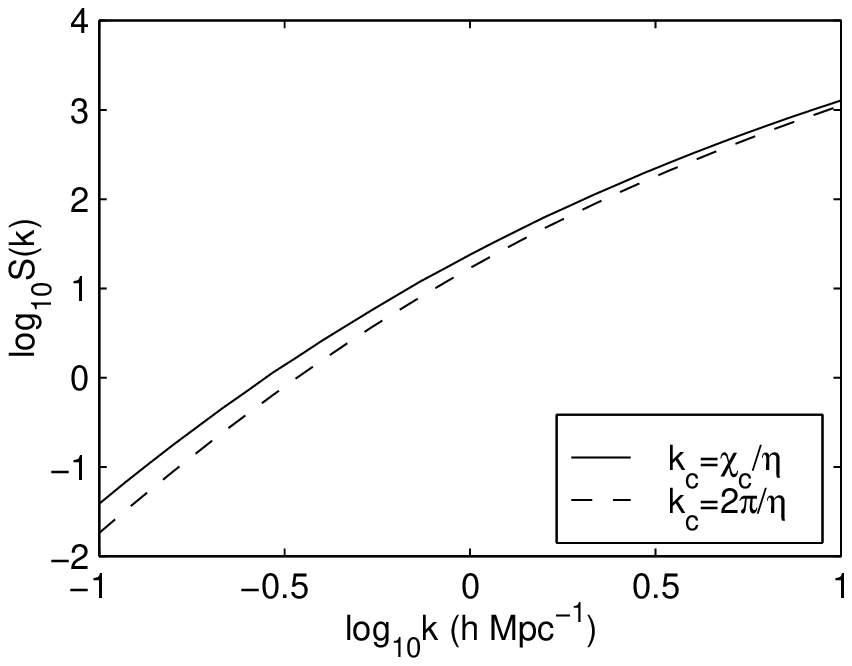, width=4in}\\
\vspace*{13pt}
  \fcaption
  {A comparison of spectra using different compensation scales.
    The solid line used the definition in Ref.\cite{painless},
    where $\chi_{\rm c}$ varies from $\sqrt{6}$ in the radiation epoch
    to $\sqrt{18}$ in the matter epoch;
    the dashed line used a compensation scale suggested in Ref.\cite{AScdm}.
    The dynamical range is $\eta=(0.01, 50)\eta_{\rm eq}$.
    }
  \label{k_cfig}
\end{figure}

Several
proposals for $k_{\rm c}(\eta)$ have been discussed in the
literature. In Ref.\cite{AScdm}, $k_{\rm c}=2\pi\eta^{-1}$ was suggested as
physically plausible, but not seriously justified.
More recently, however, in Ref.\cite{painless} the efficacy
of the approximation (\ref{delta_tilde}) has been demonstrated
by studying multi-fluid compensation back-reaction effects in greater detail.
It is claimed that the compensation scale arises naturally and uniquely from
an algebraic identity in
the perturbation analysis.  For the present study we have adopted the
analytic fit for $k_{\rm c}(\eta)$ presented in Ref.\cite{painless}:
\begin{equation}
  \label{k_c}
  k_{\rm c}(\eta)=\frac{\sqrt{6(3A^2\eta^2+12A\eta+16)}}{(A\eta+4)\eta}\, ,
\end{equation}
where $A$ is defined by (\ref{aABC}).
This $k_{\rm c}(\eta)$ smoothly interpolates from $k_{\rm c}
=\sqrt{6}\eta^{-1}$ in the radiation era to $k_{\rm c}=\sqrt{18}\eta^{-1}$
in the matter era.
Of course, there is some dependence of the perturbation amplitude on
this choice.
Fig.\ \ref{k_cfig} shows
the effect of different compensation scales on the resulting power spectrum.
Here we have used the semi-analytical model which will be described in
section \ref{sema} and verified in section \ref{semf}.
Evidently,
a larger compensation scale (at the same conformal time) will give
significantly more power at larger scales.
For example, with the smaller matter era $k_{\rm c}$
of Ref.\cite{AScdm}, the amplitude is approximately 40\% smaller
around length-scales of $60h^{-1}{\rm Mpc}$ and larger.

Clearly, the quantitative implementation of compensation effects is one
of the key uncertainties in all previously published work on gauged
cosmic strings.
The new analytic compensation factor approximation which we use here
should improve this situation, but the problem ultimately requires
a full-scale numerical treatment in which all strings and background fluid
components are accurately evolved through to the present day.\cite{AWSA}

\subsection{{\bfit Hot dark matter}}
\label{hdm}
\begin{figure}[htbp]
\vspace*{13pt}
  \centering\epsfig{figure=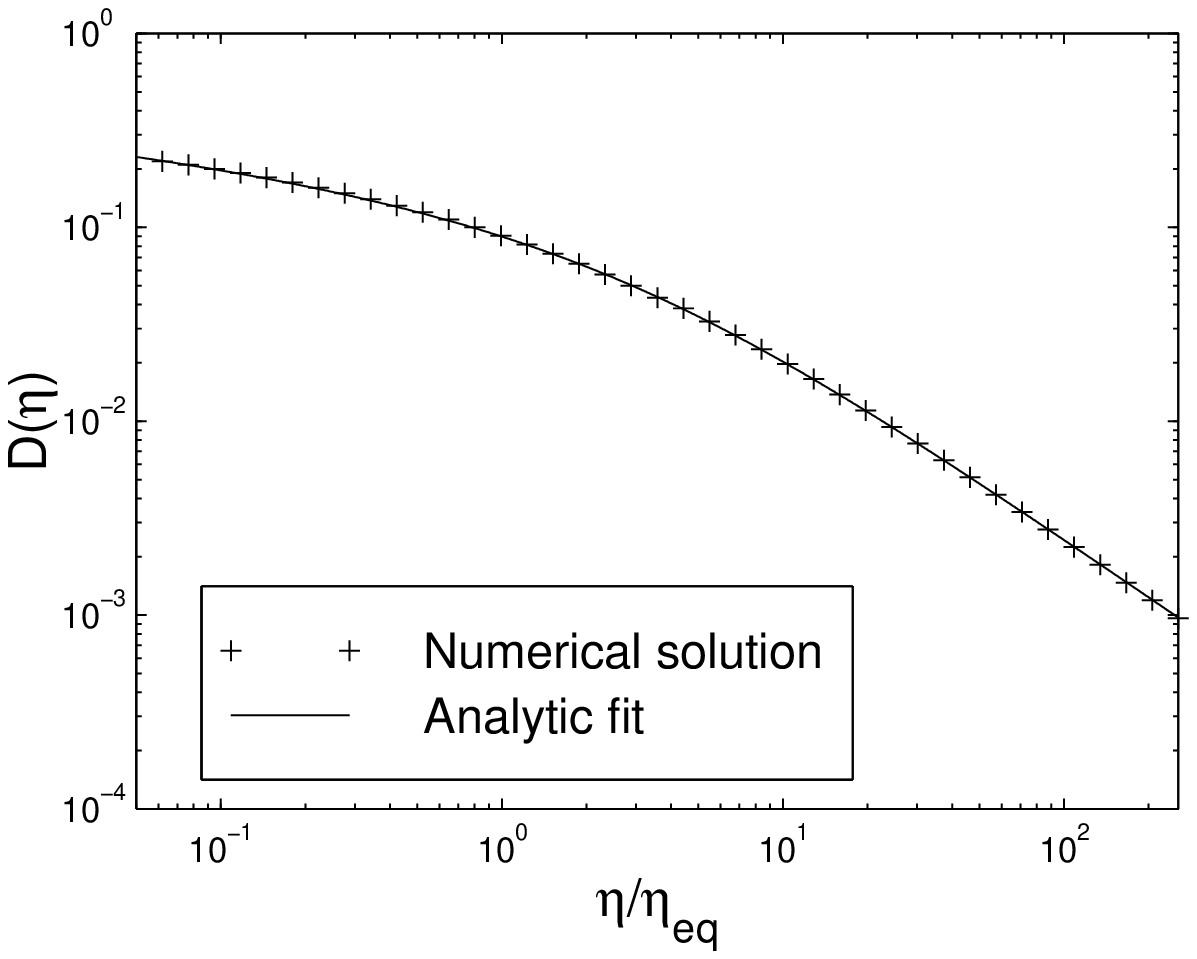, width=3.8in}\\
\vspace*{13pt}
  \fcaption
  {Neutrino free-streaming length $D(\eta)$ with
    $T_{\nu 0}=1.6914 \times 10^{-13} \, {\rm GeV}$ and
    $m_\nu=91.5 \, \Omega_{\rm h0} h^2 \, {\rm eV}$.
    The analytic fit (\ref{D}) agrees with 
    the numerical solution within a maximum of $2\%$ error.
    }
  \label{freestreaming}
\end{figure}

\noindent
In order to study the formation of structures with cosmic strings
in HDM models we use a straightforward
modification to the perturbation source
similar to that employed in Ref.\cite{AShdm}. This
method is a reasonably accurate alternative to much
more elaborate calculations using the
collisionless Boltzmann equation with defect sources.
We simply multiply the Fourier transform of the string source term
$\widetilde\Theta_+(k, \eta)$
by a damping factor ${\widetilde G}(k,\eta)$ given by
\begin{equation}
  {\widetilde G}(k,\eta)=
  \left[
    {1 \over {1+(0.435 k
        D(\eta))}^{2.03}}
  \right]^{4.43},
  \label{G}
\end{equation}
where $D(\eta)$ is the comoving distance traveled by a neutrino with momentum
$T_\nu/m_\nu$ from time $\eta$ to $\infty$.
The factor ${\widetilde G}(k,\eta)$ is a fit to numerical calculations of
the transfer function
of a Fermi-Dirac distribution of non-relativistic neutrinos, and accounts for
the damping of small-scale perturbations due to neutrino
free-streaming.\cite{AShdm}
We calculated $D(\eta)$ numerically and found
an excellent fit (within $2\%$ error, see Fig.~\ref{freestreaming}) 
for $T_{\nu 0}=1.6914 \times 10^{-13} \, {\rm GeV}$ and $m_\nu=91.5 \,
\Omega_{\rm h0} h^2 \, {\rm eV}$:
\begin{equation}
  \label{D}
  D(\eta)=\frac{1}{20}\log{\big(\frac{5\eta_{\rm eq}}{\eta}+1\big)}.
\end{equation}

\subsection{{\bfit A semi-analytical model}}
\label{sema}

\noindent
The other key difficulty facing defect simulations is their
limited dynamic range.  At any one time, an evolving string network sources
significant power over a
length-scale range which exceeds an order of magnitude.
Hence, even for simulations with a dynamic range of two orders
of magnitude in conformal time, the power spectrum will only be
reliable over one order, even before taking box-size limitations into account.
Fortunately, however, we can invoke a semi-analytic model to compensate
for this missing power,\cite{newpicture,AScdm} which proves to be fairly
accurate in the scaling regimes away from the radiation-matter transition.
The procedure is essentially to square the expression (\ref{delta_tilde})
and then average over directions
to obtain the power spectrum ${\cal P}(k)$.  This becomes a Green function
integral over the UETC's
$\langle
\widetilde{\Theta}_+({\bf k},\eta)
\widetilde{\Theta}_+({\bf k'},\eta')
\rangle$:\cite{AScdm}
\begin{eqnarray}
  (2\pi)^3 {\cal P}(k) \delta^{(3)} ({\bf k}-{\bf k'})=
&
  16\pi^2 G^2\!\!
  \int_{\eta_{\rm i}}^{\eta_0}\!\!\!
  \int_{\eta_{\rm i}}^{\eta_0}\!\!
    \widetilde{\cal G}_{\rm c}(k; \eta_0, \eta)
    \widetilde{\cal G}_{\rm c}(k'; \eta_0, \eta')\nonumber\\
&
    \langle
    \widetilde{\Theta}_+({\bf k},\eta)
    \widetilde{\Theta}_+({\bf k'},\eta')
    \rangle
    \widetilde{F}(k, \eta)
    \widetilde{F}(k', \eta')
  d\eta
  d\eta'.
  \label{P_k_delta}
\end{eqnarray}
This can be simplified by noting that although there are two time integrals,
the only significant contributions come from times
when $\eta$ and $\eta'$ are reasonably close.
That is because the strings' configurations are uncorrelated when separated
by a sufficiently long time,
which is the correlation time $\eta_{\rm c}(k,\eta)$ as defined in 
(\ref{eta_c}).
Thus (\ref{P_k_delta}) can be simplified as
\begin{equation}
  \label{semi-ana}
  {\cal P}(k) =
  16\pi^2 G^2 \mu^2
  \int_{\eta_{\rm i}}^{\eta_0}
  |{\cal G}_{\rm c}(k;\eta_0,\eta)|^2 {\cal F}(k, \eta)
  \widetilde{F}^2(k, \eta)
  d\eta
  \,,
\end{equation}
where
\begin{eqnarray}
\!\!\!\!\!\!\!
  (2\pi)^3 \mu^2 {\cal F}(k, \eta) \delta^{(3)}({\bf k}-{\bf k'})
  \!\!\!\!& = & \!\!\!\!
  \frac{
    \int_{\eta_{\rm i}}^{\eta_0}
    \langle
    \widetilde{\Theta}_+({\bf k},\eta)
    \widetilde{\Theta}_+({\bf k'},\eta')
    \rangle
    \widetilde{\cal G}_{\rm c}(k'; \eta_0, \eta')
    \widetilde{F}(k', \eta')
    d\eta'
    }{
    \widetilde{\cal G}_{\rm c}(k; \eta_0, \eta)
    \widetilde{F}(k, \eta)
    }
  \label{FTR_UETC} \\
  & \approx &
  \int_{-\eta_{\rm c}}^{\eta_{\rm c}}
    \langle
    \widetilde{\Theta}_+({\bf k},\eta)
    \widetilde{\Theta}_+({\bf k'},\eta+\eta'')
    \rangle
  d\eta''
  \,.
  \label{FTR_pre}
\end{eqnarray}
The second step in eqn.~(\ref{FTR_pre}), is a reasonable approximation 
for the following reasons.
First,
the correlation time $\eta_{\rm c}$ scales with $\eta$ and is at most
$\eta$,
while the UETC is mainly contributed from a dynamic range of
$(1 \pm 0.2) \eta$ taking the half-maximum threshold
as mentioned in subsection \ref{coss}.
Second,
around $\eta_{\rm eq}$,
$\widetilde{\cal G}_{\rm c}(k'; \eta_0, \eta')$ changes only within a factor of
$2$ when going from $0.05 \eta_{\rm eq}$ to $5 \eta_{\rm eq}$
on sub-horizon scales.

The structure function ${\cal F}(k,\eta)$
can then be obtained as discussed below:
\begin{romanlist}
  \item The form of  ${\cal F}(k,\eta)$:
    Cosmic strings are line-like objects which move relativistically and 
    so the trajectories they sweep out will be two-dimensional.
    According to the study in section \ref{funa},
    we know sheet-like objects have power spectra
    ${\cal P}(k)\propto k^{-2}$ at small scales and
    ${\cal P}(k)\propto k^{0}$ at large scales.
    For cosmic string wakes,
    we get ${\cal P}(k)\propto k^{-2.25}$ on small scales
    because of their wiggliness (see subsection~\ref{coss}).
    According to (\ref{FTR_UETC}),
    because both $\widetilde{F} (k',\eta')$ and
    $\widetilde{\cal G}_{\rm c}(k'; \eta_0, \eta')$
    are scale-independent on small scales (see Fig.~\ref{Gkfig}),
    the slope of ${\cal F}(k,\eta)$ on these scales should be
    identical to that of (\ref{FTR_pre}).
    We have seen this slope to be $-2.25$ in Fig.~\ref{integ_UETC}.
    Although in the radiation-dominated regime,
    strings are more wiggly and therefore have a steeper
    small-scale slope than $-2.25$,
    we found it sufficient to use this value throughout for
    the current purposes.
    In addition,
    because cosmic strings scale with the horizon size,
    the turnover scale from ${\cal P}(k)\propto k^{-2.25}$ to
    ${\cal P}(k)\propto k^{0}$ scales with the conformal time.
    Adding all these factors together,
    we can schematically write down the structure function in the form:
    \begin{equation}
      {\cal F}(k,\eta) =
      \frac{{\cal E}(\eta)}{\left[
          1+\left({k}/{k_{\rm to}}\right)^n
        \right]^{2.25/n}} =
      \frac{{\cal E}(\eta)}{\left[
          1+\left(Ik\eta\right)^n
        \right]^{2.25/n}}\,,
      \label{ftr_form}
    \end{equation}
    where ${\cal E}(\eta)$ is the overall normalization against time,
    $I$ is a constant,
    $k_{\rm to}=(I\eta)^{-1}$ the turnover wave number,
    and the power $n$ controls the sharpness of the turnover behavior.
    We notice that the broad peak seen in Fig.~\ref{integ_UETC} is not
    modeled in (\ref{ftr_form}).
    However, we will see in subsection~\ref{semf}
    that such a simplification has a negligible effect on the final
    power spectrum once the phenomenological structure function
    ${\cal F}(k,\eta)$ is accurately calibrated by high-resolution simulations
    over a large dynamic range.
  \item The time dependence:
    A convenient way to investigate the time dependence ${\cal E}(\eta)$
    is to look at the horizon scale $k\approx \eta^{-1}$,
    at which ${\cal F}(k,\eta)|_{\rm k\eta=1}\approx {\cal E}(\eta)$.
    Scaling implies that:
    \begin{equation}
      \left.\widetilde{\Theta}_+\right|_{\rm hor}
      \equiv a^2\delta\rho_{\rm s}
      \propto a^2(\eta^3\rho_{\rm bg}^2)^{1/2}
      \propto \eta^{-1/2} \,.
    \end{equation}
    Thus by (\ref{FTR_pre}) the structure function is
    \begin{equation}
      \begin{array}{lll}
      \left.{\cal F}(k, \eta)\right|_{\rm hor}
      & \approx &
      \frac{1}{(2\pi)^3}
      \int_{-\eta_{\rm c}}^{\eta_{\rm c}}
      \left.\langle
      \widetilde{\Theta}_+({\bf k},\eta)
      \widetilde{\Theta}_+({\bf k},\eta+\eta'')
      \rangle\right|_{\rm hor}
      d\eta'' \vspace*{2mm}\\
      & \propto &
      \int_{-\eta_{\rm c}}^{\eta_{\rm c}}
        \eta^{-1/2}(\eta+\eta'')^{-1/2}
        d\eta''
      =
      {2\eta_{\rm c} \over \eta}+{\rm O}^2({\eta_{\rm c} \over \eta}) \,.
      \end{array}
      \label{FTR_real}
    \end{equation}
    Since cosmic strings scale with the horizon size in the scaling regime,
    the ratio ${\eta_{\rm c} / \eta}$ is a constant and therefore:
    \begin{equation}
      {\cal E}(\eta) \approx
      {\cal F}(k,\eta)|_{\rm hor} \approx
      {\rm constant}\, ,
    \end{equation}
\end{romanlist}
deep in the radiation or matter era.
This result is consistent with that presented in
Ref.\cite{AScdm}, but here we give a clear physical reason for the
behavior of the structure function and obtain an explicit expression for
the relationship between ${\cal F}(k, \eta)$ and the correlation time.
The longer the correlation time,
the larger the normalization of ${\cal F}(k, \eta)$.
Therefore an insufficient dynamic range when calculating the UETC's
or ${\cal F}(k,\eta)$ will result in a severe under-estimate
of the final power spectrum.

Instead of taking the approximation (\ref{FTR_pre}),
we use simulations to calibrate ${\cal F}(k, \eta)$.
This is a more accurate procedure specially in the matter era because 
$\widetilde{\cal G}_{\rm c}(k'; \eta_0, \eta')\equiv 
\widetilde{\cal G}_{\rm c}(\eta_0, \eta')$ decays linearly with $\eta'$.
By fitting the shape and amplitude
of the power spectrum calculated from (\ref{semi-ana})
with simulations of
limited dynamic range deep in the matter and radiation eras,
we estimated ${\cal F}(k, \eta)$ in these two regimes.
We were then
able to use an interpolation based on the actual behavior of
the string density during
the transition era (see Fig.\ \ref{str_density})
to provide the normalization of ${\cal F}(k, \eta)$ which interpolates
smoothly from the radiation to the matter era.
This turns out to fit well the shape of the actual evolution of ${\cal E}(\eta)$
in (\ref{ftr_form}).
We found $I=1/20$ and $n=8$ in (\ref{ftr_form}), and
the resulting structure function is:
\begin{equation}
  \label{FTR}
  {\cal F}(k, \eta)=\frac{{\cal E}(\eta)}{\left[1+(k\eta/20)^8\right]^{0.28125}}
  \;\;{\rm with}\;\;
  {\cal E}(\eta)=0.028\left[1-\frac{0.37322}{1+(4\eta_{\rm eq}/\eta)^{1.1}}\right]^2.
\end{equation}
In (\ref{FTR}), we note first that the turnover scale is the same for both
the radiation and the matter eras, i.e.\ $k_{\rm to}=20/\eta$.
This corresponds to exactly the correlation length of long strings
which we have seen in subsection~\ref{coss}.
Second,  ${\cal E}(\eta)$ in (\ref{FTR}) interpolates smoothly
from $0.028$ deep in the radiation era to $0.011$ deep in the matter era.
Now
with $\widetilde{\cal G}_{\rm c}(k; \eta_0, \eta)$
numerically calculated from eqns.~(\ref{two}) and (\ref{three}),
we can use the eqns.~(\ref{semi-ana}) and (\ref{FTR})
to extrapolate the lost power due to the limited
dynamical range in our numerical simulation.
We notice that we have not considered the effect from loops here which 
will be discussed in subsection~\ref{final_loops}.

\subsection{{\bfit The dependence of $S(k)$ on $\Omega_{\rm m0}$, $\Lambda$, and $h$}}
\label{depome}
\noindent
Since an open universe is strongly favoured by many observations,
and the
existence of a cosmological constant is studied 
in many inflationary models,\cite{Peter,Liddle}
it is natural to explore the spectrum in these two regimes.
To do this, we first introduce a simple rescaling scheme
to extrapolate the simulated $\Omega_{\rm m0}=1$ and $\Lambda=0$ 
spectrum to open and $\Lambda$-models,
and then verify its accuracy.
The rescaling scheme is (adapted from Ref.\cite{AveCal1}):
\begin{equation}
  \label{Sopen}
  S(k,h,\Omega_{\rm m0},\Omega_{\Lambda 0}) = S(k,1,1,0) \times
  \Omega_{\rm m0}^2 {h}^4 \times f^2(\Omega_{\rm m0},\Omega_{{\Lambda 0}}) \times
  g^2(\Omega_{\rm m0},\Omega_{{\Lambda 0}}),
\end{equation}
where $k$ is in units of $\Omega_{\rm m0}h^2 \, {\rm Mpc}^{-1}$,
$f(\Omega_{\rm m0},\Omega_{{\Lambda 0}})$ and $g(\Omega_{\rm m0},
\Omega_{{\Lambda 0}})$ are given by
\begin{eqnarray}
  f(\Omega_{\rm m0},\Omega_{{\Lambda 0}})
  & = &
  \left\{
    \begin{array}{ll}
      \Omega_{\rm m0}^{-0.3} & {\rm for\ } \Omega_{\rm m0}\leq 1,\
\Omega_{\Lambda 0} = 0 \\
      \Omega_{\rm m0}^{-0.05} & {\rm for\ } \Omega_{\rm m0}+\Omega_{\Lambda 0} = 1
    \end{array}
  \right.\ ,
  \label{fomega}
  \\
  g(\Omega_{\rm m0}, \Omega_{\Lambda 0})
  & = &
  \frac{5\Omega_{\rm m0}}
  {2\left[\Omega_{\rm m0}^{4/7}-\Omega_{\Lambda 0} +
      (1+\Omega_{\rm m0}/2)(1+\Omega_{\Lambda 0}/70)\right]}\ .
  \label{gomega}
\end{eqnarray}
In (\ref{Sopen}), the leading factor $\Omega_{\rm m0}^2 h^4$ results from the
fact that the ratio of scale factors $a_0/a_{\rm eq}$ is
proportional to $\Omega_{\rm m0} h^2$ and
that
the Green function (\ref{T_k}) is proportional to this ratio.
The middle factor
$f(\Omega_{\rm m0},\Omega_{\Lambda 0})$ reflects the dependence of
the COBE normalization of $G\mu$ on the
cosmological parameters $\Omega_{\rm m0}$ and $\Omega_{\Lambda 0}$.
It changes upwards as we decrease the matter density
in an open or $\Lambda$-universe.
The last factor $g(\Omega_{\rm m0}, \Omega_{\Lambda 0})$ takes into account
the fact that in an open or
$\Lambda$-universe the linear growth of density perturbations is
suppressed relative to an $\Omega_{\rm m0}\!=\!1$ and $ \Omega_{\Lambda 0}\!=\!0$
universe\cite{Carroll}
(also verified in Ref.\cite{Einstein} for primordial perturbations).
There is also a rescaling of $k$ implicit in (\ref{Sopen})
(the unit of $k$ here is in $\Omega_{\rm m0}h^2 \, {\rm Mpc}^{-1}$
rather than $h{\rm Mpc}^{-1}$).
This rescaling is
due to the fact that
the horizon size at radiation-matter equality is proportional to
$(\Omega_{\rm m0} h^2)^{-1}$.
Hence the physical grid spacing in our simulation
should be rescaled by this factor,
since our length scale is in the unit of $\eta_{\rm eq}$.

\begin{figure}[htbp]
\vspace*{13pt}
  \centering\epsfig{figure=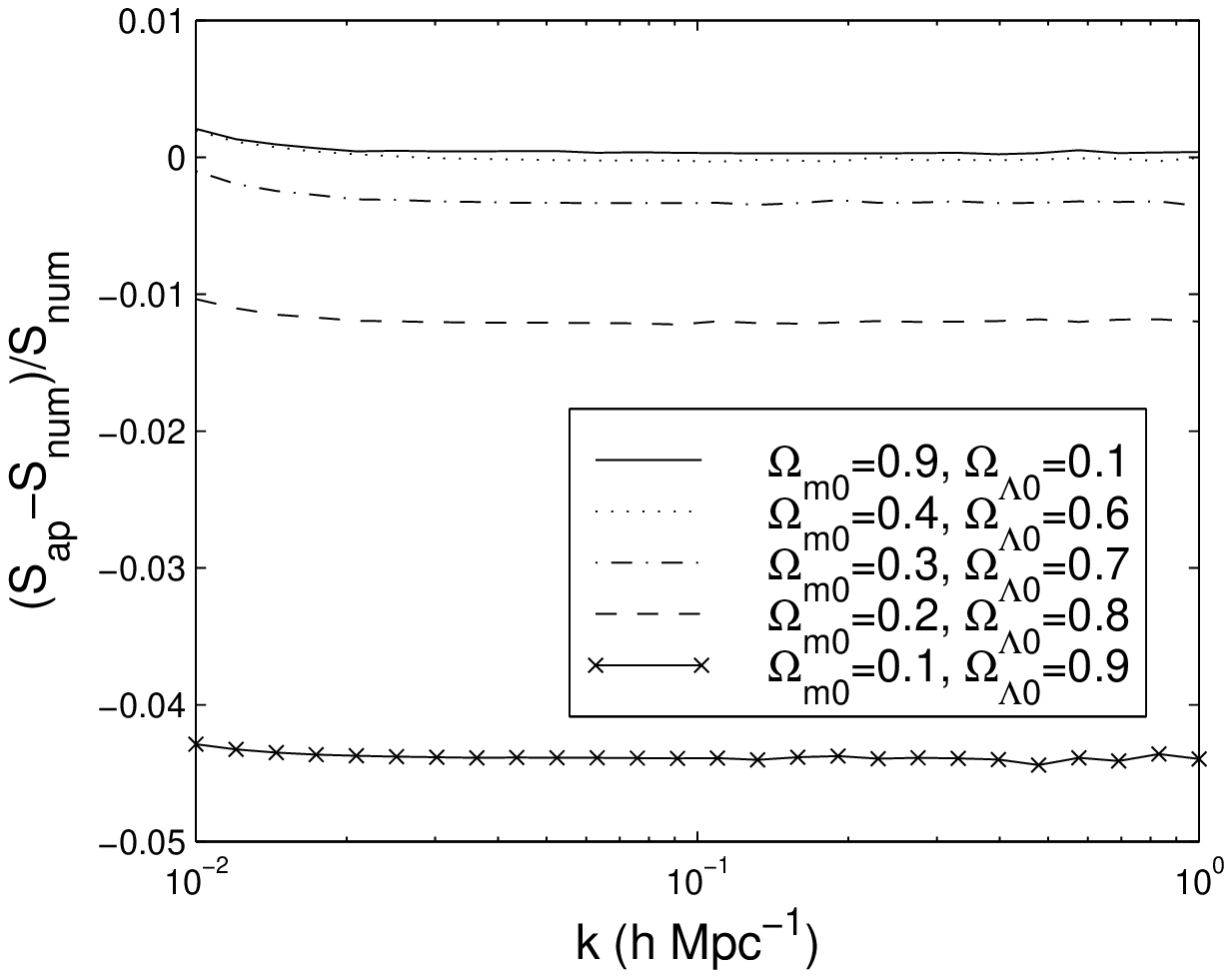, width=4.5in}\\
  \centering\epsfig{figure=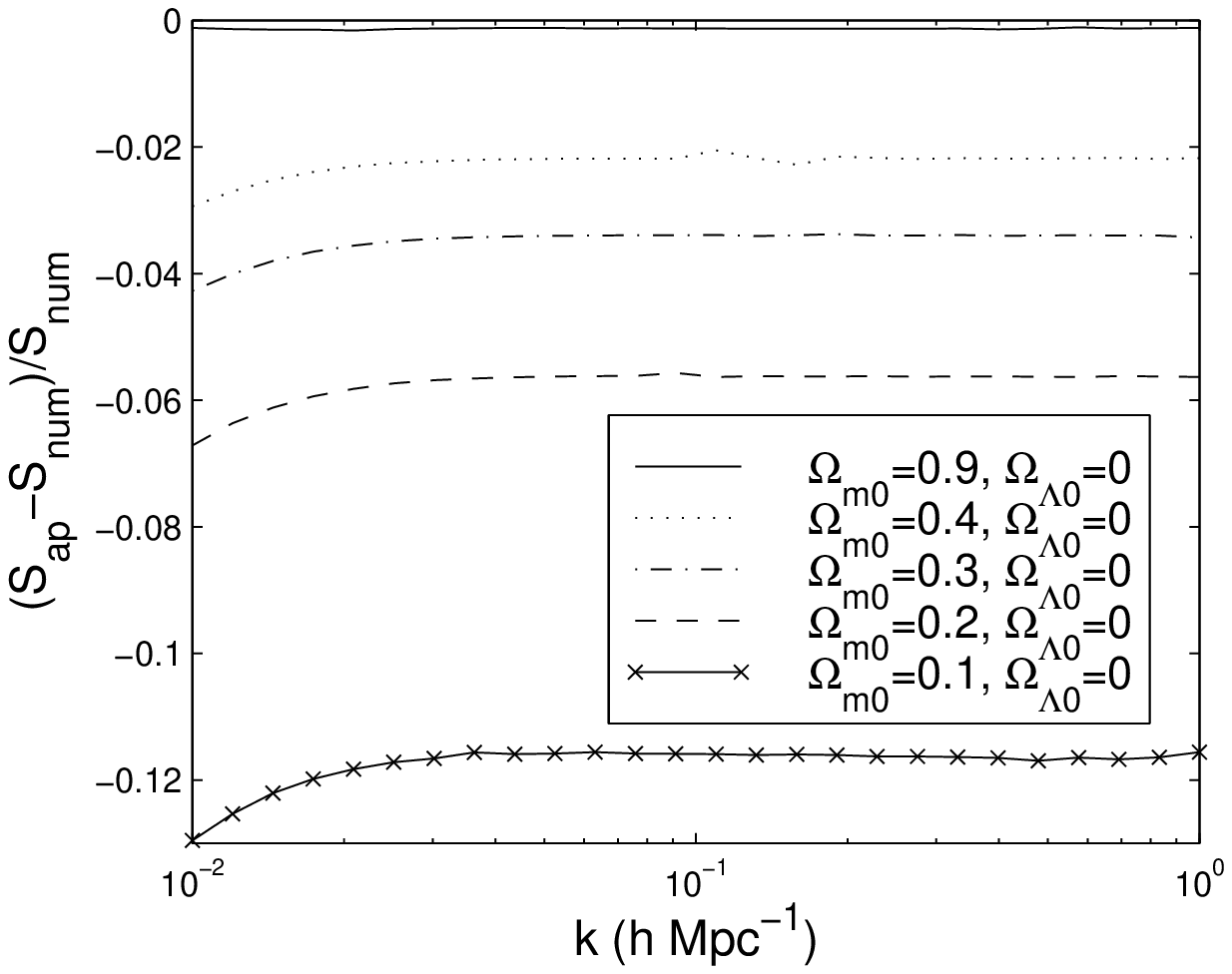, width=4.5in}\\
\vspace*{13pt}
 \fcaption
  {The accuracy of the rescaling scheme (\ref{Sopen}) for cosmic string models.
    $S_{\rm ap}\equiv 4\pi k^3 {\cal P}(k)$ is 
    obtained from a flat $\Lambda=0$ model using (\ref{Sopen}),
    while $S_{\rm num}$ is the numerical result by solving perturbation
    equations with various choices of
    $\Omega_{\rm m0}$ and $\Omega_{\Lambda 0}$.
    }
  \label{Sopen_acc}
\end{figure}

To verify the accuracy of this scheme, we use (\ref{semi-ana}) together with
(\ref{FTR}) and the Green functions obtained from solving (\ref{two})
and (\ref{three}) numerically,
to get the spectrum $S_{\rm num}(k)$
for various choices of $\Omega_{\rm m0}$ and $\Omega_{\Lambda 0}$.
We then compare this with $S_{\rm ap}(k)$ which is
extrapolated from a $K=\Lambda=0$ model using (\ref{Sopen}).
The initial time of the integral is set to $\eta_{\rm i}=0.1$,
at which both the curvature and $\Lambda$ terms are negligible,
and the final time is today.
Fig.\ \ref{Sopen_acc} shows that on the scales of interest,
the rescaling scheme (\ref{Sopen}) is accurate
within few percent error for reasonable choices of $\Omega_{\rm m0}$
and $\Omega_{\Lambda 0}$.
A similar independent work regarding the accuracy of (\ref{Sopen})
in generic defect models is also done in Ref.\cite{AveCar}.

\begin{figure}[htbp]
\vspace*{13pt}
  \centering\epsfig{figure=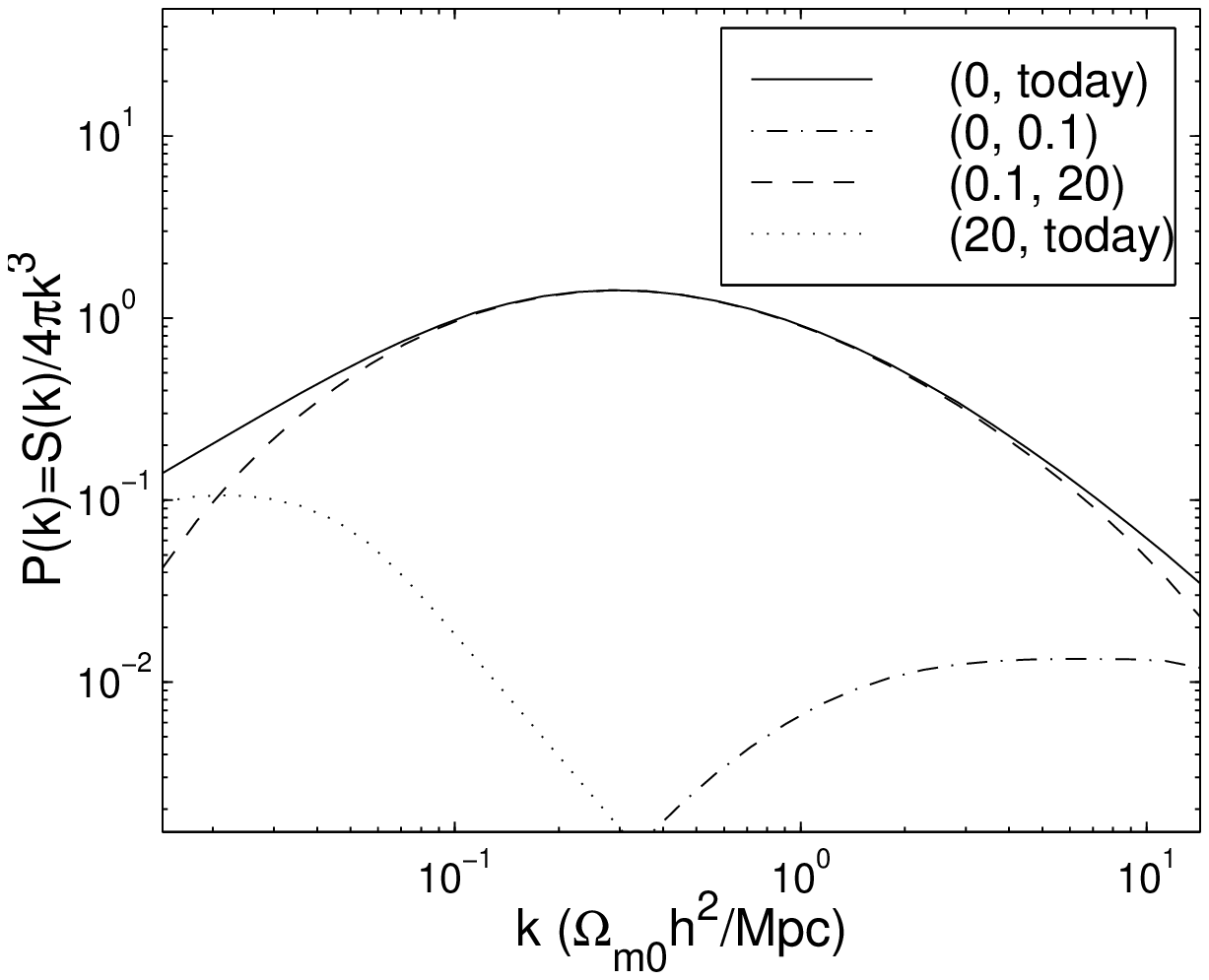, width=3.6in}\\
\vspace*{13pt}
 \fcaption
  {The relative contribution in the power spectrum from strings at different
    epochs: the total (solid line), the early times (dot-dashed line),
    around $\eta_{\rm eq}$ (dashed line), and
    the late times (dotted line).
    The numbers in brackets specify in $\eta_{\rm eq}$ the initial
    and final time of each epoch.
    }
  \label{Peml}
\end{figure}

Fig.~\ref{Peml} shows the relative contribution of different epochs to
the total power spectrum.
We can see that most power on the scales of interest is seeded around
$\eta_{\rm eq}$,
especially for models with $\Omega_{\rm m0}<1$.
As illustrated in section~\ref{perequ},
this is because at early times in the radiation-dominated regime,
the growth of small-scale perturbations are suppressed by pressure, while
at late times in the matter-dominated regime,
the large-scale perturbations have less time to grow.
This implies that the late time non-scaling behavior of 
cosmic strings has a negligible effect on matter density perturbations on 
scales of interest. We verified that for wavenumbers larger than 
$k\sim 0.01 h{\rm Mpc}^{-1}$ the late time non-scaling strings contribute 
less than $3 \%$ to the total power spectrum for any reasonable choices 
of the cosmological parameters (see also Ref.\cite{AveCar}). 


\section{Results and Discussion}
\label{resdis}

We first perform string simulations with a string sampling comoving spacing of
$1/1000$ of the simulation box sizes.
The dynamic ranges cover from $0.05$ to $300$ $\eta_{\rm eq}$,
with each single run having a dynamic range of at least 10 in conformal time.
We then perform the structure formation simulations
with comoving box sizes ranging from $5$--$130h^{-1}$Mpc today,
and a resolution of $128^3$--$512^3$.
Our simulations were carried out on
the UK Computational Cosmology Consortium supercomputer COSMOS,
a Silicon Graphics Origin2000 with 20 Gbytes main memory.
The code was parallelized to enhance performance.
We also used the SGI math libraries to implement the
Fast Fourier Transform.

\subsection{{\bfit The semi-analytic fit}}
\label{semf}
\begin{figure}[htbp]
\vspace*{13pt}
  \centering\epsfig{figure=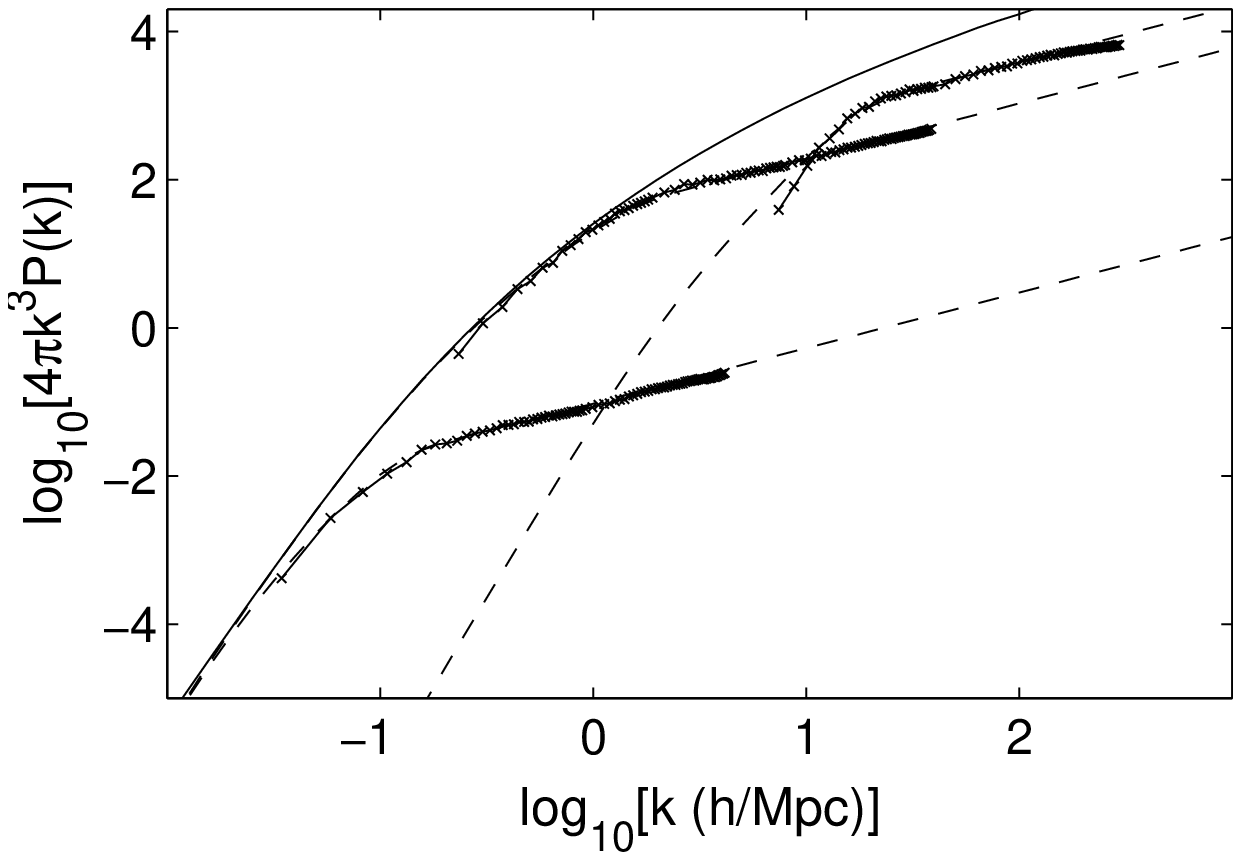, width=3.7in}\\
\vspace*{13pt}
 \fcaption
  {The comparison of our semi-analytical fit and our simulation result
    for the CDM model excluding loops.
    The top-right, central, and bottom-left solid lines with crosses are the
    simulation results in the deep radiation,
    transition through to deep into the matter,
    and deep matter eras respectively.
    The dashed lines are our semi-analytical fits corresponding to the same
    dynamical ranges of those simulations. They show a good agreement with
    each other.
    The solid line is the semi-analytical fit with a full dynamic range from
    $\eta_{\rm i}=0$ to today.
    }
  \label{open_fig1}
\end{figure}

In Fig.~\ref{open_fig1} we plot the CMBR normalized linear power
spectrum
($G\mu_6 = G\mu \times 10^6 = 1.7$,\cite{ACDKSS}
the most recent COBE normalization for strings)
induced by long strings in an $\Omega_{\rm c0}=1$ CDM cosmology
with $h=0.7$.
The
central set of numerical points
was sourced by string network simulations beginning at $\eta=0.4\eta_{\rm
eq}$ which were continued for 1318 expansion times in simulation boxes
ranging from 32--128$h^{-1}$Mpc, with a maximum resolution of $512^3$
grid points.    Given the dynamic range limitations we
have also plotted the semi-analytic fit (\ref{semi-ana}) over the full
range of wave-numbers.  The
good agreement with the semi-analytic model is illustrated by the
dashed line fits to central points, as well as to the
short normalization runs in the matter and radiation eras (also shown in
Fig.\ \ref{open_fig1}).
We can see
the little discrepancy between the semi-analytic fits and simulations
on both the large-scale and the small-scale ends.
This resulted from the fact that towards the beginning and the end of
the simulations, the UETC does not fully contribute to the simulation
power spectrum (see eq.~(\ref{FTR_pre})).
Nevertheless, at large scales ($k\sim 0.03$)
the full-dynamic-range run of the semi-analytic fit (the solid line)
is well constrained by the deep-matter-era simulation.
Given this close correspondence, which was also exhibited in
HDM simulations,
we have considerable confidence
that this approximation can reproduce the correct shape and amplitude of
the string simulation power spectrum.

\begin{figure}[htbp]
\vspace*{13pt}
  \centering\epsfig{figure=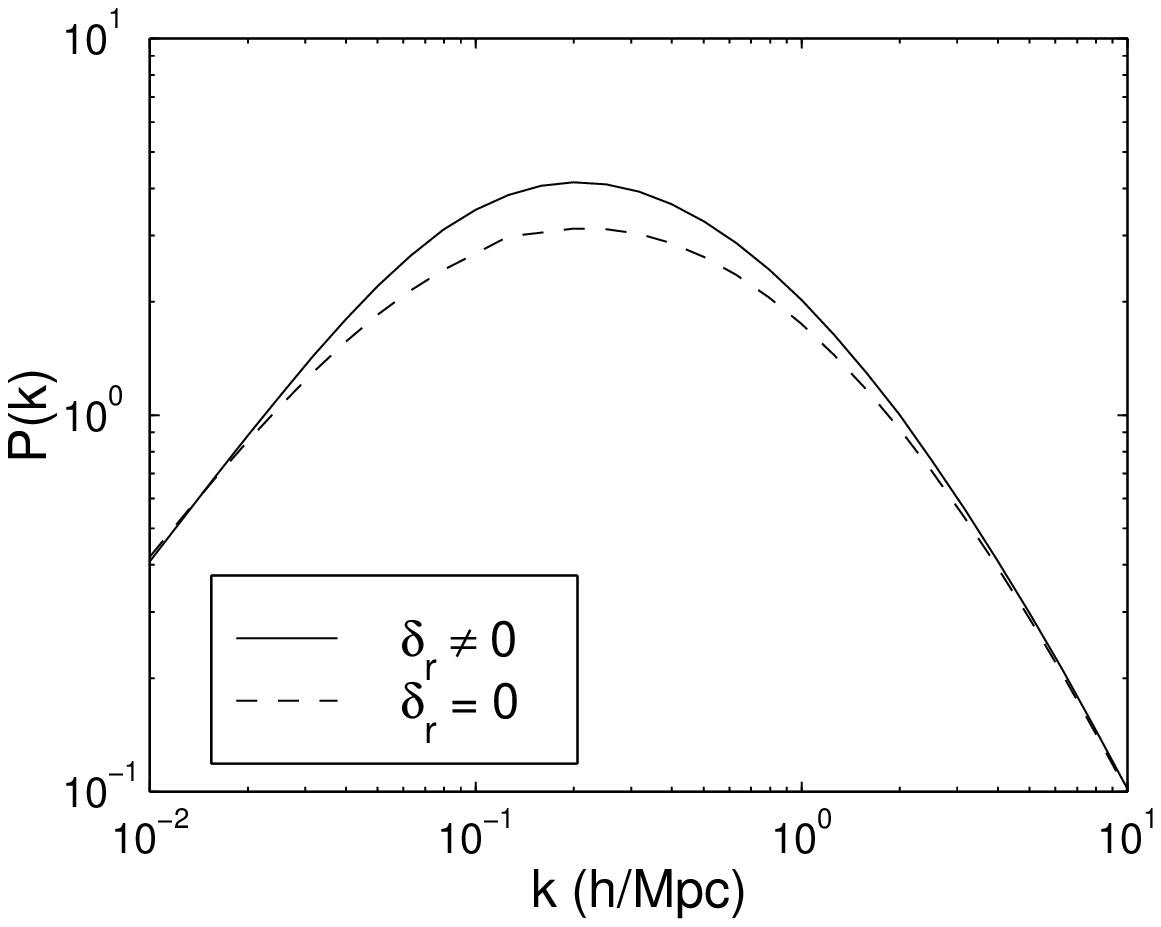, width=4in}\\
\vspace*{13pt}
 \fcaption
  {The effect of radiation perturbations in the cosmic string-seeded
    CDM spectrum. The dynamic range goes from $\eta=0$ to today.}
  \label{rad_imp}
\end{figure}

Fig.\ \ref{rad_imp} shows
how the radiation perturbations affect the matter perturbation power spectrum.
The solid line is obtained  using eqns.~(\ref{two}) and (\ref{three})
with $\Lambda=K=0$,
while
the dashed line is obtained by setting $\delta_{\rm r}=0$ 
in eqn.~(\ref{two}) and hence ignoring eqn.~(\ref{three}).
We can see that the radiation perturbations have
effect only on intermediate scales.
They boost the power spectrum by about $30\%$ at most on scales
$k\sim 0.2h{\rm Mpc}^{-1}$.
This is because at very small scales,
radiation will oscillate many times per
expansion time and will have little net effect on the matter, while
at very large scales,
most power is contributed from the matter era in which the effect from
radiation is negligible.
Although ignoring $\delta_{\rm r}$ has its convenience for numerical purposes,
the assumption of a smooth radiation background
would have ignored some intermediate-scale power
expected to result from radiation perturbations during radiation-matter
transition.

\begin{figure}[htbp]
\vspace*{13pt}
  \centering\epsfig{figure=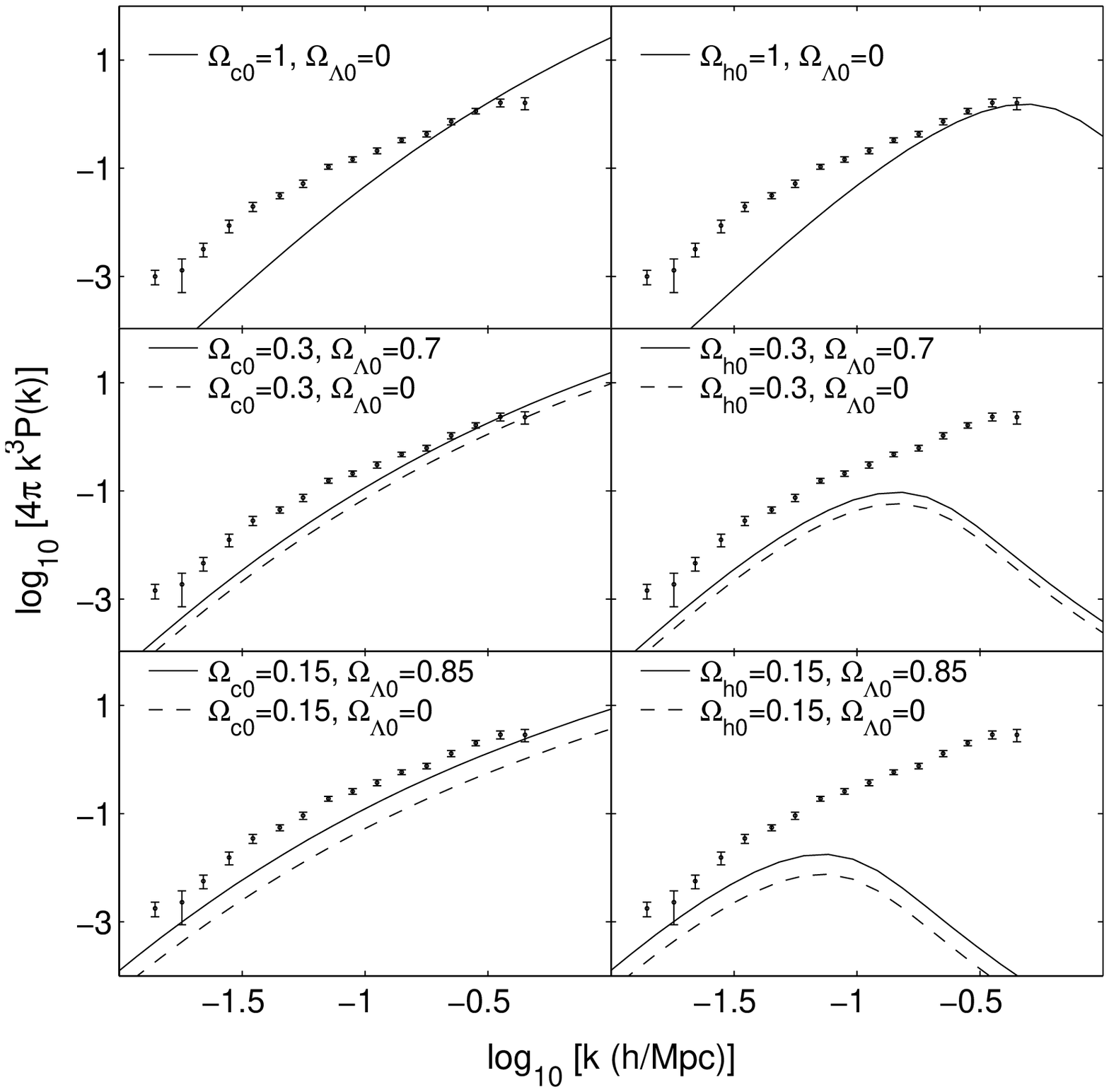, width=5in}\\
\vspace*{13pt}
  \fcaption
  { The spectra of
    cosmic strings (excluding loops)
    with CDM (left) and HDM (right)
    models in different background cosmologies.        
    The solid lines are the flat models with
    $\Omega_{\rm m0} + \Omega_{\Lambda 0}=1$;
    the dashed lines are the open models with
    $\Omega_{\Lambda 0}=0$.
    In an open universe with $\Lambda=0$, the spectrum has more large-scale
    power and less small-scale power as
    $\Omega_{\rm m0}$ decreases.
    In a flat universe made by $\Lambda$, the spectrum has
    a scale-independent boost over the previous case with the same
    $\Omega_{\rm m0}$.
    Here we use $h=0.7$ and $G\mu_6=1.7$,\cite{ACDKSS} which is the most
    recent COBE normalization for cosmic strings.
    The data points with error bars are the reconstructed linear spectrum
    by Peacock and Dodds.\cite{PD}
    }
  \label{open_fig2}
\end{figure}
The results presented here are obtained by integrating (\ref{semi-ana}),
with ${\cal F}(k,\eta)$ provided by (\ref{FTR})
and $\widetilde{\cal G}_{\rm c}(k; \eta_0, \eta)$
numerically obtained from (\ref{two}) and (\ref{three}).

\subsection{{\bfit The $\Omega_{\rm m0}=1$ and $\Lambda=0$ model}}
Consider first the $\Omega_{\rm c0}=1$ CDM model (I).
An empirical formulae which can reproduce our result
excluding loops within a maximum of
$10\%$ error for $k=0.01$--$100 h{\rm Mpc}^{-1}$ is obtained as:
\begin{equation}
  \label{Pkc_empirical}
  S_{\rm CDM}(k)=(G\mu_6)^2 41.6 (0.7q)^{p(q)}\,,
\end{equation}
where $q=k/\Gamma$ and
\begin{equation}
  \label{Pkc_empirical_power}
  p(q)=3.9-\frac{2.7}{1+(2.8q)^{-0.44}}\,.
\end{equation}
This fit is obtained using $h=1$.
The $\Gamma$ here is called the `shape parameter',
which accounts for the rescaling in $k$ due to different choices of
cosmological parameters.
In the $\Omega_{\rm B0}=0$ cosmologies,
$\Gamma$ is simply $\Omega_{\rm m0}h$,
which reflects the scaling in the horizon size at $\eta_{\rm eq}$.
As for the $\Omega_{\rm B0}\neq 0$ models,
the shape parameter $\Gamma$ will be also a function of $\Omega_{\rm B0}$
as we shall discuss later.

We calculated the standard deviation $\sigma_8$ by convolving
our CDM perturbations with a spherical window of radius $8 \, h^{-1} \,$Mpc
to find
$\sigma_{\rm 8(sim)}(h=0.5)=0.32 G\mu_6$,
$\sigma_{\rm 8(sim)}(h=0.7)=0.39 G\mu_6$ and
$\sigma_{\rm 8(sim)}(h=1.0)=0.47 G\mu_6$.
A comparison with the observational data
points is shown at the top panel of Fig.~\ref{open_fig2}.
It indicates that strings appear to induce an excess of
small-scale power and
a shortage of large-scale power, that is, the $\Omega_{\rm c0}=1$ string
model excluding loops requires a strongly scale-dependent bias.
This is not necessarily a fatal flaw for the model on small scales
because, as the corresponding HDM spectrum indicates, such excess power
can be readily eliminated in a mixed dark matter model.  However,
the problem is less tractable on large scales where
biases up to $\sigma_{\rm 100(obs)}/\sigma_{\rm 100(sim)}\approx 3.9$
around 100$h^{-1}$Mpc
might be inferred from the data points
(using $G\mu_6= 1.7$ and $h=0.7$).
Should we, therefore, rule out string models on this basis%
?\fnm{a}\fnt{a}{In Ref.\cite{against},
  the authors used a numerical fit for the observational data
  to obtain $\sigma_{100{\rm (obs)}}=3.5\times 10^{-2}$,
  and compared this with
  their simulated spectrum to yield $b_{100}=5.4$. They concluded that cosmic
  string theory is ruled out because of this high bias factor. However,
  according to the observational data\cite{PD} (see Fig.\ \ref{open_fig2}),
  there are only a few data points (with big error bars) below the scale of
  $100 h^{-1}$Mpc, so this seems a potentially unreliable way to estimate
  $\sigma_{100{\rm (obs)}}$.
  }
While acknowledging that the
model (I) linear spectrum looks unattractive, there are still three
important mitigating factors.  First, the determination of the
power spectrum on large scales around 100$h^{-1}$Mpc remains uncertain
(e.g. the observational data in Fig.~\ref{open_fig2}
has at least a 40\% normalization uncertainty).
Even with the normalization based on new galaxy surveys\cite{HTP,Einasto}
and cluster abundance\cite{Henry00},
this uncertainty still remains as large as 40\% 
at 95\% confidence level\cite{mycluster01}.
Nevertheless, it will be superseded by much larger more reliable
data sets in the near future.
Secondly, the immediate nonlinearity of
string wakes indicates that strong biasing mechanisms might operate, as
illustrated on large scales in the post-recombination hydrodynamic
simulations of Ref.\cite{SorBra}.
Thirdly, cosmic string loops have not been included here,
and we will see how they release this problem later.
Finally, unlike inflation,
defect models have never been wedded to an $\Omega=1$ cosmology.

\begin{figure}[htbp]
\vspace*{13pt}
  \centering\epsfig{figure=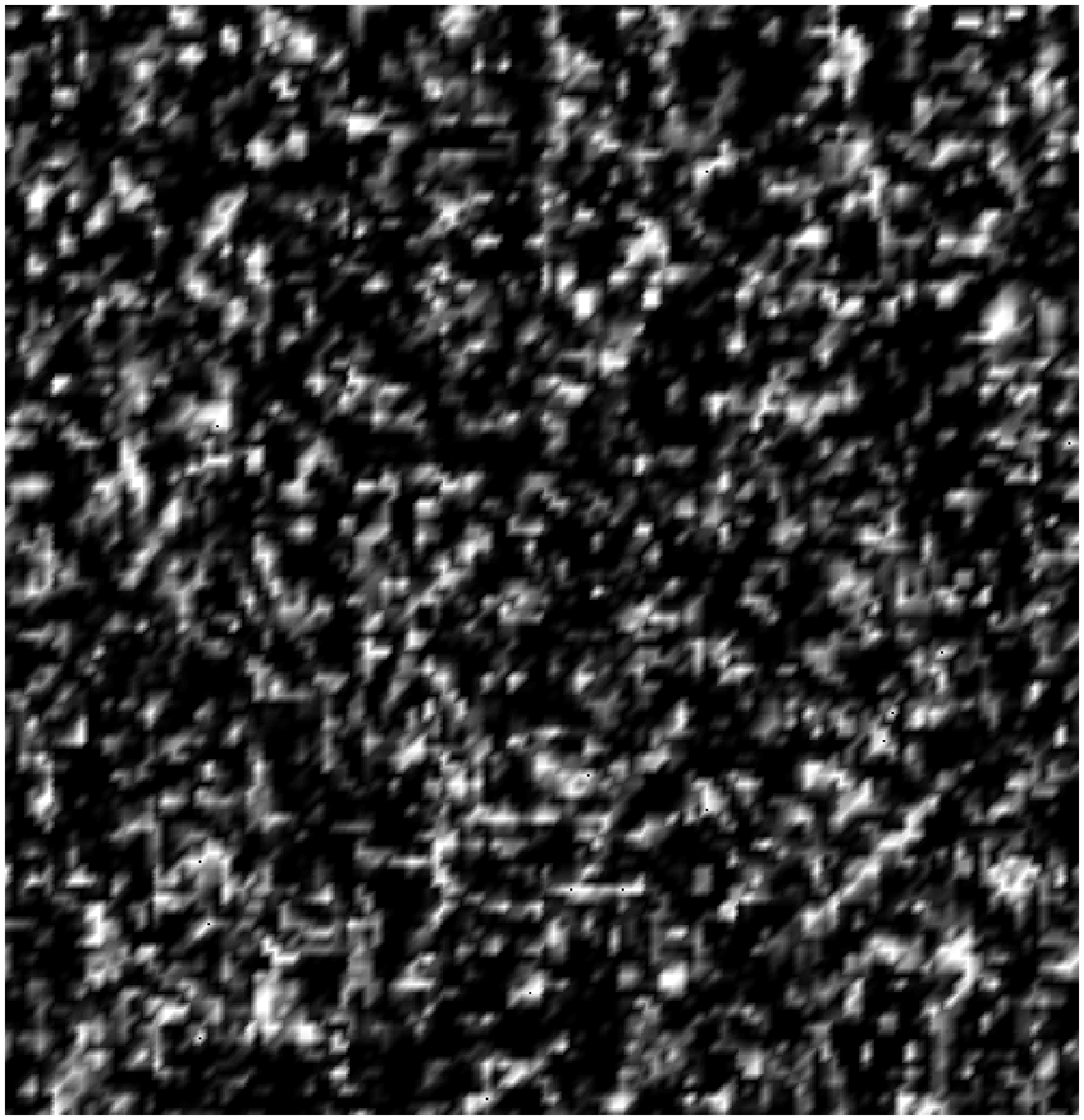, width=3.5in}\\
  \centering\epsfig{figure=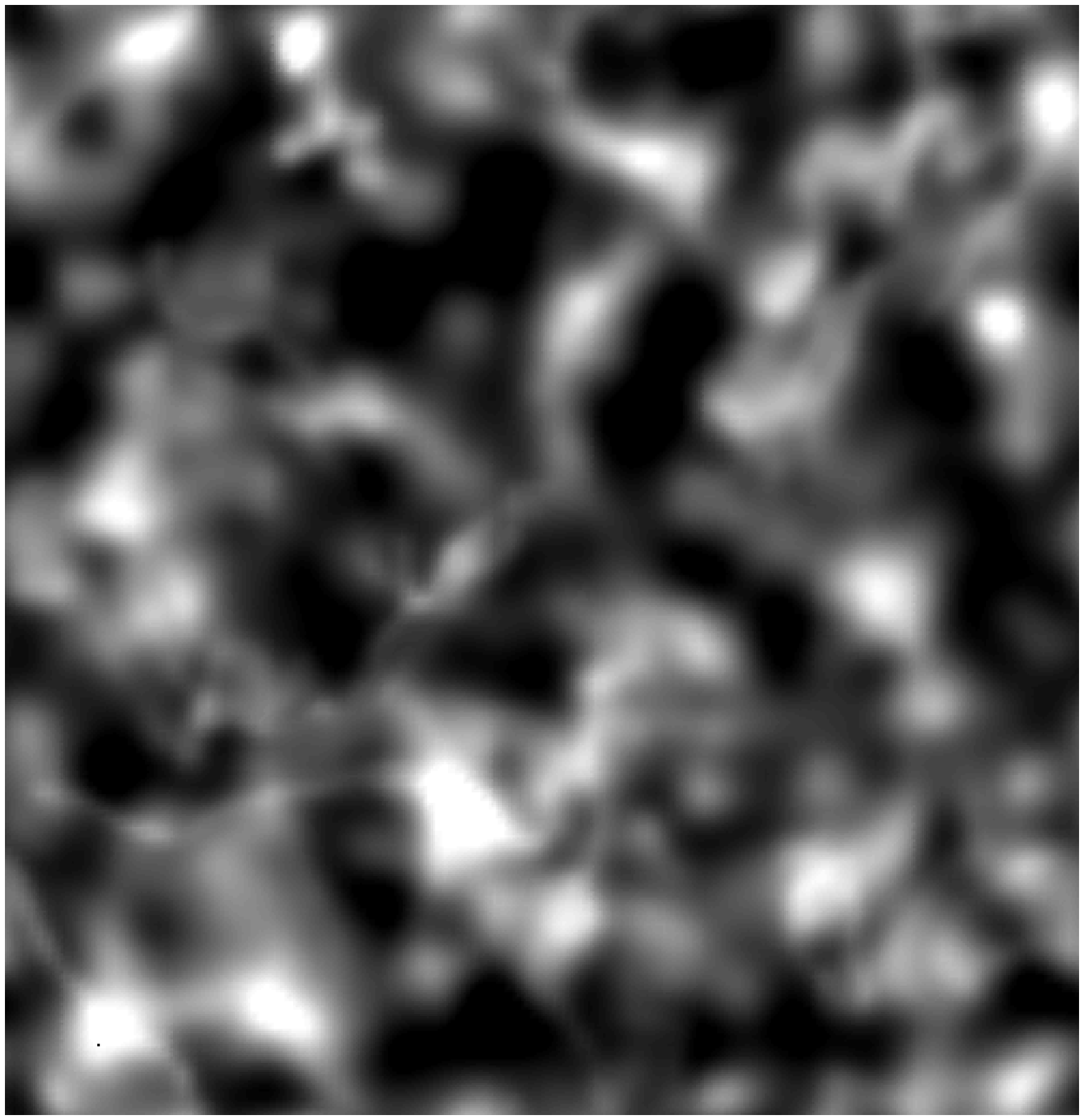, width=3.5in}\\
\vspace*{13pt}
  \fcaption
  {A comparison between CDM (top) and HDM (bottom) results.
    They are taken from slices in simulation box of size$(128 h^{-1}{\rm Mpc})^3$,
    with exactly the same string source.
    They look the same on large scales, but the HDM model has no
    fine structure.
    The color scheme is chosen so that
    the non-Gaussian feature can be clearly distinguished.
    }
  \label{CDMHDM}
\end{figure}

As for the HDM results, the small-scale power ($k \gsim 0.5 h{\rm Mpc}^{-1}$)
is removed by the neutrino free-streaming,
while on large scales it remains the same as that of the CDM model.
An empirical formulae (excluding the effect of loops and baryons)
within a maximum of
$10\%$ error for $k=0.01$--$100 h{\rm Mpc}^{-1}$ is given by:
\begin{equation}
  \label{Pkh_empirical}
  S_{\rm HDM}(k)=(G\mu_6)^2 
  \frac{708}{\left[(0.7q)^{-1.2}+5.3q\right]^{3.3}}\,,
\end{equation}
where again $q=k/\Gamma$ and $\Gamma$ is the shape parameter.
This fit again has used $h=1$.
Fig.~\ref{CDMHDM} shows a comparison between the mass distributions of
CDM and HDM models.
They are slices of a simulation box
of size $(128 h^{-1}{\rm Mpc})^3$,
with exactly the same string source.
They look the same on larger scales, but the
HDM model has no fine structures on smaller scales
due to the neutrino free-streaming.
Their non-Gaussian features can also be seen.
Ref.\cite{AveShe4} has investigated their non-Gaussian properties in detail,
and concluded that
on scales 
smaller than $1.5{(\Omega h^2)}^{-1}$Mpc, perturbations seeded by
cosmic strings are very non-Gaussian.
These scales may still be in a linear or mildly non-linear regime
in an open or $\Lambda$-universe with $\Gamma=\Omega h \lsim 0.2$.

\subsection{{\bfit The cases $\Omega_{\rm m0}<1$ and $\Lambda\neq 0$}}
\label{tca}
\begin{figure}[htbp]
\vspace*{13pt}
  \centering\epsfig{figure=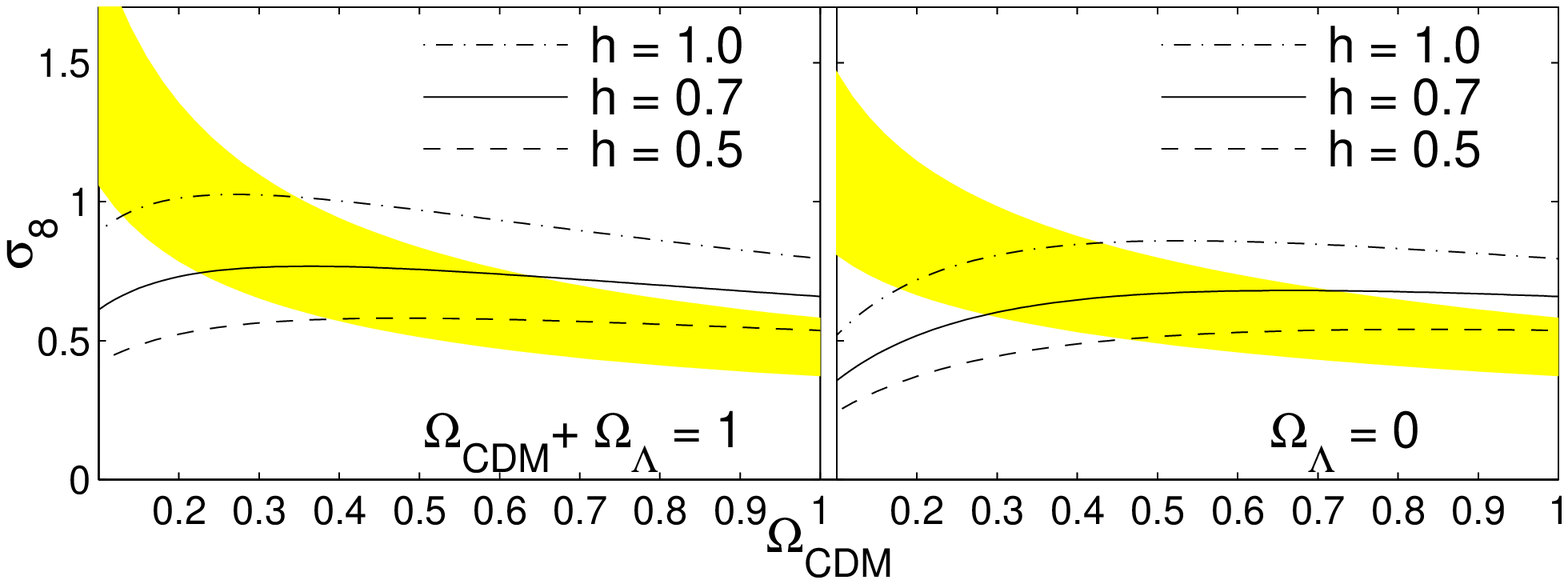, width=5in}\\
\vspace*{13pt}
  \fcaption
  {The comparison of the observationally inferred standard deviation at the
    scale 8 $h^{-1}$Mpc, $\sigma_{\rm 8(obs)}$, and
    that induced from our simulation, $\sigma_{\rm 8(sim)}$.
    We have used the COBE normalization
    $G\mu_6(\Omega_{\rm c0}=1,\Omega_{\Lambda 0}=0)=1.7$.\cite{ACDKSS,AveCal1}
    $\sigma_{\rm 8(obs)}$ is shown as the shaded area;\cite{mycluster01}
    $\sigma_{\rm 8(sim)}$ is plotted as
    dot-dashed ($h=1.0$),
    solid ($h=0.7$) and
    dashed($h=0.5$) lines. 
    }
  \label{open_fig3}
\end{figure}

We can observe from Fig.\ \ref{open_fig2}, that for open or
$\Lambda$-cosmologies
with $\Omega_{\rm m0}\approx 0.1$--$0.3$, the string $+$ CDM power
spectrum is much more encouraging.
We find that the bias on large scales
is much less scale-dependent.
For example, in model (IV)  ($\Omega_{\rm m0} =0.15$,
$\Omega_{\Lambda 0} =0.85$), the relative bias remains
$\sigma_{\rm 100(obs)}/\sigma_{\rm 100(sim)}\approx 1.4 \pm 0.2$
at $100h^{-1}$Mpc.
In Fig.\ \ref{open_fig3}, we plot $\sigma_8$ for
the full gamut of open and $\Lambda$ models.
$\sigma_{\rm 8(sim)}$ induced from our simulations with $G\mu_6$
normalized by COBE\cite{ACDKSS,AveCal1} is presented as
the dot-dashed ($h=1.0$),
the solid ($h=0.7$) and
the dashed ($h=0.5$) lines;
$\sigma_{\rm 8(obs)}$ is the shaded area (95\% confidence region), 
which is recommended by Wu,\cite{mycluster01} 
and which incorporates the latest techniques 
(e.g., the inclusion of non-spherical-collapse models as 
an improved alternative to the Press-Schechter formalism)
and large-scale data (new galaxy surveys and cluster data)
to improve earlier results.\cite{VL,ECF,WEF}
We can see from Fig.\ \ref{open_fig3} that
the consistency test between
$\sigma_{\rm 8(obs)}$ and $\sigma_{\rm 8(sim)}$ indicates,
for $h=1.0, 0.7, 0.5$ respectively,
that
$0.15\gsim\Omega_{\rm m0}\gsim 0.35$,
$0.2\gsim\Omega_{\rm m0}\gsim 0.6$ and
$\Omega_{\rm m0}\gsim 0.4$ when $\Omega_{\rm m0}+\Omega_{\Lambda 0}=1$,
and that
$0.2\gsim\Omega_{\rm m0}\gsim 0.45$,
$0.3\gsim\Omega_{\rm m0}\gsim 0.7$ and
$\Omega_{\rm m0}\gsim 0.45$ when $\Omega_{\Lambda 0}=0$.
For $h=0.7$ in both $\Lambda$ and open models,
the ratio 
$\sigma_{\rm 8(obs)}/\sigma_{\rm 8(sim)}\lsim 2$
for all $\Omega_{\rm m0}\gsim 0.1$.
Combined with an analysis similar to Fig.\ \ref{open_fig2}, we found that
for
$\Gamma=\Omega_{\rm m0} h=0.1$--$0.2$,
both $\sigma_{\rm 8(sim)}$ and the shape of spectrum induced from cosmic
string model match observations within acceptable uncertainties.
Hence, an open or $\Lambda$-cosmology in the context of string $+$ CDM model
seems to show remarkable agreement between observations and
the results of our numerical simulations.
Indeed, similar comparisons
made in an inflationary context with models like I--V appear to require
a more strongly scale-dependent bias.\cite{Peter}

As for the HDM results, the comparison with observation
seems to require a strongly scale-dependent bias for any choice of the
cosmological parameters (models I--V). However,
the lack of small-scale power may be partially overcome, at least,
if baryons are properly included in the analysis. Further investigation
using a hydrodynamical code will be required to investigate
if galaxies form early enough.

\subsection{{\bfit Loop effect}}
\label{final_loops}

\begin{figure}[htbp]
\vspace*{13pt}
\centering 
\leavevmode\epsfxsize=3.8in \epsfbox{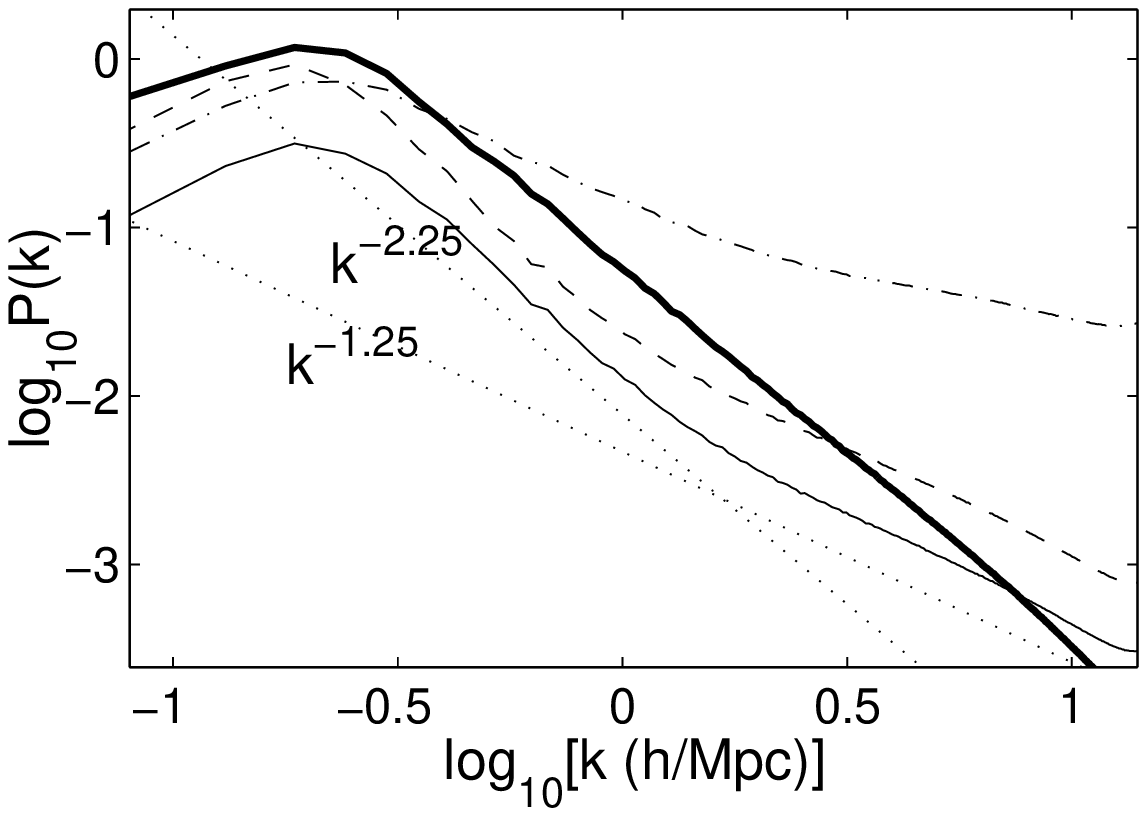}\\ 
\vspace*{13pt}
 \fcaption
 {Small dynamic range power spectra of density perturbations seeded
   by long strings (thick solid),
   by loops with initial velocities $v_*$ switched to zero (dot-dashed), and
   by loops with $v_*$ determined by string network evolution
   (dashed and thin solid).
   The thin solid line includes the effect of
   gravitational decay of the loop energy,
   while the other two loop lines don't but with loops removed 
   after a period of time $\tau_*=t_*$.
   }
 \label{figure2}
\end{figure}

Now let us include the effect from loops.
Figure \ref{figure2} shows the power spectrum of density perturbations
induced by long strings and by cosmic string loops for $f=1$ with a small 
dynamic range from $2.5$ to $5\eta_{\rm eq}$.
We can see that
when compared with the spectrum induced by static loops (dot dashed),
the amplitude of small-scale perturbations induced by moving loops (dashed)
is clearly reduced by their motion.  However, their large-scale power is 
higher because of the dependence of the gravitational interaction 
on the loop velocities,
especially when they are relativistic.
We also see that the gravitational decay of loop energy (thin solid)
damps the overall amplitude of the power spectrum (dashed)
by about a factor of 3.
We notice that between the long-string correlation scale
$k_\xi \approx 20/\eta$ (see subsection~\ref{coss})
and the scale $k_{\rm L}\approx 10 k_\xi$,
the slope of the loop spectrum (thin solid) is exactly the same as that of
the long-string spectrum $n \approx -2.25$.
We have shown in subsection~\ref{cosl}
that this close correspondence is due to copious loop
production being strongly correlated with long 
string intercommuting events and the collapse of highly curved long-string
regions,\cite{AS1} that is, near the strongest long-string perturbations.
Moreover, these correlations persist in time with the subsequent 
motion of loops and long strings lying preferentially in the same directions,
a phenomenon which  has been verified by observing animations of
string network evolution.
These correlations between loops and long strings, however, have a lower 
cutoff represented by the mean loop spacing $d_{\rm L}\sim k_{\rm L}^{-1}$.
Below $d_{\rm L}$, the effects of individual filaments swept out by moving 
loops can be identified. In terms of the power spectrum,  for $k<k_{\rm L}$
the loops are 
strongly correlated with the 
long strings and therefore reinforce the wake-like perturbations, while 
for $k>k_{\rm L}$ their filamentary perturbations increase the spectral 
index by about one to $n \approx -1.25$; this change is expected on geometrical 
grounds.

\begin{figure}[htbp]
\vspace*{13pt}
\centering 
\leavevmode\epsfxsize=4in \epsfbox{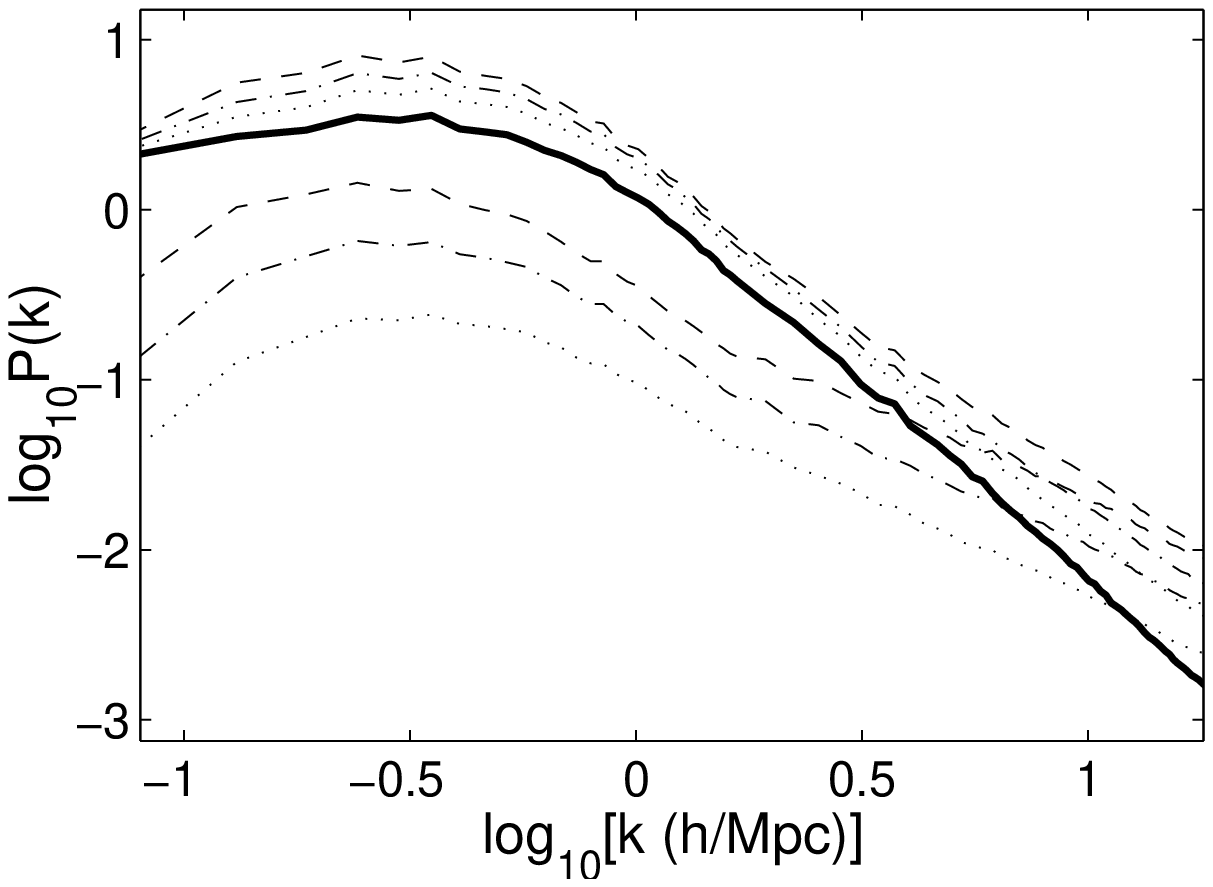}\\ 
\vspace*{13pt}
 \fcaption
 {The lower set of 3 lines are ${\cal P}_{\rm L}(k)$ for 
  $f=0.5$ (dotted), $1$ (dot-dashed) and $2$ (dashed).
  ${\cal P}_\infty (k)$ is plotted as a solid line.
  The upper set of lines are ${\cal P}_{\rm tot}(k)$
  with corresponding line styles and $f$ values
  to the lower set of lines.
  }
  \label{figure3}
\end{figure}

In figure \ref{figure3} we plot the power spectra of
density perturbations seeded
by long strings ${\cal P}_\infty (k)$,
by small loops ${\cal P}_{\rm L}(k)$,
and by both loops and long strings ${\cal P}_{\rm tot}(k)$.
The dynamic range here extends from  $0.6$ to $7.5\eta_{\rm eq}$.
As expected
${\cal P}_{\rm L}(k)$
scales more moderately than $f^2$
but more strongly than $f$ (see (\ref{rhol})).
It is also apparent that
the perturbations induced by long strings and by loops
are positively correlated with
${\cal P}_{\rm tot}(k) > {\cal P}_{\rm L}(k)+{\cal P}_\infty (k)$
throughout the whole scale range.
This positive correlation between loops and long strings
boosts the large-scale ${\cal P}_\infty (k)$ by a factor of
$1.5$, $1.8$ and $2.2$ to reach ${\cal P}_{\rm tot}(k)$
for $f=0.5, 1$ and $2$ respectively,
even if ${\cal P}_{\rm L}(k)$ is a relatively small
fraction of ${\cal P}_\infty (k)$ on these scales.

\begin{figure}[htbp]
\vspace*{13pt}
\centering 
\leavevmode\epsfxsize=4in \epsfbox{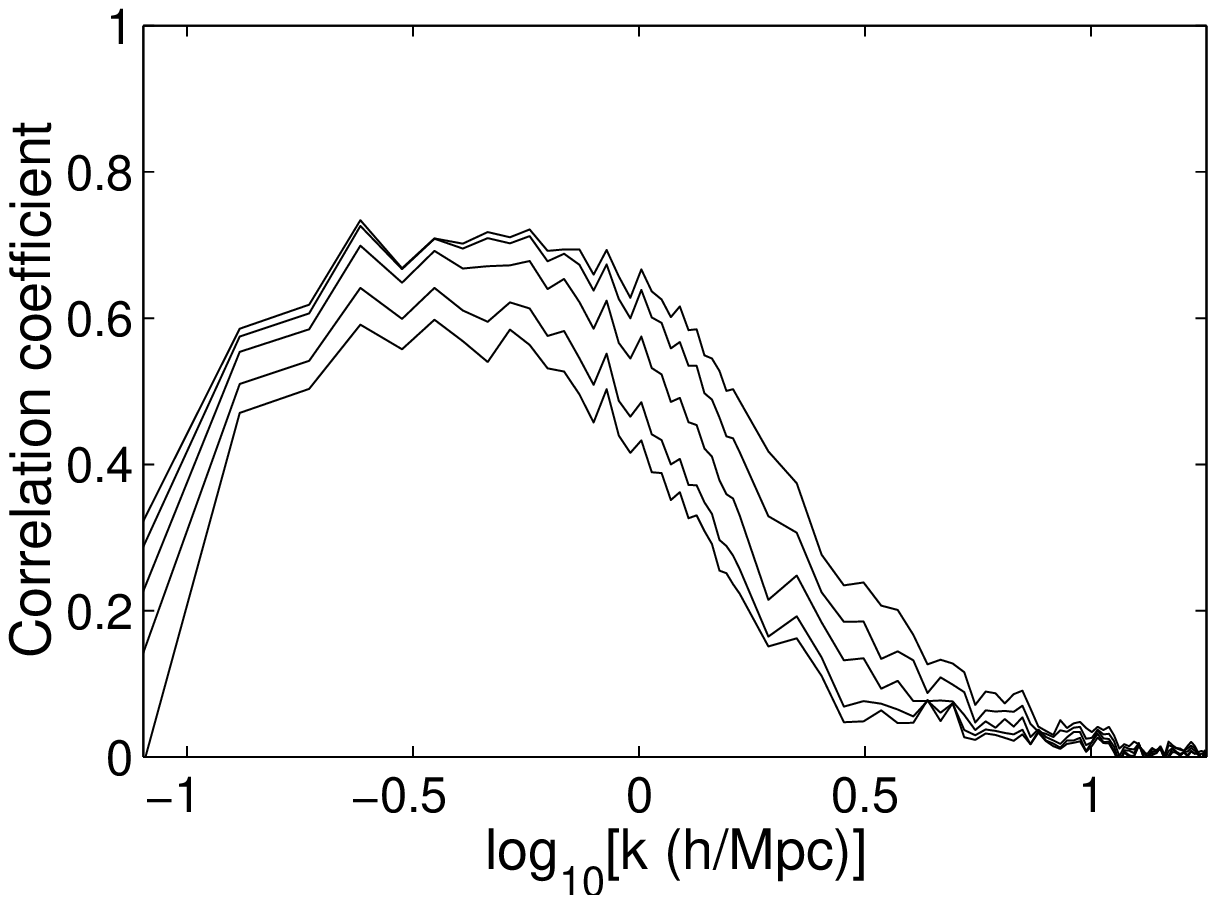}\\ 
\vspace*{13pt}
 \fcaption
 {The correlation coefficient between the long-string and
   loop induced perturbations,
   with $f=0.5, 1, 2, 4, 6$ (downwards).
   }
  \label{figure5}
\end{figure}

Figure \ref{figure5} shows the correlation coefficient
between the long-string and loop induced perturbations:
\begin{equation}
  \label{corrcoef}
  {\cal K} = 
  \frac{\langle \delta_\infty \delta_{\rm L} \rangle}
  {\langle \delta_\infty^2 \rangle^{1/2}
    \langle  \delta_{\rm L}^2  \rangle^{1/2}}\ .
\end{equation}
We see that
long strings and loops are strongly positively correlated on large scales,
but weakly correlated on small scales
where the loops dominate the perturbations (see figure \ref{figure3}).
The threshold $k_{\rm t}$ between these two regimes
must be significantly larger than $k_{\rm L}$
because, for $k < k_{\rm L}$, ${\cal P}_{\rm L}(k)$ is well below
and roughly parallel to ${\cal P}_\infty (k)$
(see figures \ref{figure2} and \ref{figure3}).
We also verify that ${\cal P}_{\rm tot}(k)/{\cal P}_\infty (k)$ is approximately
a constant for $k < k_{\rm L}\lsim k_{\rm t}$, which
 again provides strong evidence for the fact that loops behave 
as part of the long-string
network on large scales.

\begin{figure}[htbp]
\vspace*{13pt}
  \centering 
  \leavevmode\epsfxsize=4in \epsfbox{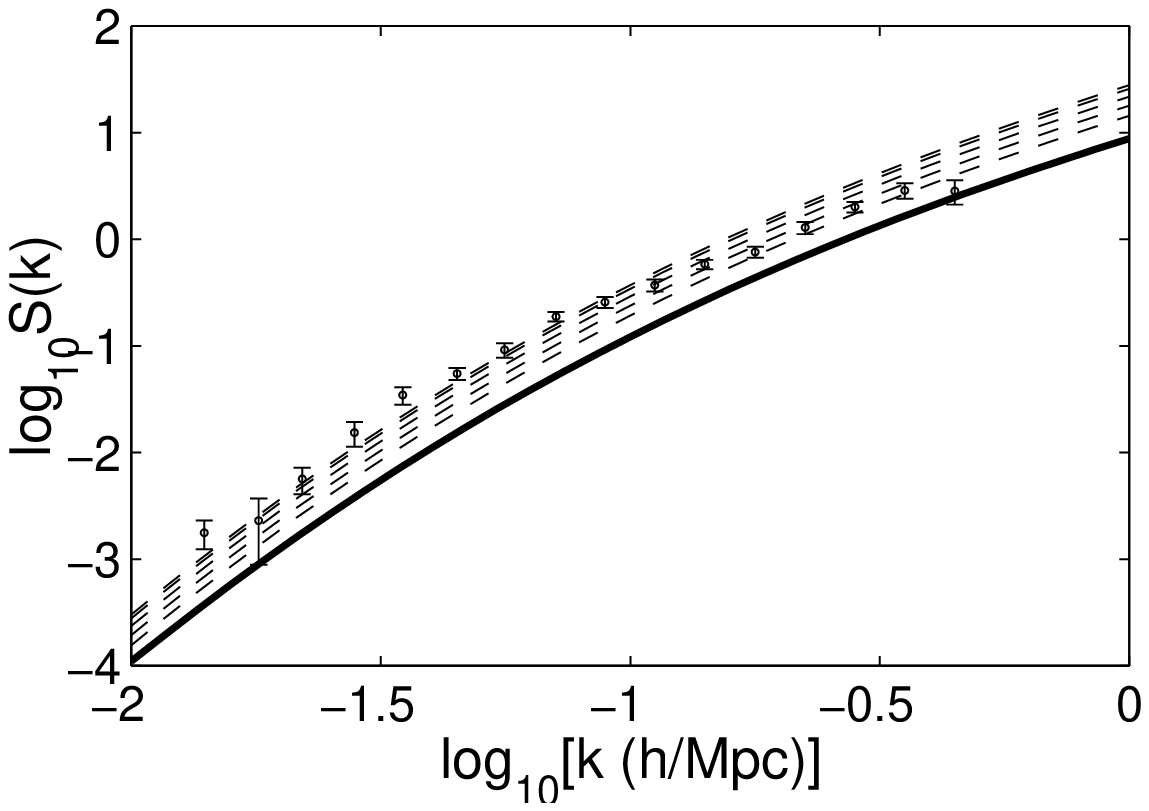}\\ 
\vspace*{13pt}
  \fcaption
  {Comparison of the observational power spectrum\cite{PD} with
    ${\cal P}_\infty(k)$ (solid), and
    ${\cal P}_{\rm tot}(k)$ for $f=0.5,1,2,4,6$ (dashed, upwards),
    with a full dynamic range.
    }
  \label{figure4}
\end{figure}

Given these properties of the string power spectra, one can easily 
construct a semi-analytic model for
${\cal P}_{\rm tot}(k)$ as for ${\cal P}_\infty (k)$.\cite{AveShe2,AveShe3}
We first multiply the structure function
${\cal F}(k,\eta)$ of ${\cal P}_\infty (k)$ by ${\cal J}(\eta, f)$
to account for the boost ${\cal P}_{\rm tot}(k)/{\cal P}_\infty (k)$
on large scales ($k < k_{\rm t}$):
\begin{equation}
  \label{J}
  {\cal J}(\eta, f) =
  \bar{\cal J}(f)
  \left[
    1.4+\frac{0.6}{1+\left(\frac{\eta}{7\eta_{\rm eq}}\right)^{1.5}}
  \right]\,,
\end{equation}
where $\bar{\cal J}(f)= 0.806,\, 1,\, 1.22,\, 1.44$ and $1.56$
for $f=0.5,\, 1,\, 2,\, 4$ and $6$ respectively.
We then multiply it again by a numerically verified form
\begin{equation}
  \label{H}
  {\cal H}(k,\eta,f)=\left[1+\left(\frac{k}{k_{\rm t}}\right)^4\right]^{1/4},
\end{equation}
to account for the turnover for $k>k_{\rm t}(\eta,f)$.
${\cal J}(\eta, f)$ is calibrated phenomenologically from simulations
deep in the
radiation era through to those deep in the matter era.
In the pseudo-scaling regime for the loop size,
$k_{\rm t}$ is revealed to be at least $10k_\xi\approx 200/\eta$
depending on $f$.
Thus we can carry out a full-dynamic-range integration to obtain
${\cal P}_{\rm tot}(k)$.
In figure \ref{figure4}
we compare this ${\cal P}_{\rm tot}(k)$ and ${\cal P}_\infty(k)$
with observations.\cite{AveShe2,AveShe3}
The background cosmology is $\Omega_{\rm c}=0.15$, $\Omega_\Lambda=0.85$ and
$h=0.7$, and
we have used the COBE normalization
$G \mu =1.7 \times 10^{-6}$  throughout.\cite{ACDKSS}
Since loops are point-like and they have little impact through the
Kaiser-Stebbins effect on COBE-scale CMBR anisotropies,
we expect this normalization to be very weakly dependent on the value of $f$;
indeed, loops were found to produce a negligible effect in Ref.\cite{ACSSV}.
Thus we see from figure \ref{figure4} that
for $f \gsim 0.5$, loops can contribute significantly to the total power 
spectrum and ease the large-scale bias problem 
seen previously.\cite{AveShe2,AveShe3,against}
Definite conclusions, therefore, 
about biasing in cosmic string models will need further advances in 
determining the magnitude of the parameter $f$, while all future large-scale
structure simulations will now require the 
the inclusion of loops.

These additional complications in modeling cosmic string structure 
formation are most obvious on small scales, where even higher resolution 
and large dynamic range simulations will be required.  
Within the present pseudo-scaling regime for loop size,
we know that $k_{\rm t}\gsim k_{\rm L} \gsim 10k_\xi$
as shown in figure~\ref{figure2} and discussed previously.
Taking this extreme minimum $k_{\rm t}=10k_\xi$, then, we find that
the semi-analytic model over the full dynamic range gives
at most a $2\%$ difference in ${\cal P}_{\rm tot}(k)$ 
for $k < 1h{\rm Mpc}^{-1}$ when the filament term ${\cal H}(k,\eta,f)$ is 
excluded from 
${\cal F}(k,\eta)$
(for $\Omega_{\rm c}=0.15$, $\Omega_\Lambda=0.85$ and $h=0.7$).
This means that 
although the simulations described in this letter are already on 
the verge of present computer capabilities,
a further detailed study on small scales will improve only the overall
normalization of ${\cal P}_{\rm tot}(k)$
but not the shape revealed here if $f$ is constant throughout.
We notice that although $f$ should be constant in both deep radiation
and deep matter eras due to the scaling of string network,
it may vary over the transition era.
This time dependence in $f$ could slightly alter the shape of
${\cal P}_{\rm tot}(k)$ given here, and
this deserves further investigation on cosmic string evolution.
We also note that advances in understanding loop formation mechanisms 
will also be crucial in quantifying the importance of 
the gravitational radiation background 
emitted by a cosmic string network and its effect on large-scale 
structure and CMBR anisotropies.\cite{AveCal2}

\subsection{{\bfit CMBR anisotropies}}
A key feature of all these string-induced power spectra is the influence
of the slow relaxation to the matter era string density from the much
higher radiation string density, which has an effective
structure function ${\cal F}(k,\eta)$ in (\ref{FTR})
with approximately 2.5 times more power than the matter era version.
Even by recombination in an $\Omega_{\rm m0}=1$ ($h=0.7$) cosmology, the string
density $\zeta(\eta)$ is more than twice its asymptotic matter era value to which we
normalize on COBE scales (see Fig.~\ref{str_density}).
This implies that the string model provides
higher than expected large-scale power around 100$h^{-1}$Mpc and below.
Interestingly, this can also be expected to produce a significant
Doppler-like peak on small angle CMBR scales, an effect noted in
Ref.\cite{ACDKSS} but not observed because matter era strings were employed.
In Ref.\cite{PST}, global strings were evolved through the
transition era without a strong Doppler peak
emerging; however, this depended on a field theory simulation of limited
dynamic range.  Since global strings on cosmological scales behave
more like local strings, the linear power spectrum approach presented here
potentially should more accurately represent the actual
global string power spectrum.
Recent work in Ref.\cite{ABR2} using a phenomenological semi-analytic approach
confirms that such Doppler-like features can
result from significant non-scaling effects during the transition
era.
Especially in the light of seeing the first Doppler peak 
in the recent observations from MAXIMA-1 \cite{MA1} and BOOMERANG \cite{B98},
the cosmic-string-predicted first peak requires 
more careful and detailed study.
We should also note that the currently observed CMB power spectrum
has quite a narrow first peak and no evident secondary peaks \cite{Jaffe2001},
and this may imply a hybrid scenario of structure formation,
which is a mix of both inflation and topological defects
\cite{Bouchet2000}.

\subsection{{\bfit Uncertainties}}
Finally, it is appropriate to comment on the key uncertainties affecting
these calculations.  To summarize at the outset, these uncertainties
primarily affect the amplitude of the string power spectrum, rather
than its overall shape which is a more robust feature.  The most recent COBE
string normalization is $G\mu_6 \approx 1.7$,\cite{ACDKSS}
which is at variance with a previous COBE normalization $G\mu_6\approx 1$
using the same string simulation.\cite{ACSSV}  This remains to be
satisfactorily resolved at the time of writing, but systematic
relativistic effects seem likely to have affected the earlier result.
Next there is uncertainty implementing the effective compensation scale at
which the perturbations are cut-off on large scales, which we have discussed
already.
Thirdly, there are uncertainties in the long-string energy density as we have
addressed in section~\ref{coss}. This causes an uncertainty of about $10\%$ in
the resulting power spectrum.
All these uncertainties add up to a factor of about 2 in the final power
spectrum while remaining the overall shape largely unchanged.
Finally, the high correlation between loops and long strings always boosts
the power spectrum up to a factor of 3,
if the loop lifetime is not much smaller than the Hubble time.
Therefore, to include the effect from loops has become necessary for any
further development in studying structure formation by cosmic string models.


\section{{Conclusion}}
\label{conclu}

In this paper we have described the results of high-resolution
numerical simulations of structure formation
seeded by a local cosmic string network
with a large dynamical range reporting at length,
for the first time,
on the effect of loops and
modifications due to the radiation-matter transition.
In the regime of large-scale structure formation, the most serious problem with
both cosmic strings and the standard CDM model (inflation) is that
they produce too much small-scale power and
insufficient large-scale power.
From the results and discussion in section
\ref{tca}, we see for cosmic strings this problem can be relaxed a great deal
by considering an open universe or the existence of a small cosmological
constant.
There is another alternative which can improve the situation.
Mixed dark matter scenarios can be employed by using
(\ref{G}) to smear out part of the small-scale perturbations,
although we see the pure HDM power spectrum requires a strongly
scale-dependent bias either on small or large scales.

As to the shortage of the overall amplitude in the string power spectrum when
compared with observations,
we have seen that this can be overcome by including cosmic string loops
with a lifetime comparable to the Hubble time or greater.
We have shown that on large scales the loops behave 
like part of the long-string network and
can therefore contribute significantly to the 
total power spectrum of density perturbations.
At present,
the typical size and lifetime of loops formed by a string network 
remains to be studied in more detail; the problem is both computationally 
and analytically challenging.
However, within the scale range of interest further developments in this area
have the potential to affect the overall amplitude of the spectrum,
while leaving the shape largely unchanged.

We conclude that although more work needs to be done, 
notably in improving the implementation of compensation
and in the study of loops,
the picture which emerges for the large-scale structure power
spectrum is encouraging.



\nonumsection{Acknowledgments}
\noindent
We would like to thank Neil Turok, Carlos Martins,
Robert Caldwell, Albert Stebbins, Richard Battye and Pedro Viana 
for useful conversations.
J.~H.~P.~W.\ is funded by 
NSF KDI Grant (9872979) and
NASA LTSA Grant (NAG5-6552).
P.\ P.\ A.\ is funded by JNICT (Portugal) under the `Program PRAXIS XXI'
(PRAXIS XXI/BPD/9901/96).
B.\ A.\ acknowledges support from NSF grant PHY95-07740.
This work was performed on COSMOS, the Origin2000 owned by the UK
Computational Cosmology Consortium, supported by Silicon Graphics/Cray
Research,
HEFCE and PPARC.

\nonumsection{References}




\appendix{. Conventions and Cosmological Background Dynamics}
\label{conbac}

\noindent
We assume that the universe is homogeneous and isotropic,
and is filled with two fluids, radiation and dark matter, whose stress-energy
tensors are also homogeneous and isotropic on average.
We will use subscripts ${\rm m,\ c,\ h,\ r}$
to denote dart matter, cold dark matter (CDM), hot dark matter (HDM), and
radiation respectively,
a subscript ``eq'' to denote the epoch of radiation-matter density equality,
and a subscript ``0'' to denote the epoch today.
We will ignore the contribution of the defect field stress energy,
which is always much smaller than
the total energy density of radiation and matter.
We use Greek letters as space-time indices (e.g.\ $\mu=0,1,2,3$),
and Latin letters as spatial indices (e.g.\ $i=1,2,3$).
The metric signature is ($- + + +$) and
units used are normalized to $\hbar=c=k_B=1$.

\begin{figure}[t]
  \centering\epsfig{figure=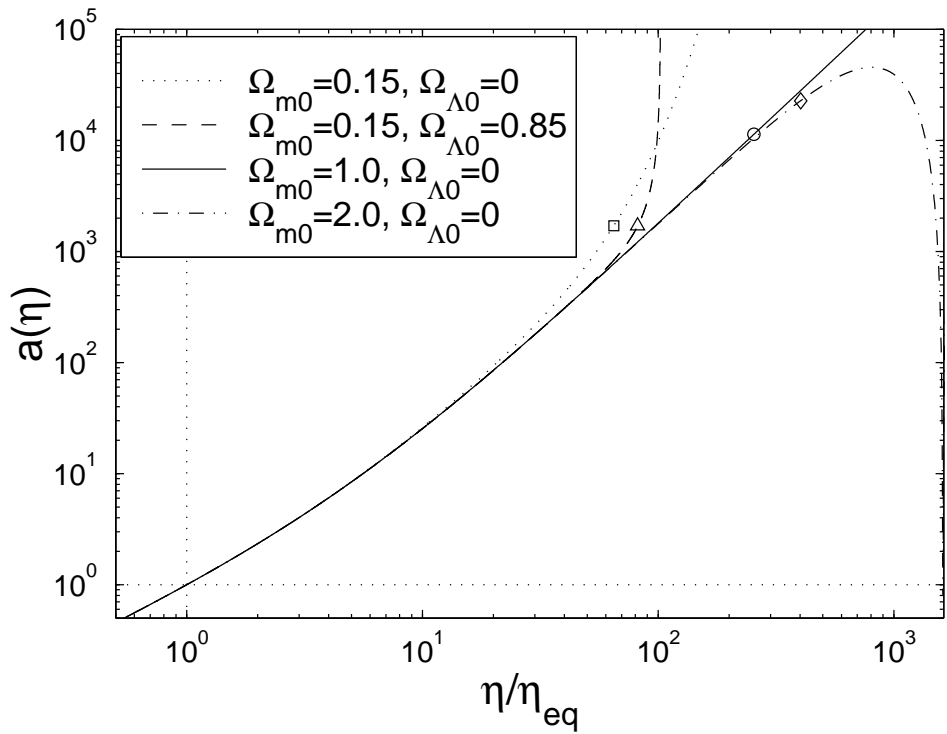, width=5in}
  \caption[]
  {The effect of the curvature and the cosmological constant in
    the background cosmology. Plotted are the exact solutions
    of scale factor $a(\eta)$ (see text for details).
    The square, triangle, circle and diamond mark the epoch today
    (with $h=0.7$) for different models.
    }
  \label{scalefactor}
\end{figure}

In a flat Friedmann-Robertson-Walker (FRW) universe with only radiation
and matter components which evolve independently and adiabatically,
the scale factor $a(\eta)$ is determined by the unperturbed Einstein equation
\begin{equation}
  \label{fried}
  \dot{a}^2 + K a^2 =
  \frac{8\pi G \rho_{\rm m0} a_0^3}{3} (1+a) + \frac{\Lambda}{3}a^4,
\end{equation}
where
a dot represents a derivative with respect to the conformal time $\eta$,
$K$ is the curvature, $\rho_{\rm m}$ is the matter energy density,
$\Lambda$ is the cosmological constant,
and we have normalized $a_{\rm eq}=1$.
If we define
$\Omega_{\rm m}=8\pi G \rho_{\rm m}/3H^2$, 
$\Omega_{\rm r}=8\pi G \rho_{\rm r}/3H^2=8\pi G \rho_{\rm m}/3aH^2$, 
$\Omega_\Lambda=\Lambda/3H^2$, and
$\Omega_K=-K/a^2H^2$, where $H=\dot{a}/a^2$ is the Hubble parameter,
then we have from (\ref{fried}) that
$\Omega_{\rm m}+\Omega_{\rm r}+\Omega_\Lambda+\Omega_K=1$ and
\begin{equation}
  \label{omega}
  \frac{\Omega_{\Lambda 0}}{\Omega_{\rm m0}} =
  \frac{\Lambda}{8\pi G \rho_{\rm m0}}, \quad
  \frac{\Omega_{K0}}{\Omega_{\rm m0}} =
  \frac{-3K}{8\pi G \rho_{\rm m0}a_0^2}.  
\end{equation}
We notice that $\Omega_{\rm r0}/\Omega_{\rm m0}=a_0^{-1}\ll 1$.
We also define
\begin{equation}
  \label{aABC}
  A=\frac{2(\sqrt{2}-1)}{\eta_{\rm eq}}\,, \quad
  B=\frac{\Omega_{K0}}{\Omega_{\rm m0}a_0}\,, \quad
  C=\frac{\Omega_{\Lambda 0}}{\Omega_{\rm m0}a_0^3}\,,
\end{equation}
where $B,C\ll 1$ according to the current observational results.
Thus we can rewrite (\ref{fried}) as
\begin{equation}
  \label{eta_a}
  \left(\frac{da}{d\eta}\right)^2 =
  \bar{A}^2 ( 1 + a +  B a^2 + C a^4 )\,,
\end{equation}
where
\begin{equation}
  \label{barA}
  \bar{A} = \frac{1}{\eta_{\rm eq}}\int_0^1\frac{da}{(1+a+Ba^2+Ca^4)^{1/2}}
  \approx A\,.
\end{equation}
(\ref{eta_a}) can then be numerically evaluated with certain choice of
$\Omega_{\rm m0}$, $\Omega_{\Lambda 0}$ and $\Omega_{K0}$.
Assuming three species of neutrinos and
because at $\eta_{\rm eq}$ both the curvature and the cosmological constant
terms are negligible in (\ref{fried}),
we obtain
$a_0=23219 \Omega_{\rm m0}h^2$,
$\eta_{\rm eq}= 16.3098\ (\Omega_{\rm m0}h^2)^{-1} {\rm Mpc}$ today,
and the physical time
$t_{\rm eq}=3.4058 \times 10^{10} (\Omega_{\rm m0}h^2)^{-2}{\rm sec}$.
In certain cases, (\ref{eta_a}) can be exactly solved:
\begin{romanlist}
\item $K=\Lambda=0$ (i.e.\ $\Omega_{\rm m0}=1, \Omega_{\Lambda 0}=0$):
  \begin{eqnarray}
    a(\eta) & = & A^2\eta^2/4 + A\eta \,,
    \label{aflat} \\
    t(\eta) & = & A^2\eta^3/12 + A\eta^2/2 \,,
    \label{tflat}
  \end{eqnarray}
  which give $\eta_{\rm eq}=3t_{\rm eq}/\sqrt{2}$.
\item $K<0,\ \Lambda=0$ (i.e.\ $\Omega_{\rm m0}<1, \Omega_{\Lambda 0}=0$):
  \begin{eqnarray}
    a(\eta) & = &\frac{1}{2B}\left[
      \cosh{(\bar{A}\sqrt{B}\eta)}
      + 2\sqrt{B}\sinh{(\bar{A}\sqrt{B}\eta)} -1
    \right] \,,
    \label{aopen}    \\
    t(\eta) & = & \frac{1}{\bar{A}B}\left[
      \cosh{(\bar{A}\sqrt{B}\eta)}
      + \frac{1}{2\sqrt{B}}\sinh{(\bar{A}\sqrt{B}\eta)}
      - \frac{\bar{A}\eta}{2} - 1
    \right] \,.
    \label{topen}
  \end{eqnarray}
\item $K>0,\ \Lambda=0$ (i.e.\ $\Omega_{\rm m0}>1, \Omega_{\Lambda 0}=0$):
  \begin{eqnarray}
    a(\eta) & = & \frac{1}{2B}\left[
      \cos{(\bar{A}\sqrt{-B}\eta)}-2\sqrt{-B}\sin{(\bar{A}\sqrt{-B}\eta)}-1
    \right] \,,
    \label{aclose}\\
    t(\eta) & = & \frac{1}{\bar{A}B}\left[
      \cos{(\bar{A}\sqrt{-B}\eta)}
      + \frac{1}{2\sqrt{-B}}\sin{(\bar{A}\sqrt{-B}\eta)}
      - \frac{\bar{A}\eta}{2} -1
    \right] \,.
    \label{tclose}    
  \end{eqnarray}
\end{romanlist}
We notice that
at early times (\ref{aopen},\ref{topen}) and (\ref{aclose},\ref{tclose})
decay to (\ref{aflat},\ref{tflat}) exactly.
At late times
(\ref{aflat},\ref{tflat}), (\ref{aopen},\ref{topen}) and (\ref{aclose},\ref{tclose}) give the
asymptotic forms $a \propto \eta^2$, $\exp(\bar{A}\sqrt{B}\eta)$ and
$1-\cos(\bar{A}\sqrt{-B}\eta)$ (before the recollapse), or
$a \propto t^{2/3}$, $a \propto t$ and $a \propto 1-\cos[2\bar{A}(-B)^{3/2}t]$
(before the recollapse) respectively.
Figure \ref{scalefactor} shows some examples of these solutions.

\appendix{. Power Spectrum Conventions}
\label{conpow}
\noindent
We define the Fourier transform as:
\begin{equation}
  \widetilde{f}({\bf k})=\frac{1}{V}\int d^3{\bf x} f({\bf x}) e^{i{\bf k \cdot x}}\,,
\end{equation}
where ${\bf k}$ is the wave vector,
${\bf x}$ the physical coordinates,
and the integration taken over a large volume $V$.
We will use
a tilde~$\;\widetilde{ }\;$~to denote the Fourier transform of a function.
The power spectrum ${\cal P}(k)$ for a mass density distribution
$\rho({\bf x})$
is then defined as the spherically symmetric Fourier transform of the
autocorrelation function $\xi(r)$ of density contrast
$\delta({\bf x})=\delta\rho/\rho_{\rm av}$:
\begin{equation}
\label{powerspectrum}
  {\cal P}(k)
  = \frac{1}{2\pi^2} \int \xi(r)\frac{\sin{kr}}{k}rdr
  = {V\langle|\widetilde{\delta}_{\bf k}|^2\rangle \over 8\pi^3}\,,
\end{equation}
where $k=|{\bf k}|$,
$r$ is the correlation distance, and
$\widetilde{\delta}_{\bf k}$ is the Fourier transform of the perturbations
$\delta({\bf x})$.
A useful quantity when comparing numerical results with observation is the
dimensionless ``spectrum of the matter density contrast'', which is given by
contributions of $\widetilde{\delta}_k$ over a logarithmic interval $\triangle k/k\sim 1$:
\begin{equation}
\label{spectrum}
S(k)
= \left\langle\left(\frac{\delta\rho}{\rho_{\rm av}}\right)^2\right\rangle_k
\approx \frac{Vk^3}{2\pi^2}\langle|\widetilde{\delta}_k|^2\rangle
= 4\pi k^3 {\cal P}(k) \,.
\end{equation}
This quantity is dimensionless and
gives the rms density fluctuation on a particular
length-scale $l=2\pi/k$.
Another useful statistic is the mass fluctuation amplitude at
certain length-scale $\lambda$,
i.e.\ the standard deviation of the mass density distribution $\sigma_\lambda$.
We convolve the density contrast $\delta({\bf x})$
with a spherical window of radius $\lambda$
to obtain the variance:
\begin{equation}
  \label{sigma_lambda}
  \sigma_\lambda^2
  =
  4\pi \int |w(k\lambda)|^2 k^2 {\cal P}(k) dk
  \;\;{\rm with}\;\;
  w(x)=\frac{3(\sin x-x\cos x)}{x^3}\,.
\end{equation}
We note that $w(x)$ is the Fourier transform of a spherical window
and ${\cal P}(k)$ in (\ref{sigma_lambda}) can be calculated from
(\ref{powerspectrum}).


\end{document}